%% file: MM_P009a_v2.tex
\documentclass[12pt]{article}

\input{PackagesDefinitionsSettings}

\newcommand{\authorname}{Marko Mlikota}
\newcommand{\authorinstitution}{Geneva Graduate Institute}

\begin{document}

\input{part_TitlepageToC}


\input{part_Intro}
\input{part_Model}

\input{part_Inference}

\input{part_App1}

\input{part_Conclusion}


\begin{spacing}{1.0}
	\bibliography{../../P009a}
\end{spacing}




\begin{appendix}
	\markright{This Version: \today }
	\renewcommand{\theequation}{A.\arabic{equation}}
	\setcounter{equation}{0}

	\renewcommand*\thetable{A-\arabic{table}}
	\setcounter{table}{0}
	\renewcommand*\thefigure{A-\arabic{figure}}
	\setcounter{figure}{0}

	\addtocontents{toc}{\protect\setcounter{tocdepth}{1}}









	\input{part_appendix_model}

	\input{part_appendix_inference}

	\input{part_appendix_app1}

\end{appendix}

\end{document}

%% file: PackagesDefinitionsSettings.tex
\usepackage{geometry} 	
\usepackage{setspace}   	
\usepackage{amsmath,amssymb,amsfonts,amsthm}		
\usepackage[T1]{fontenc}
\usepackage{libertinus}
\usepackage{libertinust1math}
\usepackage{bm}                                 

  \everydisplay{%
    \abovedisplayskip=10pt%
    \belowdisplayskip=10pt%
    \abovedisplayshortskip=6pt%
    \belowdisplayshortskip=6pt%
  }
\usepackage{enumitem}

\usepackage{graphicx}
\usepackage{subcaption}
\usepackage{float}      

\usepackage{fancyhdr}			

\usepackage{xcolor}				
\usepackage{soul} 				

\usepackage{hyperref}			
	\hypersetup{colorlinks=true, citecolor = black, linkcolor = black, urlcolor=black}

\usepackage{tocloft}			

\usepackage{natbib}
\usepackage[USenglish]{babel}
\usepackage[style=iso]{datetime2} 

\usepackage[capitalise]{cleveref}

\usepackage{dutchcal} 

\usepackage[textsize=tiny, disable]{todonotes} 
\usepackage{marginnote}
\newcommand{\mytodo}[2][]{{
 \let\marginpar\marginnote
 \reversemarginpar
 \todo[#1]{#2}}}

\usepackage{booktabs} 
\usepackage{multicol}
\usepackage{multirow}

\usepackage{array}
\newcolumntype{N}{c@{}S}
\newcolumntype{H}{>{\setbox0=\hbox\bgroup}c<{\egroup}@{}}
\newcolumntype{L}[1]{>{\raggedright\let\newline\\\arraybackslash\hspace{0pt}}m{#1}}
\newcolumntype{C}[1]{>{\centering\let\newline\\\arraybackslash\hspace{0pt}}m{#1}}
\newcolumntype{R}[1]{>{\raggedleft\let\newline\\\arraybackslash\hspace{0pt}}m{#1}}

\usepackage[round-mode=places, table-number-alignment=center]{siunitx}

\usepackage{lscape} 

\usepackage[raggedright]{titlesec} 
\makeatletter
\renewcommand*{\@seccntformat}[1]{\csname the#1\endcsname\hspace{20pt}} 
\makeatother

\titleformat*{\section}{\fontsize{16}{16}\bfseries}

\titlespacing*{\paragraph}
  {0pt}
  {0.25em}
  {0.5em}

\titlespacing*{\section}
  {0pt}
  {1em}
  {0.5em}

\titlespacing*{\subsection}
  {0pt}
  {0.5em}
  {0.25em}

\titlespacing*{\subsubsection}
  {0pt}
  {0.25em}
  {0.125em}

\definecolor{darkpowderblue}{rgb}{0.0, 0.2, 0.6}
\definecolor{carmine}{rgb}{0.59, 0.0, 0.09}
\definecolor{cadmiumgreen}{rgb}{0.0, 0.42, 0.24}
\definecolor{burntorange}{rgb}{0.8, 0.33, 0.0}

\newcommand{\cbr}[1]{\left\{ {#1} \right\}}
\newcommand{\sbr}[1]{\left[ {#1} \right]}
\newcommand{\br}[1]{\left( {#1} \right)}

\newcommand{\myquote}[1]{``{#1}"}

\newcommand{\mean}{\mathbb{E}}
\newcommand{\var}{\mathbb{V}}

\newcommand{\reals}{\mathbb{R}}
\newcommand{\integers}{\mathbb{Z}}
\newcommand{\naturals}{\mathbb{N}}
\newcommand{\rationals}{\mathbb{Q}}
\newcommand{\complexnumbers}{\mathbb{C}}

\newcommand{\wq}{\omega_q}

\newcommand{\ii}{\mathrm{i}}

\newcommand{\one}[1]{\mathbf{1}\cbr{ #1 }}

\newcommand{\bigabslines}[1]{  \left| #1 \right| }

\newcommand{\calF}{\mathcal{F}}
\newcommand{\calG}{\mathcal{G}}
\newcommand{\calI}{\mathcal{I}}

\newcommand{\convdist}{\overset{d}{\rightarrow}}

\newcommand{\covar}{\text{Cov}}

\newcommand{\bmat}[1]{\begin{bmatrix} #1 \end{bmatrix}}

\newcommand{\vecop}{\mathrm{vec}}

\newcommand\blfootnote[1]{%
  \begingroup
  \renewcommand\thefootnote{}\footnote{#1}%
  \addtocounter{footnote}{-1}%
  \endgroup
}

\newcommand{\citetwo}[2]{%
  \citet{#1} and \citet{#2}}
\newcommand{\citethree}[3]{%
  \citet{#1}, \citet{#2} and \citet{#3}}
\newcommand{\citefour}[4]{%
  \citet{#1}, \citet{#2}, \citet{#3} and \citet{#4}}

\newcommand{\thmspaceAbove}{5pt plus 2pt minus 2pt}
\newcommand{\thmspaceBelow}{5pt plus 2pt minus 2pt}
\newtheoremstyle{myplain}%
  {\thmspaceAbove}
  {\thmspaceBelow}
  {\itshape}
  {}
  {\bfseries}
  {.}
  { }
  {}
\theoremstyle{myplain}

\newtheorem{corollary}{Corollary}

\newtheorem{proposition}{Proposition}

\newtheorem{lemma}{Lemma}
\newtheorem{algo}{Algorithm}

\newtheorem{conjecture}{Conjecture}
    \crefname{conjecture}{Conjecture}{Conjectures}

\makeatletter
\newtheorem*{rep@theorem}{\rep@title}
\newcommand{\newreptheorem}[2]{%
\newenvironment{rep#1}[1]{%
 \def\rep@title{#2 \ref{##1}}%
 \begin{rep@theorem}}%
 {\end{rep@theorem}}}
\makeatother

\newreptheorem{theorem}{Theorem}
\newreptheorem{proposition}{Proposition}
\newreptheorem{corollary}{Corollary}
\newreptheorem{lemma}{Lemma}

\newcommand{\myproposition}[3]{
\noindent 
\begin{proposition}[#1] \label{#2} \hspace*{0cm}\\
	#3
\end{proposition}
}

\newcommand{\myrepproposition}[3]{
\noindent 
\begin{repproposition}{#2}[#1] \hspace*{0cm}\\
	#3
\end{repproposition}
}

\newcommand{\propositionmarker}[1]{%
  \refstepcounter{proposition}%
  \label{#1}%
}

\newcommand{\mylemma}[3]{
\noindent
\begin{lemma}[#1] \label{#2} \hspace*{0cm}
	#3
\end{lemma}
}

\newcommand{\myproof}[1]{
    \noindent {\bf Proof:} #1 
}

\newcommand{\myproofof}[2]{
    \noindent {\bf Proof of \cref{#1}:} #2
}

\newcounter{propositionpart}[proposition]

\makeatletter
\renewcommand{\p@propositionpart}{\theproposition}

\makeatother
\crefalias{propositionpart}{proposition}

\newcommand{\propositionpartmarker}[1]{%
  \refstepcounter{propositionpart}%
  \label{#1}%
}

\newtheorem{assumptionARpq}{Assumption}
\newtheorem{assumptionNVARpq}{Assumption}
\newtheorem{assumptionNVARpqWC}{Assumption}

\crefname{assumptionARpq}{Assumption}{Assumptions}
\Crefname{assumptionARpq}{Assumption}{Assumptions}

\crefname{assumptionNVARpq}{Assumption}{Assumptions}
\Crefname{assumptionNVARpq}{Assumption}{Assumptions}

\crefname{assumptionNVARpqWC}{Assumption}{Assumptions}
\Crefname{assumptionNVARpqWC}{Assumption}{Assumptions}

\newcommand{\myassumptionARpq}[3]{%
    \vspace{-2pt}
    \noindent
    \begin{assumptionARpq}[#1]\label{#2}
        #3
    \end{assumptionARpq}
    \vspace{-5pt}
}

\newcommand{\myassumptionNVARpq}[3]{%
    \vspace{-2pt}
    \noindent
    \begin{assumptionNVARpq}[#1]\label{#2}
        #3
    \end{assumptionNVARpq}
    \vspace{-5pt}
}

\newcommand{\myassumptionNVARpqWC}[3]{%
    \vspace{-2pt}
    \noindent
    \begin{assumptionNVARpqWC}[#1]\label{#2}
        #3
    \end{assumptionNVARpqWC}
    \vspace{-5pt}
}

\setlist[itemize]{itemsep=2pt, topsep=3pt, parsep=0pt, partopsep=0pt}
\setlist[enumerate]{itemsep=2pt, topsep=3pt, parsep=0pt, partopsep=0pt}

\newcommand*{\plotPath}{Plots}%

\newcommand{\plotSizeSingle}{0.5}
\newcommand{\plotSizeDouble}{0.35}
\newcommand{\plotSizeTriple}{0.3}
\newcommand{\plotSizeQuadruple}{0.24}

\newcommand{\figspaceBeforePlotArea}{\vspace*{-12pt}}
\newcommand{\figspaceAfterPlotArea}{\vspace*{-16pt}}
\newcommand{\figspaceBetweenCaptionAndNotes}{\\[3pt]}
\newcommand{\figspaceBetweenPlotRows}{\vspace*{0pt}}

\newcommand{\tabspaceAfterRule}{\\[-10pt]}      
\newcommand{\tabspaceAfterHeaderRow}{\\[2pt]}
\newcommand{\tabspaceAfterBodyRow}{\\[0pt]}

%% file: part_TitlepageToC.tex

\title{\vspace*{-0.5 in} \fontsize{18}{20}\selectfont 
	\textbf{Dynamic Innovation Transmission Through Networks: Theory, Large $T$-Inference, and Business Cycles by Lagged Input-Output Conversion}}


\author{\fontsize{12}{12}\selectfont 
		\hspace*{1pt}\authorname\blfootnote{
		\setlength{\baselineskip}{4mm} \fontsize{9}{11}\selectfont Correspondence:  Department of International Economics, Geneva Graduate Institute (IHEID), 1202 Geneva, Switzerland. 
		Email: marko.mlikota@graduateinstitute.ch. 
		This paper is a revised version of a part of my dissertation at the University of Pennsylvania,
		and parts of it circulated in a previous, broader working paper titled \myquote{Cross-Sectional Dynamics Under Network Structure: Theory and Macroeconomic Applications}.
		%
		I am highly indebted to my advisors, Frank Diebold and Frank Schorfheide for their continued support and guidance.
		I thank Shuxiao Zhou for research assistance. 
		For helpful comments and discussions I am also grateful to 
		Eric Auerbach, Xu Cheng, Marco Del Negro, Thorsten Drautzburg, Simon Freyaldenhoven, Wayne Gao, Domenico Giannone, Bo Honor\'e, Leon Huetsch, Daniel Lewis, Laura Liu, Julien Martin, Geert Mesters, Aaron Mora, Serena Ng, Michael Pollmann, Giorgio Primiceri, Hélène Rey, Joao Ritto, Esteban Rossi-Hansberg, Elisa Rubbo, Adrien Wicht and seminar- and conference-participants at Penn, 7th Lindau Meeting on Economic Sciences, SNDE Symposium 2023, Philadelphia Fed, and NBER Summer Institute 2023 (\myquote{Forecasting \& Empirical Methods}), among others.
		I gratefully acknowledge financial support by the Swiss National Science Foundation under Project Grant 10.003.235.} \\[-10pt]
	{\em \small \authorinstitution} }

\date{\vspace*{0pt} {\normalsize This Version: \today} }
\maketitle


\vspace*{-25pt}

\begin{spacing}{1.1}

\begin{abstract}

	\noindent
	I develop an econometric framework 
	that rationalizes the dynamics of a cross-sectional variable by lagged transmissions of innovations along bilateral links between units.
	While nesting the Spatial Autoregression and Spatial Error Model as limits and producing equivalent impulse-responses in the long run,
	the Network-Vector-Autoregression (NVAR) I propose can accommodate general transmission patterns over time and yields ``networked'' transition dynamics distinct from those implied by autocorrelated innovations.
	I discuss large $T$-inference conditional on a network.
	I then estimate an NVAR for US sectoral output, as derived under a Real Business Cycle economy with lagged input-output conversion. 
	Under the preferred specification, lagged transmissions of productivity shocks along supply chains account for 85\% of the persistence in aggregate output growth
	and reduce shock-variances relative to an economy with contemporaneous input-output conversion by 73\% on average across sectors.

	\todo{BEFORE SHARING: disable todo + delete ToC}

\end{abstract}



\vspace*{5pt}

\thispagestyle{empty}







\end{spacing}

%% file: part_Intro.tex

\newpage

\clearpage
\setcounter{page}{1}


\section{Introduction}
\label{sec_intro}

In economics, we often encounter a cross-section of units linked by a network of bilateral ties, 
such as sectors connected through supply chains 
or individuals acquainted with each other. 
A large theoretical and empirical literature documents that 
networks amplify idiosyncratic shocks and 
generate comovement in cross-sectional variables at a given point in time \citep{BramoulleGaleottiRogers2016,Carvalho-TahbazSalehi2019}. 
%
%
How network-induced comovements unfold over time, however, is less well understood.
The literature predominantly assumes contemporaneous network interactions (e.g. \citet{BramoulleDjebbariFortin2009, AcemogluEtAl2012, ElliottGolubJackson2014}) as embodied by the Spatial Autoregressive (SAR) and Spatial Error Models (SEM), 
which assume that innovations contemporaneously transmit along network connections of all orders.
The resulting static framework is useful for \emph{\myquote{networked} steady state comparisons}.
Among the few studies that consider lagged network interactions, most posit that network effects materialize exactly one link per period (e.g. \citet{LongPlosser1983, GolubJackson2010,Korniotis2010}).
While useful for analyzing the qualitative implications of the laggedness of network interactions, this assumption can be restrictive for empirical work concerned with \emph{\myquote{networked} transition dynamics}.
%
%


I construct an econometric framework that derives the dynamics of a cross-sectional variable from the lagged innovation transmission along bilateral links between cross-sectional units: the Network-Vector-Autoregression (NVAR).\footnote{
    This label is used by \citet{ZhuPanLiLiuWang2017}, whose process I generalize as described in the following.
}
Like existing studies, it assumes uni-directional transmission
and time-invariant links.
Unlike existing studies, it can accommodate general patterns on how innovations travel along a direct link over time and, consequently, how network connections of higher order accumulate as time progresses.
To obtain this generality, I differentiate the frequency of network interactions from the frequency of observation.
As their ratio $q \in \naturals$ diverges to infinity, the impulse-response to contemporaneous high-frequency disturbances approaches its counterpart from the SAR and SEM. 
Holding this relative frequency fixed, the same impulse-response is obtained as the long-term effect of a permanent innovation.
By taking a stance on the timing of network effects, the NVAR goes beyond such steady state comparisons and characterizes transition dynamics.
Thereby, lagged network interactions and autocorrelations in innovations lead to qualitatively different dynamics; the former formalize the idea that innovations take time to feed through the network, while the latter prolong the effects of each round of transmission along a network-connection.

In the NVAR, the interdependence of observations $y_{it}$ and $y_{j,t-h}$ arises as the interplay of the temporal distance between periods $t$ and $t-h$ and the cross-sectional distance between units $i$ and $j$ in the network.
Correspondingly, stationarity depends on roots of a scalar-autoregressive process defined by the timing of innovation transmission along a single link and eigenvalues of the network adjacency matrix $A$.
Distinguishing these temporal and spatial dimensions is key for characterizing the restricted VAR process I define at network-interaction frequency as a restricted VARMA process at observational frequency. 
This observed process illustrates the gradual emergence of networked contemporaneous correlation among errors under $q>1$ 
and shows that their Normality in the limit $q \rightarrow \infty$ with contemporaneous network effects can be motivated by the accumulation of temporally independent high-frequency innovations. 
Accounting for covariates shows that the SAR and SEM are limits of two different specifications: the former assumes that covariates propagate through the network along with innovations to $y_{it}$, whereas the latter assumes that they do not.


The NVAR can be used to estimate dynamic network effects, i.e. the timing at which innovations (and covariates) transmit dynamically through a network $A$ and shape the dynamics of a cross-sectional variable $y_t=(y_{1t}, ..., y_{nt})'$.
For fixed $A$ and $q$, this timing is determined by $\alpha \in \reals^p, p \in \naturals$, which shows how the transmission along a direct link in $A$ is spread out over $p$ network-interaction periods.
I discuss large $T$-inference conditional on $A$.
Under mild assumptions, $\alpha$ is point-identified for $q=1$, for any $q>1$ if $y_t$ is a flow variable and for odd $q>1$ if $y_t$ is a stock variable, while it is identified up to alternating sign -- i.e. its identified set contains two points, $\alpha$ and $\alpha^- = (-\alpha_1,\alpha_2,-\alpha_3,\dots,(-1)^p\alpha_p)'$ -- for even $q$ if $y_t$ is a stock variable.
This characterization is based on my identification analysis of an imperfectly observed scalar-autoregressive process in \citet{P017}, which captures the temporal dimension of the identification problem for $\alpha$ and which I can link to provided that $A$ has at least one real and non-zero eigenpair.
Under additional, technical assumptions, the variance-covariance-matrix of innovations $\Sigma$ is point-identified for any $p$ and $q$ regardless of the nature of $y_t$.
Covariates' effects $Z_t \beta + A Z_t \gamma$ are point-identified under mild additional assumptions.
If $q=1$, then data is observed at the network-interaction frequency and inference on $(\alpha,\beta,\gamma,\Sigma)$ boils down to a linear regression with regressors that summarize lagged observations by direct links.
For $q>1$, estimation can be conducted numerically with likelihood evaluations via the Kalman filter. 
In either case, the Gaussian Maximum Likelihood estimator is consistent and asymptotically Normal.
Interestingly, estimation precision does not necessarily decrease with $q$, but for persistent processes it may first increase with it.


I then apply the NVAR to quantify the potential for the laggedness of firms' input-output conversion (IOC) to endogenize business cycles, as suggested by \citet{LongPlosser1983} in a Real Business Cycle (RBC) production economy with one period-lagged IOC.
I generalize the timing of IOC in their economy by assuming that inputs from the past $p$ periods are imperfectly substitutable in production
and by allowing the model frequency to differ from the observational frequency.
Observed sectoral output then follows an NVAR($p,q$), where 
$A$ is the input-output matrix, 
$\alpha \in [0,1]^p$ with $\sum_{l=1}^p \alpha_l = 1$ shows firms' time profile of input sourcing, 
and the innovations are sectoral productivity (TFP) processes, which I assume are difference-stationary with persistent common and sector-specific shocks.
Using the input-output table and monthly industrial production across 23 manufacturing sectors in the US economy, I estimate $\alpha$ and TFP-autocovariances while allowing for general $p$ and $q$.
In an otherwise equivalent economy with contemporaneous IOC, 
sectoral output evolves as an SAR.
While long-run output responses to TFP shocks are the same in both economies, the role of IOC in transition dynamics differs; 
in the contemporaneous-IOC economy, output growth persistence is due to autocorrelation in exogenous TFP, whose effects are proportionately amplified at all horizons by input-output connections of all order,
whereas in the lagged-IOC economy, this amplification is spread-out over time, hence relating output growth persistence to input-output connections of relevant order, as determined by the exact timing of lagged IOC.

Model selection criteria indicate that lagged-IOC specifications generally fit the data better than the model with contemporaneous IOC. 
Under the preferred specification, inputs are converted to outputs at a frequency of 10 days and the transmission of productivity shocks along a direct supplier-customer-link takes 40 days.
The estimated NVAR reveals the networked impulse-responses of sectoral and aggregate output to idiosyncratic and common TFP shocks.
It suggests that sectoral positions in the supply chain network not only mediate the strength of sectoral shocks' effects on aggregates -- as established by \citet{AcemogluEtAl2012} --, but also its timing, with an unclear relationship between the two; for example, owing to its position at the top of supply chains, a TFP shock in the nonmetallic minerals sector materializes more slowly than a shock to furniture production, despite similar long-term effects.

The estimated NVAR allows me to decompose the autocovariance function of aggregate output growth through the lens of the RBC economy.
%
%
This reveals that lagged IOC can account for around 85\% of the persistence in aggregate output growth.
Most of the remainder is accounted for by autocorrelation in a single common TFP shock,
leaving a small role for the autocorrelations in idiosyncratic sectoral TFP shocks, whose estimates under the preferred lagged-IOC specification are substantially reduced relative to the economy with contemporaneous IOC.
Even more pronounced is the reduction in estimated TFP-variances, which amounts to 73\% on average across sectors.
Overall, my results strongly support the hypothesis of \citet{LongPlosser1983}; accounting for the laggedness of firms' conversion of inputs into outputs can substantially reduce the \myquote{measure of our ignorance} -- the variation and autocorrelation in exogenous TFP shocks. 
Moreover, under the preferred lagged-IOC specification, a single common TFP process accounts for 90\% of the aggregate output growth variance and for around 85\% of its persistence.
My results thus attribute a smaller share of aggregate fluctuations to sector-specific shocks than do \citetwo{LongPlosser1987}{FoersterSarteWatson2011}, who identify common shocks from the contemporaneous correlation of sectoral innovations.
My analysis suggests that abstracting from the laggedness of IOC may confound sector-specific shocks with sectoral heterogeneity in the temporal exposure to common shocks.


\paragraph*{Contribution}






There is a large and long-standing econometric literature on spatial and network-interactions that focuses on identifying network effects and accounting for heterogeneity in cross-sectional and panel data-regressions \citep{Anselin1988,Anselin2003,Manski1993,Lee2007,BramoulleDjebbariFortin2009,ShaliziThomas2011,Elhorst2012,KuersteinerPrucha2020,DePaula-Rasul-Souza2024}.\footnote{
    A separate strand works with dyadic data, i.e. observed edges (rather than nodes) (see \citet{Graham2020}).
}
Studies predominantly assume contemporaneous network interactions as in the canonical SAR or SEM.
A few studies consider lagged interactions and typically assume a single lag (see work surveyed in \citetwo{LeeYu2011}{Elhorst2012}).
\citethree{ZhuPanLiLiuWang2017}{Schennach2018}{KnightEtAl2020} cast them in an explicit time series model and generalize the lag length.
My setup is closest to \citet{ZhuPanLiLiuWang2017}.
I characterize Granger causality in their environment as the cross-sectional innovation transmission along network connections of different order, formalizing the idea that network effects materialize dynamically over time.
I then generalize this mapping between network connections and dynamics by separating the network-interaction- and observational frequencies,
which allows me to derive the SAR and SEM as limits of a process driven by lagged network interactions that evolve at an ever higher frequency.\footnote{
    \citethree{Maung2022}{ArmillottaFokianos2023}{BarigozziCavaliereMoramarco2025} extend \citet{ZhuPanLiLiuWang2017} along different dimensions than I do.
}
For this generalized model, I characterize the observed process, derive the identified set and discuss estimation.

Many studies incorporate networks in time series analyses.
Some \emph{interpret} quantities of a given time series model -- typically a VAR -- using network analysis (e.g. \citet{DieboldYilmaz2009,DieboldYilmaz2014,BillioGetmanskyLoPelizzon2012}).
Others \emph{restrict} time series dynamics using networks (e.g. \citet{CarvalhoWest2007,ChudikPesaran2011,CaporinErdemliogluNasini2021,MartinPassinoCucuringuLuati2025}).\footnote{
    In particular, \citefour{HollyPetrella2012}{DahlhausSchaumburgSekhposyan2021}{DeGraeve-Schneider2023}{ChodorowReich-Gabaix-Viviano2025} exploit networks for shock-identification in VARs.
    \citefour{AhelegbeyBillioCasarin2016JAE}{BarigozziBrownlees2018}{BillioCasarinRossini2019}{ChenLiLiLinton2025} build time series models to infer a network.
    \citet{Bykhovskaya2023} models the evolution of a network observed over time.
}
Among the latter, the widely used Global VAR (GVAR) of \citet{PesaranSchuermannWeiner2004} restricts contemporaneous and lagged interdependencies across units in a panel-VAR using exogenous network-weights (e.g. bilateral trade) and enables the estimation of several interconnected VAR models under a weak exogeneity assumption.\footnote{
    See \citet{ChudikPesaran2014} for an overview of GVAR-applications and \citet{CanovaCiccarelli2013} for an overview of panel-VAR approaches more generally.
}
The NVAR also belongs to this second category: it restricts innovations in a (cross-sectional) VAR to transmit along bilateral links among variables (units), which leads to chained causality patterns as in \citet{DufourRenault1998}.
Different from other \myquote{networked} time series approaches above and the GVAR in particular, and in line with the spatial- and network interaction-literature discussed above, it features a single variable per cross-sectional unit and a single type of connections among them (see \citet{ElhorstEtAl2021} for a comparison of the two literatures).
Entertaining the assumption that innovations transmit along bilateral links, I generalize the timing of network effects.
This results in an interpretable model of dynamic network effects that can be motivated structurally with an RBC production economy and that provides a more explicit bridge between the two literatures, as demanded by \citet[Conclusion]{ChudikPesaran2014}.

The spatial panel literature has developed inference for models with lagged network interactions \citep{YuEtAl2008,Korniotis2010,Li2017,KuersteinerPrucha2020}.
%
These studies take $T$ as fixed or let $n$ and $T$ grow jointly and must therefore contend with the incidental-parameter bias induced by unit fixed effects, whereas I take $n$ as fixed and let $T \rightarrow \infty$, which allows me to leave $\Sigma$ unrestricted.
Moreover, they implicitly set $q=1$ and estimate the strength of one period-lagged network interactions,\footnote{
    \citet{ZhuPanLiLiuWang2017} are an exception: they set $q=1$ but have general $p$.
} whereas I estimate the time profile of innovation transmission along a direct link over $p$ periods, whose identification is non-trivial for $q>1$.
My identification analysis applies any finite $q \in \naturals$ and shows that the reflection problem of \citet{Manski1993} only appears in the limit $q \rightarrow \infty$ with contemporaneous effects, where only the spatial lag parameter $\sum_{l=1}^p \alpha_l$ is identified under well-known conditions \citep{KelejianPrucha1999,KelejianPrucha2010JoE,BramoulleDjebbariFortin2009,LeeYu2011}.
Moreover, I indirectly show that for a known $A$, $\alpha$ and innovation-autocorrelations are separately identified by characterizing their qualitatively different transitional dynamics and by tracing out the likelihood in my application by means of marginal posteriors under uniform priors.
%



My application contributes to the series of efforts to microfound sectoral and aggregate dynamics using production networks \citep{Horvath1998,Horvath2000,Dupor1999,Shea2002,Carvalho2010,FoersterSarteWatson2011,AcemogluEtAl2012,DiGiovanni-Levchenko-Mejean2014,BaqaeeFarhi2019}.
Mirroring the prevalence of contemporaneous interactions in econometrics and related spatial economic analyses \citep{DesmetParro2025}, most existing studies rely on contemporaneous IOC, which reveals that the production network contributes to sectoral comovement and aggregate volatility (see work surveyed in \citet{Carvalho-TahbazSalehi2019}).
A notable exception is \citet{LongPlosser1983}, who assume one period-lagged IOC and show that it can endogenize business cycles -- i.e. persistence in aggregates -- without relying on autocorrelation in exogenous shocks, non-rational expectations, or other frictions.
Their analysis points to a more prominent role for production networks than as mere amplifiers of existing dynamics. 
%
%
\citet{CarvalhoReischer2021} characterize persistence in that economy and show that its evolution implied by observed changes in the US input-output network 
accounts well for empirical measures of the changing persistence of aggregate output growth,
which suggests that the endogenization of business cycles through lagged IOC has empirical merit.
%
I quantify the empirical salience of \citet{LongPlosser1983}'s hypothesis by taking an RBC economy with general IOC-lags to the data and conducting inference on the timing of IOC along with the autocorrelations and variances of exogenous TFP processes.
Under the preferred model, I compute their respective contribution to the entire autocovariance function of aggregate output growth, supplementing the decomposition of \citet{FoersterSarteWatson2011}, who considered its variance alone.
My treatment of the timing of IOC as an object of inference is motivated by their conclusion that the importance of idiosyncratic shocks \myquote{depends critically on details of the specification and the values of parameters that govern intra- and intertemporal linkages} (p. 28) and by recurring discussions on the appropriate observational frequency for a given model specification (e.g. \citet[p. 335--336]{LongPlosser1987}, \citet[p. 73--74]{Horvath2000}, \citet[p. 428]{Shea2002}, \citet[p. 5--6]{FoersterSarteWatson2011}).\footnote{
    In particular, \citet[p. 335--336]{LongPlosser1987} note that \myquote{the appropriate time interval [of observation] is the one that most closely corresponds to the production interval}.
   The timing of IOC I estimate indeed fits with production- and inventory-lengths reported by \citetwo{Ramey1989}{AntrasTubdenov2025}.
}
Because I abstract from other mechanisms that delay the propagation of sectoral shocks 
-- alternative time-to-build-mechanisms \citep{KydlandPrescott1982}, capital accumulation over a sectoral investment-network \citep{Horvath1998,FoersterSarteWatson2011,VomLehnWinberry2022} and staggered price setting along chains of production \citep{Blanchard1982,Bouakez-Cardia-RugeMurcia2014} --,
the contribution of lagged IOC to output persistence I estimate should be read as an upper bound.


\paragraph*{Outline}

%
I discuss the model and its properties in \cref{sec_model}. 
\cref{sec_inference} treats identification and inference. 
In \cref{sec_app1}, I study the role of input-output conversion in US business cycles. 
\cref{appsec_model,appsec_inference,appsec_app1} accompany these sections, respectively. 
\cref{sec_conclusion} concludes.

%% file: part_Model.tex
\section{Lagged Network Effects: Theory}
\label{sec_model}







\subsection{Bilateral Connections in Networks}
\label{subsec_model_networkessentials}

A network is represented by an $n \times n$ adjacency matrix $A$ with elements $a_{ij}$. 
I consider a directed, weighted and possibly signed network; $a_{ij} \in [-1,1]$ shows the sign and strength of the link from unit $i$ to unit $j$, with $a_{ij} \neq a_{ji}$ possibly. 
If $a_{ij}=0$, I say unit $i$ is not connected to unit $j$. 
Self-links $a_{ii} \neq 0$ are permitted.
The set of bilateral links $\{a_{ij}\}_{i,j=1:n}$ give rise to a myriad of higher-order (hyper-dyadic) links among units, referred to as walks.\footnote{
	Whenever convenient to simplify notation, I write $a:b$ for the set of integers $\{a,a+1,...,b\}$, $a\leq b$.}
%
%
A walk from $i$ to $j$ of length $k \geq 1$ is a product of bilateral links that lead from $i$ to $j$, possibly over intermediary units: $\prod_{s=1}^{k} a_{u_s,u_{s+1}}$, with $u_1 = i$ and $u_{k+1} = j$.
It is non-zero if all of these units are sequentially connected. 
Simple matrix algebra reveals that $[A^k]_{ij}$ sums all walks from $i$ to $j$ of length $k$. I refer to this quantity as the $k$th-order connection from $i$ to $j$.
%
In the following example, even though unit $3$ is not directly connected to unit $1$ ($a_{31}=0$), there exists a second-order connection via unit $2$ ($a_{32}a_{21}\neq 0$). 
In a production network, unit $1$ could be a supplier to unit $2$, who in turn is a supplier to unit $3$.
\begin{align*}
	A =
	\begin{bmatrix}
			0 & 0 & .8 \\
			.7 & 0 & .2 \\
			0 & .9 & 0 \\
	\end{bmatrix} \; , \; \;
	A^2 =
	\begin{bmatrix}
			0 & .72 & 0 \\
			0 & .18 & .56 \\
			.63 & 0 & .18 \\
	\end{bmatrix} \; , \; \;
	A^3 =
	\begin{bmatrix}
			.50 & 0 & .14 \\
			.13 & .50 & .04 \\
			0 & .16 & .50 \\
	\end{bmatrix} \; .
\end{align*}




\subsection{Lagged Innovation Transmission via Bilateral Links}
\label{subsec_model_laggednetworkeffects}

Consider a stationary cross-sectional time series $y_t = (y_{1t}, ..., y_{nt})'$ with mean zero.
Under an NVAR, the dynamics of $y_t$ are driven by the lagged transmission of innovations $u_{it}$ along bilateral links $a_{ij}$ among cross-sectional units $i,j=1:n$.
Throughout, I assume that links are fixed over time and that transmission is uni-directional: the direct link from $i$ to $j$, $a_{ij}$, transmits innovations from $j$ to $i$.\footnote{
    The distinction is only relevant for directed networks.
} 
Innovations $u_t$ may be cross-sectionally correlated.


\subsubsection{Single Lag in Innovation Transmission: NVAR(1,1)}
\label{subsubsec_model_laggednetworkeffects_NVAR11}

Consider a VAR(1) with an autoregressive matrix proportional to the adjacency matrix $A$:
\begin{align} \label{eq_NVAR11}
	y_t = \alpha A y_{t-1} + u_t \; , 
\end{align}
with $\alpha \in \reals$.
Under this process, the one step-ahead expectation of $y_{it}$ is proportional to the weighted sum of one period-lagged values of $y_{jt}$ for all units $j$ to which $i$ is directly linked: $\mean_{t-1}[y_{it}] = \alpha \sum_{j=1}^n a_{ij} y_{j,t-1}$.
For example, in the RBC input-output economy of \citet{LongPlosser1983}, the expected output in sector $i$ tomorrow is a weighted average of the production in its supplier-sectors $j$ today.
In \citet{GolubJackson2010}, individual $i$'s expected opinion tomorrow is a weighted average of their friends' opinions today. For reasons clarified below, I dub this process NVAR(1,1).
It captures the essence of dynamics in spatial panel data models with lagged network effects discussed in \citetwo{LeeYu2011}{Elhorst2012}.

\begin{figure}[t]
	\begin{center}
		\figspaceBeforePlotArea
		\begin{subfigure}[b]{\plotSizeSingle\textwidth}
				\centering
				\includegraphics[width=1\textwidth, clip]{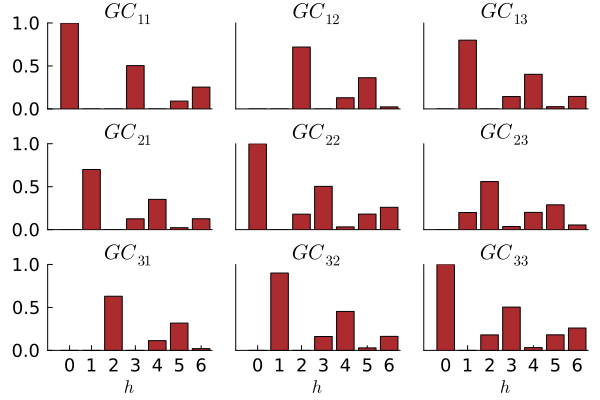}
		\end{subfigure}
		\figspaceAfterPlotArea
	\end{center}
    \caption{Example of Generalized Impulse-Responses under an NVAR($1,1$)\figspaceBetweenCaptionAndNotes
    \scriptsize {\em Notes:} Panel $(i,j)$ shows $\partial y_{i,t+h} / \partial y_{jt}=[\alpha^hA^h]_{ij}$ under $\alpha=1$. Note that $\partial y_{i,t} / \partial y_{jt} = 1$ if $i=j$ and zero otherwise, i.e. by definition, the contemporaneous responses to all but a series' own innovation are zero.}
    \label{plot_model_GIRF_NVAR11}
\end{figure}

Under \cref{eq_NVAR11}, $h$th-order network connections drive the dynamics of $y_t$ at horizon $h \geq 1$:
$\partial y_{i,t+h} / \partial y_{jt}  = \partial y_{i,t+h} / \partial u_{jt} = [\alpha^h A^h]_{ij}$ (see also \citet[Theorem 2]{CarvalhoReischer2021}).
Thus, if $\alpha > 0$ and $a_{ij}>0 \; \forall \; i,j$, $y_j$ Granger-causes $y_i$ at horizon $h$ iff there exists an $h$th-order connection from $i$ to $j$.\footnote{
	With $\alpha, a_{ij} \in \reals \; \forall \; i,j$, this holds generically and as a necessary condition.
	Also, note that
	$\partial y_{i,t+h} / \partial u_{jt}$ is also referred to as the Generalized Impulse-Response Function. It disregards shock identification and considers instead the propagation of reduced form errors $u_t$ over future time periods $t+h$, $h\geq 1$.
}
%
\cref{plot_model_GIRF_NVAR11} illustrates this Granger-causality pattern for the network from \cref{subsec_model_networkessentials} under $\alpha=1$.
It points to two insights that extend to more general processes than \cref{eq_NVAR11}.
First, lagged network interactions generate persistence: even white noise-innovations $\{u_{jt}\}_{j=1}^n$ lead to persistent reactions of $y_t$ because each $u_{jt}$ affects connected units $i \neq j$ with a lag.\footnote{
	An individual $y_{jt}$ reacts persistently to its own white noise-innovation $u_{jt}$ if there is a cycle in the network, i.e. a walk from $j$ to $j$. Notably, this can hold even without a self-link $a_{jj} \neq 0$.} 
This generates the endogenous business cycles under lagged input-output conversion (IOC) in \citet{LongPlosser1983}.
Second, under this type of persistence, network-connections affect not only the strength of impulse-responses, but also their timing.
For example, while unit 2 is directly linked to unit 1 and therefore experiences the latter's innovation with a lag of one period, unit 3 only has a second-order connection to unit 1 and therefore reacts only after two periods.
Thus, Granger-causality can take the form of a chain, as pointed out in \citet{DufourRenault1998}.\footnote{
	Specifically, they show that even if a series $x_t$ does not Granger-cause a series $y_t$ at horizon 1, under the presence of a third series $z_t$, $x_t$ might cause $y_t$ at higher horizons as the causality could run from $x_t$ to $z_t$ to $y_t$.
}
In the present case of a cross-sectional time series driven by lagged innovation transmission along bilateral links, 
their insight emerges naturally and their generally non-trivial conditions for Granger-(non)causality boil down to the presence or absence of network connections of relevant order between the relevant variables (units). 
%

The process in \cref{eq_NVAR11} reveals useful theoretical insights in both macro- \citep{LongPlosser1983} and microeconomics \citep{GolubJackson2010}, yet for some empirical analyses it may be too restrictive to assume that transmission occurs one link per period and materializes completely within a single period.
This implies, for instance, that in response to news gathered by a friend of a friend, $j$, an individual $i$ adjusts their opinion in two periods, not earlier, not later.
Similarly, an innovation to a sector $j$'s output affects the output of other sectors $i$ located two positions downstream (\myquote{customers of customers}) in exactly two periods.
In either case, earlier responses are ruled out altogether, and later adjustments only occur if $i$ has a third- or higher-order connection to $j$. 
In the following, I generalize the timing of lagged network effects by extending the simple process above along two dimensions.


\subsubsection{Multiple Lags in Innovation Transmission: NVAR($p,1$)}
\label{subsubsec_model_laggednetworkeffects_NVARp1}

Let the cross-sectional time series $y_t$ evolve according to a VAR($p$), where each autoregressive matrix is proportional to the same network adjacency matrix $A$: 
\begin{align} \label{eq_NVARp1}
	y_t = \alpha_1 A y_{t-1} + ... + \alpha_p A y_{t-p} + u_t \; ,
\end{align}
with $\alpha = (\alpha_1, ..., \alpha_p)' \in \reals^{p}$.
I dub this process NVAR($p,1$).
It simplifies the process considered in \citet{ZhuPanLiLiuWang2017} 
by not distinguishing self- and cross-links,  
by abstracting from restrictions on the rows of $A$, 
and by disregarding the intercept and covariates.
Moreover, it is nested by the process of \citet{Schennach2018} when considering the same network adjacency matrix at every lag.
This and the homogeneity of $\alpha$ are also what distinguishes it from \citet{KnightEtAl2020}.
These two simplifications are essential 
for my interpretation of the model, the generalization of its transmission-timing and its structural derivation in a lagged-IOC production economy.

\myrepproposition{NVAR($p,1$): Granger-Causality}{prop_GCGIRF_NVARp1}{
	Let $y_t$ evolve as in \cref{eq_NVARp1}.
	Then
	$ \partial y_{i,t+h} / \partial u_{j,t}  = c^{h}_{\underline{k}} \sbr{A^{\underline{k}}}_{ij} + ... + c^{h}_{h} \sbr{A^{h}}_{ij} $,
	where $\underline{k} = \text{ceil}(h/p)$, and the coefficients $\{c^h_k\}_{k=\underline{k}:h}$ are polynomials in $\{\alpha_l\}_{l=1:p}$.
}

\cref{prop_GCGIRF_NVARp1} shows that compared to \cref{eq_NVAR11}, under the NVAR($p,1$) with $p>1$, dynamics at horizon $h$ are affected by lower-order connections.
If $\alpha_l \neq 0 \; \forall \; l$, then $c^h_k \neq 0 \; \forall \; k$, and \cref{prop_GCGIRF_NVARp1} implies that $y_j$ Granger-causes $y_i$ at horizon $h$ iff there exists a connection from $i$ to $j$ of at least one order $k \in \{\underline{k}, \underline{k}+1, ..., h\}$.
The coefficient $c^{h}_{k}$ shows the influence of connection-order $k$ on the impulse-response at horizon $h$.
If $i$ has a first- and no higher-order connections to $j$, then $\partial y_{i,t+h} / \partial u_{j,t} = \alpha_h a_{ij}$ for $h=1:p$ and zero otherwise.
Hence, the NVAR($p,1$) specifies that dynamics are driven by the transmission of innovations along direct links lagged over $p$ periods.
$\alpha$ determines how it is spread out over these $p$ periods and, consequently, how the transmission along higher-order connections accrues as time progresses.
Note that transmission is assumed to be homogeneous across unit pairs $(i,j)$ and over time.

\subsubsection{Multiple Rounds of Innovation Transmission: NVAR($p,q$)}
\label{subsubsec_model_laggednetworkeffects_NVARpq}

Suppose now that we observe an NVAR($p,1$) process $x_\tau$ every $q \in \naturals$ periods, either as a snapshot (stock variable) or as a sum over the sampling interval (flow variable):
\begin{align} \label{eq_NVARpq}
    x_\tau = \alpha_1 A x_{\tau-1} + ... + \alpha_p A x_{\tau-p} + e_\tau \; , \quad
    y_{t} =
    \begin{cases}
        x_{tq} & \text{if $y_t$ is stock variable} \; , \\
        x_{tq} + ... + x_{tq-q+1} & \text{if $y_t$ is flow variable} \; . \footnotemark
    \end{cases}
\end{align}
\footnotetext{
	If the flow is instead observed as an average, $y_t = (x_{tq} + ... + x_{tq-q+1})/q$, all results for the flow case in this paper apply after rescaling: impulse-responses are divided by $q$, and $\var[u_t]$, $\gamma_\nu(h)$ and $\gamma_y(h)$ by $q^2$.
}
I dub this process NVAR($p,q$). 
$q$ indicates the relation between the frequency of network interactions at which $x_\tau$ evolves and the frequency at which data $\{y_t\}_{t=1:T}$ is observed. 
If $q=1$, the two coincide and, trivially, $y_t$ and $x_\tau$ are the same NVAR($p,1$) process.
If instead $q>1$, then network interactions occur at higher frequency than data is observed. 
For example, under yearly observations, $q=4$ implies quarterly network interactions.
As a result, multiple rounds of transmission occur in a single observational period, and dynamics at horizon $h$ are affected by higher-order connections, $k>h$.
\cref{prop_GCGIRF_NVARpq_stock} shows this for the stock case, and \cref{prop_GCGIRF_NVARpq_flow} in \cref{appsec_model} shows it for the flow case.

\myrepproposition{NVAR($p,q$), Stock Variable: Granger-Causality}{prop_GCGIRF_NVARpq_stock}{
	Let the stock variable $y_t$ evolve as in \cref{eq_NVARpq} for some $q \in \naturals \backslash\{1\}$.
	Then 
	$ \partial y_{i,t+h} / \partial y_{j,t}  = c^{h}_{\underline{k}} \sbr{A^{\underline{k}}}_{ij} + ... + c^{h}_{hq} \sbr{A^{hq}}_{ij} $,
	where $\underline{k} = \text{ceil}(hq/p)$, and the coefficients $\{c^h_k\}_{k=\underline{k}:hq}$ are polynomials in $\{\alpha_l\}_{l=1:p}$.\footnote{
		If $\alpha_l, a_{ij} \in \reals_{++} \; \forall \; l,i,j$, then $y_j$ Granger-causes $y_i$ at horizon $h$ iff there exists a connection from $i$ to $j$ of at least one order $k \in \{\underline{k}, \underline{k}+1, ..., hq\}$. With $\alpha_l, a_{ij} \in \reals \; \forall \; l,i,j$, this holds generically and as a necessary condition.
	}
}

\cref{prop_NVARpq_observeddynamics} characterizes the observed process explicitly: it follows a VARMA process with autoregressive coefficient-matrices that are polynomials in $A$.\footnote{
	The proof states them in closed-form. \cref{prop_NVARpq_observeddynamics} builds on the representation-result of \citet{P017} for an AR($p$) process observed every $q$ periods, the bridge to which is discussed in more detail in \cref{subsec_inference_ID}.
}
While one could directly restrict VARMA-coefficient matrices to be polynomials in $A$, my setup permits structural interpretations of $\alpha$ and $q$ that enable me to map the model to a lagged-IOC production economy in \cref{sec_app1}, to derive the identified set for $\alpha$ by exploiting the model's relation to regularly but infrequently observed or temporally aggregated scalar-AR($p$) processes in \cref{sec_inference}, and to characterize a range of properties in the remainder of this section.

In the stock case, we can further generalize the transmission pattern:
for any $q \in \rationals_{++}$, we can write $q=q_1q_2$ with $q_2, q_1^{-1} \in \naturals$
and model $y_t$ as an NVAR($p^*,q_2)$ with $p^* = p/q_1 \in \naturals$ and restricted parameters $\{\alpha_l\}_{l=1:p^*}$ that complies with the interpretations of $p$ and $q$.
For example, under quarterly observations, $q=3/2$ signifies that network interactions occur every two months. 
This amounts to observing every $3$ periods a snapshot of a monthly process that depends on its value in the past $p$ $2$-month-periods.
\cref{appsec_model} elaborates.

\myrepproposition{NVAR($p,q$): Observed Dynamics}{prop_NVARpq_observeddynamics}{
    Let $y_t$ evolve as in \cref{eq_NVARpq} for some $q \in
    \naturals$, with $e_\tau \sim WN(0,\Sigma)$, 
	and let $q^* = \text{floor}(d/q)$ with $d = p(q-1)$ in the stock case and $d = (p+1)(q-1)$ in the flow case.\footnote{
		Note that $q^*\leq p-1$ and $q^*\leq p$ in the two cases, respectively.
	}
	Then $y_t$ follows a VARMA($p,q^*$): $\calG_q(L)y_t = \nu_t$,
	where $\calG_q(z) = I_n - \sum_{k=1}^{p} \calG_{q,k} z^k$ is a real polynomial of degree $\leq p$,
	$\nu_t = \sum_{k=0}^{d} B_{q,k}e_{tq-k}$ has mean zero and autocovariances $\gamma_\nu(h) =
	\sum_{k=0}^{d-qh}B_{q,k+qh}\Sigma B_{q,k}'$ for $0 \leq h \leq q^*$ and zero for $h > q^*$,
	and $\{\calG_{q,k}\}_{k=1}^p$ and $\{B_{q,k}\}_{k=0}^{d}$ are polynomials in $A$.
}


\subsection{Further Properties of the NVAR}
\label{subsec_model_furtherproperties}

\paragraph*{Stationarity}

The interdependence of $y_{it}$ and $y_{j,t-h}$ in the NVAR combines the cross-sectional distance between units $i$ and $j$ in the network and the temporal distance between periods $t$ and $t-h$.
Correspondingly, \cref{prop_WS_NVARpq} characterizes weak stationarity (w.s.) in terms of eigenvalues of $A$ and roots of an AR($p$) process driven by the transmission-profile along a single link, $\alpha$.
Intuitively, it states that the NVAR($p,1$) is w.s. iff the AR($p$) process 
$(1 - \phi_1 L - ... - \phi_p L^p)x_\tau = e_\tau$ with $\phi_l = \alpha_l \lambda_i$ is w.s. for each eigenvalue $\lambda_i$ of $A$, 
though AR-coefficients must be real-valued.
This sharpens the sufficient conditions provided by \citet[Theorems 1 and 4]{ZhuPanLiLiuWang2017}.
Practically, it simplifies stationarity-checks under $p>1$, especially for large $n$.

\myrepproposition{NVAR($p,q$): Stationarity}{prop_WS_NVARpq}{%
	For some $q \in \naturals$ and $e_\tau \sim WN(0,\Sigma)$, let $y_t$ evolve as in \cref{eq_NVARpq}.
	For an eigenvalue $\lambda$ of $A$, define the scalar lag-polynomial $\phi_\lambda(z) = 1 - \lambda \sum_{k=1}^{p}\alpha_k z^k$.
	Then $x_\tau$ is w.s. iff $\phi_{\lambda_i}(z) \neq 0$ for all $|z|\leq 1$ and all eigenvalues $\lambda_i$ of $A$,
	and $y_t$ is w.s. if $x_\tau$ is.
}


\paragraph*{Relation to SAR and SEM}

The literature overwhelmingly considers contemporaneous network interactions.
Under a cross-sectional variable $\bar y_t = (\bar y_{1t}, ..., \bar y_{nt})'$, covariates $Z_t \in \reals^{n \times k}$, parameters $\beta, \gamma \in \reals^k$ and innovations $v_t$, the two predominant models state
\begin{alignat}{2}
    \bar y_t
    &= aA\bar y_t + Z_t\beta + AZ_t\gamma + v_t \; , \quad
    &&\quad \text{(SAR)} \label{eq_SAR} \\
    \bar y_t
    &= Z_t\beta + AZ_t\gamma + \bar u_t \; , \quad
    \bar u_t = aA\bar u_t + v_t \; ,
    &&\quad \text{(SEM)} \label{eq_SEM}
\end{alignat}
typically for a single period $t$ (cross-sectional data).
\cref{eq_SAR} is the Spatial Durbin Model, which nests the Spatial Autoregression (SAR) or spatial lag model for $\gamma = 0$, and \cref{eq_SEM} is the Spatial Durbin Error Model, which nests the Spatial Error Model (SEM) for $\gamma = 0$ (see \citefour{Anselin1988}{LeSagePace2009}{Elhorst2010}{Elhorst2012} for terminology).
For simplicity, I will refer to the former as SAR and the latter as SEM.

Without covariates ($\beta = \gamma = 0$), both models coincide and state $\bar y_t = aA \bar y_t + v_t$, which I refer to as the covariate-free SAR.
Provided that $|\lambda_i|<1/|a|$ for all eigenvalues $\lambda_i$ of $A$,
we can write $\bar y_t = (I-aA)^{-1} v_t = (I + aA + (aA)^2 + (aA)^3 + ...) v_t$.
Hence, the implicit assumption is that connections of all order materialize in a single period.
In line with this, \cref{prop_NVARpq_SARSEM__LTresp} establishes that the response of $\bar y_{it}$ to an innovation $v_{jt}$ in the SAR -- given by the Leontief inverse $(I- a A)^{-1}$ -- equals the long-run response of $y_{it}$ to a permanent innovation $e_{j\tau}$ in the NVAR($p,q$) if $y_t$ is a stock and $q^{-1}$ times that response if $y_t$ is a flow variable.\footnote{
	\citet[Eq. 10]{Elhorst2012} and \citet[Corollary 5(ii)]{CarvalhoReischer2021} state this result for an NVAR(1,1).
	Also, note that, as for any VAR, the long-term response to a permanent innovation is equal to the cumulative response to temporary, one-period innovations.
}
By taking a stance on the timing and frequency of network interactions, 
the NVAR shows how any such long-term effects materialize over time. 
Thereby, $\alpha$ decomposes temporally the spatial autoregressive parameter $a=\sum_{l=1}^p \alpha_l$
and reveals that strong temporary network effects are compatible with weak long-term effects.

The long run is defined in terms of the frequency of network interactions.
Therefore, rather than looking at the cumulative response to temporary innovations over an exceedingly long horizon $H$ under a given $q$, we can obtain the same response at a finite horizon $H$ when considering $y_t$ to be the sum of an underlying process $x_\tau$ (flow case) that evolves at an ever higher network interaction-frequency $q$.
\cref{prop_NVARpq_SARSEM__contempresp} formalizes this for the response at horizon $H=0$, i.e. at the observational period in which the high-frequency innovation occurs.


\myrepproposition{NVAR($p,q$): Relation to SAR and SEM Without Covariates}{prop_NVARpq_SARSEM}{%
	Let $y_t$ evolve as in \cref{eq_NVARpq} for some $q \in \naturals$, and let $\bar y_t = aA \bar y_t + v_t$ with $a=\sum_{l=1}^p \alpha_l$.
	Assume $x_\tau$ and $y_t$ are w.s., and let $\kappa_q = 1$ in stock case and $\kappa_q = q$ if $y_t$ is a flow variable.
	Define $\gamma_y(h) = \mean[y_ty_{t-h}']$, $\Theta_h = \partial x_{\tau+h}/ \partial x_{\tau}$, $\Gamma_h = \sum_{m=0}^h \Theta_m$ and $u_t \equiv y_t - \mean[y_t|\calF_{t-1}]$ with $\calF_{t-1} = \{x_{tq-q}, x_{tq-q-1}, ... \}$.
	Then
	\begin{enumerate}
		\item[(i)] $\underset{H\rightarrow \infty}{\lim}  \sbr{ \partial y_{t+H} / \partial e_{tq} + \partial y_{t+H} / \partial e_{tq+1} + ... + \partial y_{t+H} / \partial e_{(t+H)q} }  = \kappa_q \partial \bar y_t / \partial v_t  = \kappa_q (I- a A)^{-1}$;
		
		\item[(ii)] in flow case, 
		$\underset{g\rightarrow \infty}{\lim}  \partial y_{t} / \partial e_{\tau-g} = (I-aA)^{-1}$ for $g \in 0:(q-1)$;

		\item[(iii)] if $e_\tau \sim WN(0,\Sigma)$, $\var[u_t] = \Sigma + \sum_{h=1}^{q-1} \Theta_h \Sigma \Theta_h'$ in stock- 
		and $\var[u_t] = \sum_{h=0}^{q-1} \Gamma_h \Sigma \Gamma_h'$ in flow-case.
	\end{enumerate}
	If in addition $e_\tau \overset{i.i.d.}{\sim} (0,\Sigma)$ with $\Sigma$ p.d., then as $q \rightarrow \infty$
	\begin{enumerate}
		\item[(iv)] $\calG_q(z) \rightarrow I_n$;
		
		\item[(v)] in stock case, 
		$\gamma_y(0) = \sum_{h=0}^\infty \Theta_h\Sigma\Theta_h' \; \forall \; q$, and $\gamma_y(h) \rightarrow 0 \; \forall \; h \geq 1$,
		so that for fixed $\Sigma$ and Normal $e_\tau$, $y_t$ converges to an i.i.d. sequence distributed as $x_\tau$, while for $\Sigma = q^{-1}\Sigma_*$ it converges to zero in mean square;
		
		\item[(vi)] in flow case, if $\Sigma = q^{-1}\Sigma_*$ with $\Sigma_*$ p.d. and $\sqrt{q}\,e_\tau$ has a distribution that does not depend on $q$ with $\mean\|\sqrt{q}\,e_\tau\|^{2+\delta} < \infty$ for some $\delta>0$, then $u_t \convdist N(0,\Gamma \Sigma_* \Gamma')$ with $\Gamma = (I-aA)^{-1}$;
		
		\item[(vii)] in flow case, under the conditions of (vi),
		$\gamma_y(0) \rightarrow \Gamma\Sigma_*\Gamma'$,
		$\gamma_y(h) \rightarrow 0 \; \forall \; h \geq 1$, 
		and 
		$ \br{y_t', ..., y_{t+S}'}' \convdist N\br{0, \; I_{S+1} \otimes \Gamma\Sigma_*\Gamma'} \; \forall \; S \in \naturals$,
		i.e. $y_t$ converges to an i.i.d. sequence distributed as $\bar y_t$ in the covariate-free SAR with $v_t \sim N(0,\Sigma_*)$.
	\end{enumerate}
}



Contemporaneous network interactions lead to networked correlation among $\{\bar y_{it}\}_{i=1:n}$:
under the covariate-free SAR, $\var[\bar y_t] = \Gamma \Sigma_v \Gamma'$ with $\Sigma_v=\var[v_t]$ and $\Gamma = (I-aA)^{-1}$.
$\covar(\bar y_{it},\bar y_{jt})$ then reflects the bilateral exposure between $i$ and $j$ and their mutual exposure to third units, as determined by network connections of all order.\footnote{
	For $a=1$, $\var[\bar y_t]=\var[(I+A+A^2+...)v_t]$ provided that $|\lambda_i|<1$ for all eigenvalues $\lambda_i$ of $A$.
	If $\Sigma_v = \text{diag}(\sigma^2_1, ..., \sigma^2_n)$, then
	$ \covar(\bar y_{it},\bar y_{jt}) = \sigma^2_i \one{i=j} + \sigma^2_j \sum_{h=1}^\infty \sbr{A^{h}}_{ij} + \sigma^2_i \sum_{h=1}^\infty \sbr{A^{h}}_{ji} + \sum_{o=1}^n \sigma^2_o \sum_{h_i=1}^\infty \sum_{h_j=1}^\infty \sbr{A^{h_i}}_{io} \sbr{A^{h_j}}_{jo} $.
}
\citet{DeGraeve-Schneider2023} exploit this for shock identification, restricting the impact responses of sectoral output to sectoral shocks.

Networked contemporaneous correlation is obtained as soon as the network interaction-frequency exceeds the frequency of observation, even under finite values of the former.
In an NVAR($p,q$), we can define the \myquote{observable innovations} $u_t$ as in \cref{prop_NVARpq_SARSEM}.\footnote{
	Compared to \cref{prop_NVARpq_observeddynamics}, 
	\cref{prop_NVARpq_SARSEM__varobsinnov,prop_NVARpq_SARSEM__limit_flow_obsinnov} also split the observed series $y_t$ into a predictable component and news, but relative to a different information set;
	$\nu_t = \calG_q(L)y_t$ conditions on observed lags $y_{t-1}, ..., y_{t-p}$ alone, while $u_t = y_t - \mean[y_t|\calF_{t-1}]$ conditions on the history at network interaction-frequency.
	%
}
\cref{prop_NVARpq_SARSEM__varobsinnov} states $\var[u_t]$ under general $p\in \naturals$ and $q\in \naturals\backslash\{1\}$.
Under $q=1$, $u_t = e_{tq}$ inherits the variance of the exogenous innovations $e_\tau$.
Under $q>1$, $u_t$ is a linear combination of $e_{tq}, e_{tq-1}, ...$, 
and its variance reflects units' bilateral exposure and mutual exposure to third units, 
as determined by network connections of orders $k \in 1:(q-1)$
along which the underlying innovations $e_\tau$ travel within a single period of observation.
For example, if $y_t$ is a flow variable and $q=2$,
we have $u_t = e_{tq} + (I+\alpha_1 A) e_{tq-1}$,
and $\covar(u_{it},u_{jt})$ is shaped by $a_{ij}$, $a_{ji}$ and $\{a_{ik}a_{jk}\}_{k=1:n}$.\footnote{
	If $\Sigma=diag(\sigma^2_1, ..., \sigma^2_n)$, then
	$ \covar(u_{it},u_{jt}) = 2\sigma_i^2\one{i=j} + \alpha_1 a_{ij}\sigma_j^2 + \alpha_1 a_{ji} \sigma_i^2+ \alpha_1^2 \sum_{k=1}^n a_{ik}a_{jk} \sigma^2_k $.
}

\cref{prop_NVARpq_SARSEM__limit_AutoregPol,prop_NVARpq_SARSEM__limit_stock,prop_NVARpq_SARSEM__limit_flow_obsinnov,prop_NVARpq_SARSEM__limit_flow_y} characterize observed dynamics in the limit as $q\rightarrow\infty$ and make the relation to SAR and SEM explicit.
In the limit, the autoregressive matrices of the observed process from \cref{prop_NVARpq_observeddynamics} vanish and all network effects are contained in the contemporaneous covariance of innovations.
In the flow case, the observed process then becomes observationally equivalent to $\bar y_t$ from the covariate-free SAR with Normal innovations.
\cref{prop_NVARpq_SARSEM__limit_stock} clarifies that this result requires temporal aggregation, i.e. the flow- rather than stock-case; as in \cref{prop_NVARpq_SARSEM__LTresp,prop_NVARpq_SARSEM__contempresp}, it is the summation of high-frequency innovations within an observational period that turns the \emph{long-run} objects of the NVAR into the \emph{contemporaneous} objects of the SAR.


So far, the analysis without covariates reveals that the SAR and SEM implicitly assume an infinitely high frequency of network interactions (relative to observations) 
and that the assumption of Normally distributed errors in these models can be motivated by temporally independent high-frequency innovations $e_\tau$ with $\var[e_\tau] = q^{-1}\Sigma_*$.
Adding covariates, the NVAR reveals that the SAR and SEM are the limits of two different specifications.
Specifically, with $Z_t$ observed at observational frequency and $m_t = Z_t \beta + A Z_t \gamma$ as in \cref{eq_SAR,eq_SEM}, consider
\begin{alignat}{3}
    x_\tau &= \alpha_1 A x_{\tau-1} + ... + \alpha_p A x_{\tau-p} + m_{t(\tau)}/q + e_\tau \; , &\quad
    y_{t} &= x_{tq} + ... + x_{tq-q+1}  &&\quad \text{(Spec. A)} \; , \label{eq_NVARpqWC_specA} \\
    x_\tau &= \alpha_1 A x_{\tau-1} + ... + \alpha_p A x_{\tau-p} + e_\tau \; , &\quad
    y_{t} &= x_{tq} + ... + x_{tq-q+1} + m_{t}  &&\quad \text{(Spec. B)} \; , \label{eq_NVARpqWC_specB}
\end{alignat}
where $t(\tau)=\text{ceil}(\tau/q)$ so that $m_{t(\tau)} = m_t$ for $\tau \in (tq-q+1):tq$.
\cref{prop_NVARpqWC_observeddynamics} states the observed process under either specification.
Based on it, \cref{prop_NVARpqWC_SARSEM} shows that specification A leads to the SAR of \cref{eq_SAR} in the limit as $q\rightarrow\infty$, while specification B leads to the SEM of \cref{eq_SEM}.\footnote{
	The factor $1/q$ in \cref{eq_NVARpqWC_specA} spreads covariates evenly over the $q$ interaction periods of an observational period, just as $\Sigma = q^{-1}\Sigma_*$ spreads the innovation variance. A covariate held fixed over these interaction periods acts like a permanent innovation for $q$ periods, and by \cref{prop_NVARpq_SARSEM__LTresp} the flow-response is $q$ times the SAR response, while by \cref{prop_NVARpq_SARSEM__limit_flow_obsinnov} the innovations accumulate only at rate $\sqrt{q}$. Without the $1/q$, the covariate-part of $y_t$ would dominate the innovation-part, and the limit would be a regression function without an error term.
}
Hence, the SAR implicitly assumes that covariates propagate through the network, whereas the SEM assumes that they do not.
This distinction matters for applied work: from a cross-section it is hard to tell apart interaction effects among the dependent variable and interaction effects among the error terms (see \citet[pp. 164--165]{LeSagePace2009}, \citet[pp. 31--33]{Elhorst2014}), 
and \cref{prop_NVARpqWC_SARSEM} turns this specification question into a question about timing that panel data at finite $q$ may be able to answer.
Note that the two specifications are different, non-nested models for every $q$, including $q=1$.
For $p=1$ and $q=1$, they are the time-space recursive models of \citet[pp. 190--203]{LeSagePace2009} who derive the SAR and SEM as their long-run equilibria. 

\myrepproposition{NVAR($p,q$): Observed Dynamics With Covariates}{prop_NVARpqWC_observeddynamics}{%
	Let $y_t$ evolve as in \cref{eq_NVARpqWC_specA} or \cref{eq_NVARpqWC_specB} for some $q\in\naturals$, with $e_\tau\sim WN(0,\Sigma)$, $Z_t$ w.s., and $(\alpha,A)$ s.t. $x_\tau$ in \cref{eq_NVARpq} is w.s..
	Let $\calG_q(z)$, $\nu_t$ and $q^*$ be as in \cref{prop_NVARpq_observeddynamics},
	let $\Theta_h = \partial x_{\tau+h}/ \partial x_{\tau}$,
	and let $y^*_t$ be the flow variable from \cref{eq_NVARpq} driven by the same $e_\tau$.
	Then
	\begin{enumerate}
		\item[(i)] under \cref{eq_NVARpqWC_specA}, 
			$y_t = \sum_{s \geq 0} \Xi_s m_{t-s} + y^*_t$, 
			and $y_t$ follows a VARMAX($p,q^*,q^*$): $\calG_q(L)y_t = \sum_{s=0}^{q^*}\Pi_s m_{t-s} + \nu_t$, 
			where $\{\Xi_s\}_{s \geq 0}$ and $\{\Pi_s\}_{s=0:q^*}$ are polynomials in $A$, 
			$\Xi_0 = q^{-1}\sum_{j=0}^{q-1}(q-j)\Theta_j$, 
			and $\|\Xi_s\| \leq c \varrho^{(s-1)q+1}/\br{q(1-\varrho)^2}$ for $s \geq 1$ and some $c < \infty$ and $\varrho \in (0,1)$ independent of $q$;
		
		\item[(ii)] under \cref{eq_NVARpqWC_specB}, 
			$y_t = m_t + y^*_t$, 
			and $y_t$ follows a VARMAX($p,q^*,p$): $\calG_q(L)y_t = \sum_{s=0}^{p}\Pi_s m_{t-s} + \nu_t$ with $\Pi_0 = I_n$ and $\Pi_s = -\calG_{q,s}$ for $s=1:p$.
	\end{enumerate}
}

\myrepproposition{NVAR($p,q$): Relation to SAR and SEM With Covariates}{prop_NVARpqWC_SARSEM}{%
	Let $y_t$ evolve as in \cref{eq_NVARpqWC_specA} or \cref{eq_NVARpqWC_specB} for some $q\in\naturals$, with $Z_t$ w.s. and independent of $\{e_\tau\}$, and $(\alpha,A)$ s.t. $x_\tau$ in \cref{eq_NVARpq} is w.s..
	Suppose $e_\tau \overset{i.i.d.}{\sim} (0,\Sigma)$ with $\Sigma = q^{-1}\Sigma_*$, where $\Sigma_*$ and $\sqrt{q}\,e_\tau$ satisfy the conditions of \cref{prop_NVARpq_SARSEM__limit_flow_obsinnov}, and let $\Gamma = (I-aA)^{-1}$.
	Then, as $q \rightarrow \infty$,
	\begin{enumerate}
		\item[(i)] under \cref{eq_NVARpqWC_specA}, $\forall \; S \in \naturals$,
			$\br{(y_t - \Gamma m_t)', ..., (y_{t+S} - \Gamma m_{t+S})'}' \convdist N\br{0, I_{S+1}\otimes\Gamma\Sigma_*\Gamma'}$ jointly with $\{Z_t\}$, where the limit is independent of $\{Z_t\}$, 
			i.e. $(y_t,Z_t)$ converges in distribution to the SAR in \cref{eq_SAR} with $v_t \overset{i.i.d.}{\sim} N(0,\Sigma_*)$ independent of $\{Z_t\}$;
		
		\item[(ii)] under \cref{eq_NVARpqWC_specB}, 
			the same holds for $\br{(y_t - m_t)', ..., (y_{t+S} - m_{t+S})'}'$, 
			i.e. $(y_t,Z_t)$ converges in distribution to the SEM in \cref{eq_SEM} with $v_t \overset{i.i.d.}{\sim} N(0,\Sigma_*)$ independent of $\{Z_t\}$.
	\end{enumerate}
}


\paragraph*{Difference to Exogenous Persistence} 

Instead of relying on lagged network interactions, one can also introduce transitional dynamics in the SAR through autocorrelated innovations $v_t$.
This approach is predominant in macroeconomic studies of input-output-economies; 
under fairly standard assumptions (see e.g. \citet{Carvalho-TahbazSalehi2019}), sectoral production and prices evolve as a covariate-free SAR, whereby $v_t$ denotes productivity,
which is typically assumed to follow an autoregressive process.
Accounting for serial correlation in disturbances is also common in the spatial panel literature \citep{LeeYu2011,Elhorst2012}.

Lagged network interactions and autocorrelated innovations lead to different kinds of dynamics.
As first illustrated in \cref{plot_model_GIRF_NVAR11} for the NVAR(1,1) and stated by \cref{prop_GCGIRF_NVARpq_stock,prop_GCGIRF_NVARpq_flow} for the general NVAR($p,q$), lagged network interactions relate impulse-response functions (IRFs) to network-connections of different order, with the precise mapping determined by $\alpha$ and $q$.
As a result, how strongly $y_i$ reacts to an impulse to $y_j$ depends on the strength of network-connections from $i$ to $j$ of relevant order,
and how fast it reacts depends on how many lower- as opposed to higher-order connections there are.
For instance, in \citet{GolubJackson2010}, an individual adjusts their opinion faster after news shared by closer than more distant friends.
Similarly, in \citet{LongPlosser1983}, a sector contracts its production faster after a negative productivity shock to more immediate suppliers than to ones located further upstream.

In contrast, autocorrelated innovations lead to exponentially decaying impulse-responses. 
This shape is preserved when they are coupled with contemporaneous network interactions:
if $\bar y_t = aA \bar y_t + v_t$ and $(1-\rho_jL)v_{jt} = \epsilon_{jt}$, then
\begin{align} \label{eq_GCGIRF_SAR_AR1innov}
	\frac{\partial \bar y_{i,t+h}}{\partial \epsilon_{jt}} = \sbr{(I-aA)^{-1}}_{ij} \rho_j^h \; .
\end{align}
In this case, the strength of connections from $i$ to $j$ (of all order) merely scales the exponentially decaying impulse-response; network connections merely amplify exogenous dynamics.

\begin{figure}[t]
	\captionsetup[subfigure]{justification=centering}
	\begin{center}
		\figspaceBeforePlotArea
		\subcaption*{\footnotesize{$\alpha = (.8,.2)', \rho_j = .6$}}
		\begin{subfigure}[b]{\plotSizeTriple\textwidth}
				\centering
				\subcaption*{\scriptsize{Transmission of Single Innovation\\($c^h_{\underline{k}(h)}:c^h_h$ for $h=1:6$)}}
				\includegraphics[width=1\textwidth, clip]{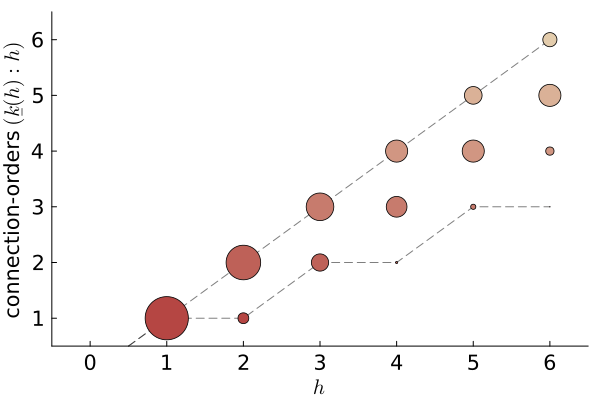}
		\end{subfigure}
		\begin{subfigure}[b]{\plotSizeTriple\textwidth}
				\centering
				\subcaption*{\scriptsize{$IRF_{ij}$ to Persistent Shock,\\$(A^k)_{ij}=\one{k=1}$}}
				\includegraphics[width=1\textwidth, clip]{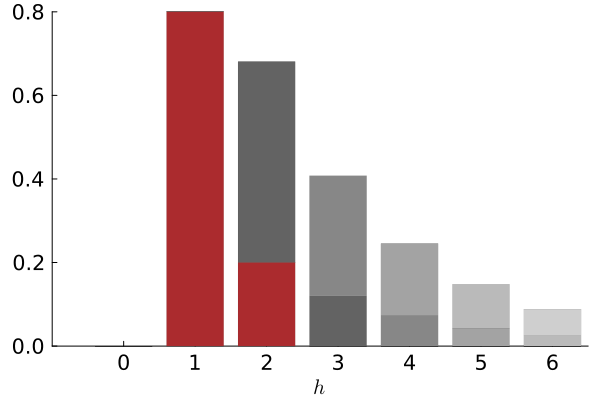}
		\end{subfigure}
		\begin{subfigure}[b]{\plotSizeTriple\textwidth}
				\centering
				\subcaption*{\scriptsize{$IRF_{ij}$ to Persistent Shock,\\$(A^k)_{ij}=\one{k=2}$}}
				\includegraphics[width=1\textwidth, clip]{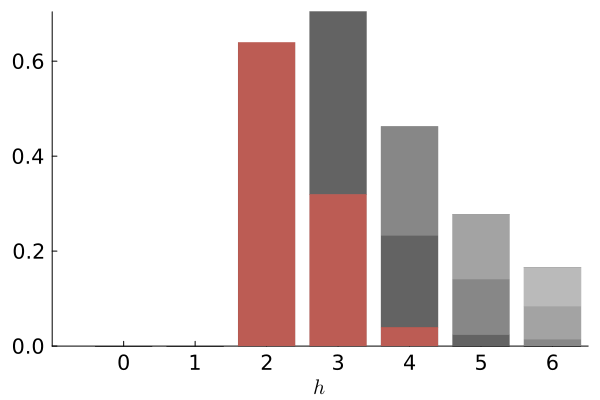}
		\end{subfigure}

		\figspaceBetweenPlotRows

		\subcaption*{\footnotesize{$\alpha = (.2,.8)', \rho_j = .6$}}
		\vspace*{5pt}
		\begin{subfigure}[b]{\plotSizeTriple\textwidth}
				\centering
				\includegraphics[width=1\textwidth, clip]{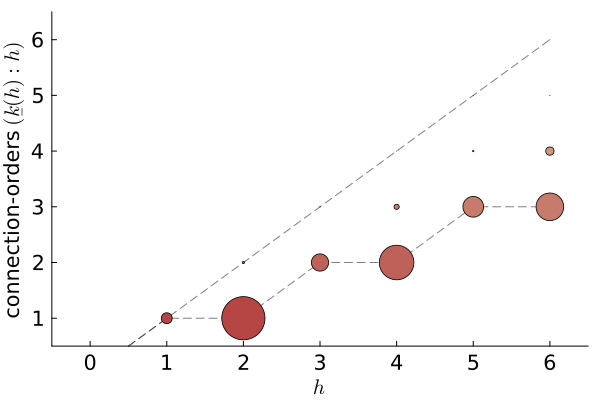}
		\end{subfigure}
		\begin{subfigure}[b]{\plotSizeTriple\textwidth}
				\centering
				\includegraphics[width=1\textwidth, clip]{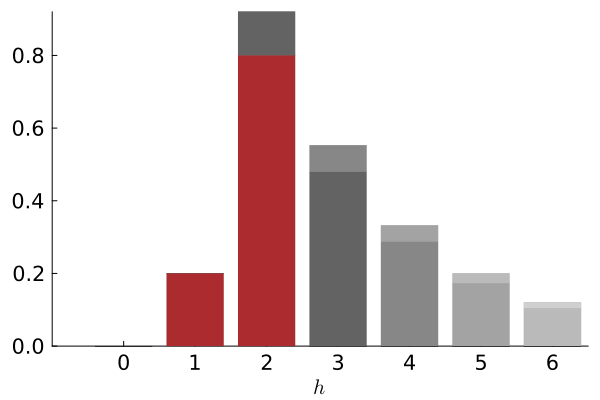}
		\end{subfigure}
		\begin{subfigure}[b]{\plotSizeTriple\textwidth}
				\centering
				\includegraphics[width=1\textwidth, clip]{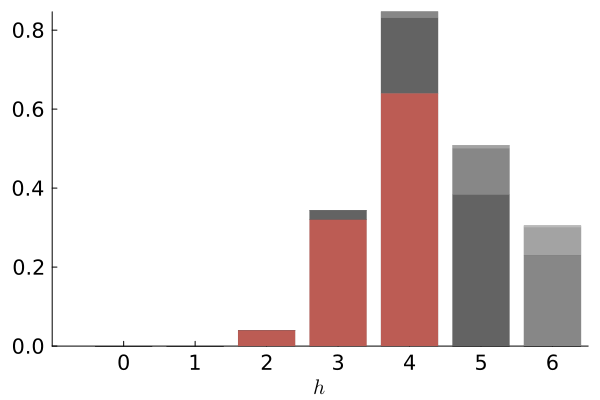}
		\end{subfigure}
		\figspaceAfterPlotArea

	\end{center}
	\caption{Composition of Impulse-Responses under NVAR($2,1$) with AR($1$)-Innovations\figspaceBetweenCaptionAndNotes
	\scriptsize {\em Notes:} 
	The top row refers to $\alpha=(.8,.2)'$, the bottom row to $\alpha=(.2,.8)'$. 
	The left panel depicts the coefficients $c^h_{\underline{k}(h)}:c^h_h$ from \cref{prop_GCGIRF_NVARp1} for $h=1:6$, which illustrate how an innovation transmits along connection-orders (y-axis) over time (x-axis; horizons). 
	The dashed lines bound the relevant connection-orders in an NVAR($2,1$): $k\in \underline{k}(h):h$. 
	The middle panel plots the IRF from \cref{eq_GCGIRF_NVARp1_AR1innov} under $\rho_j=.6$ for a pair $(i,j)$ with only a first-order connection. 
	The right panel repeats this for a pair with only a second-order connection.}
	\label{plot_model_example_HorizonsOrdersIRFs}

\end{figure}

When the two sources of dynamics are combined, their qualitatively distinct contributions remain.
Consider an NVAR($p,1$), and let $(1-\rho_jL)u_{jt} = \varepsilon_{jt}$. 
Then, using \cref{prop_GCGIRF_NVARp1},
\begin{align} 
	\frac{\partial y_{i,t+h}}{\partial \varepsilon_{jt}} 
	= \sum_{l=0}^h \frac{\partial y_{i,t+h}}{\partial u_{j,t+h-l}} \frac{\partial u_{j,t+h-l}}{\partial \varepsilon_{j,t}} 
	= \sum_{l=0}^h \cbr{ c^{l}_{\underline{k}(l)}(\alpha) \sbr{A^{\underline{k}(l)}}_{ij} + ... + c^{l}_{l}(\alpha) \sbr{A^{l}}_{ij} } \rho_j^{h-l} \; , \label{eq_GCGIRF_NVARp1_AR1innov}
\end{align}
with $\underline{k}(l) = \text{ceil}(l/p)$.
%
Under $p=1$ and $\alpha_1 = 1$, this simplifies to $\sum_{l=0}^h \sbr{A^l}_{ij} \rho_j^{h-l}$.
While $\alpha$ determines the timing of innovation transmission along direct links and therefore also the timing of transmission along higher-order connections from $i$ to $j$, $\rho_j$ determines the persistence of $y_i$'s response after each round of transmission. 
\cref{plot_model_example_HorizonsOrdersIRFs} illustrates for an NVAR($2,1$) and $\rho_j=.6$, once under $\alpha = (.8,.2)'$ (top row), once under $\alpha = (.2,.8)'$ (bottom row).
The left panel depicts how connection-orders accumulate over horizons. 
The middle and right panels show the resulting impulse-responses for two sectors that share only a first- or only a second-order connection, respectively.
Under $\alpha=(.8,.2)'$, the innovation travels faster through the network than under $\alpha = (.2,.8)'$:
in the former case, the transmission along a direct link (darkest red dots and bars) is strongest after one period, in the latter after two periods.
Consequently, the transmission along second-order connections (lighter red dots and bars) is strongest in the second period under $\alpha=(.8,.2)'$ and in the fourth period under $\alpha = (.2,.8)'$.
Exogenous persistence carries the fraction $\rho_j$ of any transmission at horizon $h$ over to horizon $h+1$, the fraction $\rho_j^2$ to horizon $h+2$, etc. (ever lighter gray bars).
%
%
%
\cref{prop_LTresp_NVARpq_AR1innov} shows that the NVAR and SAR yield the same long-run responses even under AR(1)-innovations.

\myrepproposition{NVAR($p,q$): Long-Term Response to Permanent AR(1)-Innovations}{prop_LTresp_NVARpq_AR1innov}{%
	Let $y_t$ evolve as in \cref{eq_NVARpq} for some $q \in \naturals$, and let $\bar y_t = a A \bar y_t + v_t$, with $a =\sum_{l=1}^p \alpha_l$.
	Assume $x_\tau$, $y_t$ and $\bar y_t$ are w.s. and, for $\rho_j \in [0,1)$, $(1-\rho_j L) e_{j\tau} = \varepsilon_{j\tau} \sim WN$ and $(1-\rho_j L) v_{jt} = \epsilon_{jt} \sim WN$.
	Define $\kappa_q$ as in \cref{prop_NVARpq_SARSEM}.
	Then
	\begin{align*}
		\underset{H\rightarrow \infty}{\lim}  \sbr{ \frac{\partial y_{i,t+H}}{\partial \varepsilon_{j,tq}}  + ... + \frac{\partial y_{i,t+H}}{\partial \varepsilon_{j,(t+H)q}} }  
		= \kappa_q \underset{H\rightarrow \infty}{\lim} \sbr{ \frac{\partial \bar y_{i,t+H}}{\partial \epsilon_{j,t}}  + ... + \frac{\partial \bar y_{i,t+H}}{\partial \epsilon_{j,t+H}} }
		= \kappa_q \frac{1}{1-\rho_j}\sbr{(I- a A)^{-1}}_{ij} \; .
	\end{align*}
}

%% file: part_Inference.tex
\section{Lagged Network Effects: Large $T$-Inference}
\label{sec_inference}

The NVAR can be used to estimate dynamic network effects.
For a fixed $A$, and given $p$ and $q$, these are determined by $\alpha$, the timing of transmission along a direct link.
In \cref{subsec_inference_ID}, I analyze the identification of $\alpha$, $\Sigma$ and potential covariate-effects $(\beta,\gamma)$ under large $T$ and fixed $n$,
based on which I discuss estimation in \cref{subsec_inference_est}. 
\cref{appsec_inference} accompanies this section.




\subsection{Identification}
\label{subsec_inference_ID}

The parameter $\alpha$ is not necessarily identified:
under the stock case with $q=2$, 
$y_t = x_{t2}$ with $F(L)x_\tau = e_\tau$ and $F(z) = I_n - A\sum_{k=1}^p \alpha_k z^k$ 
is observationally equivalent to 
$y_t = \tilde{x}_{t2}$ with $F(-L)\tilde{x}_\tau = \tilde e_\tau$, $\tilde{x}_\tau = (-1)^\tau x_\tau$ and $\tilde e_\tau = (-1)^\tau e_\tau$ so that we cannot learn the (joint) sign of $(\alpha_1,\alpha_3,...,\alpha_{2\text{ceil}(p/2)-1})'$ in this case.
The identification problem is akin to that under an analogously defined AR($p,q$) process, i.e.  -- with slight abuse of notation -- a scalar-AR($p$) process $x_\tau = \phi_1 x_{\tau-1} + \dots + \phi_p x_{\tau - p} + e_\tau$ observed every $q \in \naturals$ periods as $y_t$ under the same relationship as in \cref{eq_NVARpq}.
The identification of $(\phi,v)$, $\phi = (\phi_1,\dots,\phi_p)' \in \reals^p$ in this environment is a long-standing question in the literature \citep{Telser1967,PalmNijman1984}.
Building on this literature, in \citet{P017} I derive the identified set of $(\phi,v)$ for general $p$ and $q$ under mild assumptions.
%
Writing it as $\calI^{s,1}_{p,q}(\phi,v)$ in the stock case and as $\calI^{f,1}_{p,q}(\phi,v)$ in the flow case, my exact but implicit expressions lead to the following explicit but conjectured statement:
$\calI^{f,1}_{p,q}(\phi,v) = \{(\phi,v)\}$,
and $\calI^{s,1}_{p,q}(\phi,v) = \{(\phi,v)\}$ if $q$ is odd and $\calI^{s,1}_{p,q}(\phi,v) = \{(\phi,v),(\phi^-,v)\}$ if $q$ is even,
where $\phi^- = (-\phi_1,\phi_2,-\phi_3,\dots,(-1)^p\phi_p)'$.
I prove this
for $q=1$, 
for $q=2$, 
under real roots of $\phi(z)$ in the stock case,
and under real roots and $q$ odd in the flow case.
For more general cases -- different $(\phi,v)$ with $p\in 1:12$ and $q\in 1:10$ --, I verify the conjecture numerically using the algorithm I provide to calculate all candidate points and determine their membership in $\calI^{s,1}_{p,q}(\phi,v)$ and $\calI^{f,1}_{p,q}(\phi,v)$.

The identified set of $(\alpha,\Sigma)$ in the NVAR($p,q$) can be characterized based on that of $(\phi,v)$ in the AR($p,q$).
I consider the parameter space
$$ \Theta_p^A = \Big\{ (\alpha,\Sigma) : \det F(z) \neq 0 \; \forall \; |z| \leq 1 \; , \; \Sigma \; \text{symmetric p.d.} \Big\}\;, $$
Two points of $\Theta_p^A$ are \emph{observationally equivalent} if they induce the same distribution of $\{y_t\}$,
and the \emph{identified set} of a true value $(\alpha,\Sigma)$ is the set of all points of $\Theta_p^A$ observationally equivalent to it,
written as $\calI^{s,A}_{p,q}(\alpha,\Sigma)$ and $\calI^{f,A}_{p,q}(\alpha,\Sigma)$ for the stock- and flow-case, respectively.
I write $\calI^{s,A}_{p,q}(\alpha)$ and $\calI^{f,A}_{p,q}(\alpha)$ for their projections on $\alpha$. 
As in \cref{prop_WS_NVARpq}, define $\phi_\lambda(z) = 1 - \lambda \sum_{k=1}^{p}\alpha_k z^k$ for an eigenvalue $\lambda$ of $A$,
and let $\alpha^- = (-\alpha_1,\alpha_2,-\alpha_3,\dots,(-1)^p\alpha_p)'$.
%

\myassumptionNVARpq{}{ass_NVARpq_ID_stationarity}{
    $x_\tau$ is w.s., i.e. $\det F(z) \neq 0 \; \forall \; |z| \leq 1$.
}

\myassumptionNVARpq{}{ass_NVARpq_ID_actualorderisp}{
    $\alpha_p \neq 0$.
}

\myassumptionNVARpq{}{ass_NVARpq_ID_distincteigvalqthpowers}{
    $A$ has a real eigenvalue $\lambda_* \neq 0$ with a real left eigenvector $w_* \neq 0$, and the roots $\eta_1,\dots,\eta_p$ of $\phi_{\lambda_*}$ satisfy $\eta_i^q \neq \eta_j^q$ for all $i \neq j$.\footnote{
        A real eigenvalue of a real matrix always has a real left eigenvector. 
        If $A$ has non-negative entries and spectral radius $\varrho(A) > 0$, the first part holds automatically with $\lambda_* = \varrho(A)$, by the Perron--Frobenius theorem. 
        The second part is \cref{ass_ARpq_ID_distincteigvalqthpowers} of \citet{P017} for $\phi_{\lambda_*}$.
    }
}

\myassumptionNVARpq{}{ass_NVARpq_ID_normality}{
    $e_\tau \overset{i.i.d.}{\sim} N(0,\Sigma)$.
}

\myassumptionNVARpq{}{ass_NVARpq_ID_Adiagonalizable}{
    $A$ is diagonalizable: it has left eigenpairs $(\lambda_j, w_j)$, $j = 1:n$, possibly complex, s.t. $W = [w_1,\dots,w_n]$ is invertible.
}

\myassumptionNVARpq{}{ass_NVARpq_ID_techassforSigma_stock}{
    For all eigenvalues $\lambda_j, \lambda_k$ of $A$, $[H_{jk}]_q \neq 0$.\footnote{
        $H_{jk}(z) = [ \phi_{\lambda_j}(z)\,\phi_{\lambda_k}(z^{-1}) ]^{-1}$, and for a Laurent function $Q(z)$, I define $[Q]_q(z) = q^{-1}\sum_{j=0}^{q-1} Q(\wq^j z)$ with $\wq = e^{2\pi \ii/q}$, which keeps only the powers of $z$ divisible by $q$. 
    }
}

\myassumptionNVARpq{}{ass_NVARpq_ID_techassforSigma_flow}{
    For all eigenvalues $\lambda_j, \lambda_k$ of $A$, $[S_q H_{jk}]_q \neq 0$.\footnote{
        I define $s_q(z) = 1 + z + \dots + z^{q-1}$ s.t. $y_t = X_{tq}$ for $X_\tau = s_q(L)x_\tau$. 
        In turn, I define $S_q(z) = s_q(z)s_q(z^{-1})$.
    }
}

Under the mild \cref{ass_NVARpq_ID_stationarity,ass_NVARpq_ID_actualorderisp,ass_NVARpq_ID_distincteigvalqthpowers,ass_NVARpq_ID_normality} -- analogous to \cref{ass_ARpq_ID_stationarity,ass_ARpq_ID_actualorderisp,ass_ARpq_ID_distincteigvalqthpowers,ass_ARpq_ID_normality} of \citet{P017} restated in \cref{appsec_inference} and linked to them by \cref{lemma_NVARtoARlink} --, 
\cref{prop_ID_NVARpq} derives the identified set for $\alpha$ based on the identified set of $(\phi,v) = (\lambda_*\alpha, w_*'\Sigma w_*)$.
The numerically verified conjecture from \citet{P017} then yields
$\calI^{s,A}_{p,q}(\alpha) = \{\alpha\}$ for $q$ odd, $\calI^{s,A}_{p,q}(\alpha) = \{\alpha,\alpha^-\}$ for $q$ even, and $\calI^{f,A}_{p,q}(\alpha) = \{\alpha\}$.
If in addition \cref{ass_NVARpq_ID_Adiagonalizable,ass_NVARpq_ID_techassforSigma_stock,ass_NVARpq_ID_techassforSigma_flow} hold, then $\Sigma$ is point-identified.
In the stock case with $q$ even, $\alpha$ is point-identified under sign-restrictions that can be specified based on the interpretation of $\alpha$ as the timing of innovation transmission along a direct link.

\myrepproposition{NVAR($p,q$): Identified Set via AR($p,q$)}{prop_ID_NVARpq}{
    Let $y_t$ evolve as in \cref{eq_NVARpq} for some $q \in \naturals$. 
    Suppose we only observe $y_t$ and \cref{ass_NVARpq_ID_stationarity,ass_NVARpq_ID_actualorderisp,ass_NVARpq_ID_distincteigvalqthpowers,ass_NVARpq_ID_normality} hold.
    Define $(\phi,v) = (\lambda_*\alpha, w_*'\Sigma w_*)$.
    %
    Then
    \begin{enumerate}
        \item[(i)] $\calI^{s,A}_{p,q}(\alpha) \subseteq \cbr{ \lambda_*^{-1}\tilde\phi : (\tilde\phi,\tilde v) \in \calI^{s,1}_{p,q}(\phi,v) }$;
        \item[(ii)] in stock case for $q$ even, $(\alpha^-,\Sigma)$ is observationally equivalent to $(\alpha,\Sigma)$, and $\alpha^- \neq \alpha$;
        \item[(iii)] if $\calI^{s,1}_{p,q}(\phi,v) = \{(\phi,v)\}$ for $q$ odd and $\calI^{s,1}_{p,q}(\phi,v)=\{(\phi,v),(\phi^-,v)\}$ for $q$ even (conjecture in \citet{P017}), then $\calI^{s,A}_{p,q}(\alpha) = \{\alpha\}$ for $q$ odd and $\calI^{s,A}_{p,q}(\alpha) = \{\alpha,\alpha^-\}$ for $q$ even;
        \item[(iv)] if $\calI^{s,A}_{p,q}(\alpha)$ is as in (iii) and in addition \cref{ass_NVARpq_ID_Adiagonalizable,ass_NVARpq_ID_techassforSigma_stock} hold, then $\calI^{s,A}_{p,q}(\alpha,\Sigma) = \{(\alpha,\Sigma)\}$ for $q$ odd and $\calI^{s,A}_{p,q}(\alpha,\Sigma) = \{(\alpha,\Sigma),(\alpha^-,\Sigma)\}$ for $q$ even;
    \end{enumerate}
    %
    \begin{enumerate}
        \item[(v)] $\calI^{f,A}_{p,q}(\alpha) \subseteq \cbr{ \lambda_*^{-1}\tilde\phi : (\tilde\phi,\tilde v) \in \calI^{f,1}_{p,q}(\phi,v) }$;
        \item[(vi)] if $\calI^{f,1}_{p,q}(\phi,v) = \{(\phi,v)\}$ (conjecture in \citet{P017}), then $\calI^{f,A}_{p,q}(\alpha) = \{\alpha\}$;
        \item[(vii)] if $\calI^{f,A}_{p,q}(\alpha)$ is as in (vi) and in addition \cref{ass_NVARpq_ID_Adiagonalizable,ass_NVARpq_ID_techassforSigma_flow} hold, then $\calI^{f,A}_{p,q}(\alpha,\Sigma) = \{(\alpha,\Sigma)\}$.
    \end{enumerate}
}

\cref{ass_NVARpq_ID_distincteigvalqthpowers} is crucial.
Its first part 
guarantees that $w_*'x_\tau$ follows an AR($p,q$) and enables me to link the identified set of $(\alpha,\Sigma)$ to that of $(\phi,v)$.
This bridge exists only because the NVAR loads a single scalar $\alpha_l$ on the same $A$ at every lag.
Its second part 
holds for all parameter values outside a Lebesgue-null set, so that my identification analysis applies at any finite $q$.
However, the separation $\min_{i\neq j}|\eta_i^q-\eta_j^q|$ shrinks
geometrically in $q$ so that identification becomes weaker as $q$ grows and is lost in the limit as $q \rightarrow\infty$.
In that limit, characterized by \cref{prop_NVARpq_SARSEM,prop_NVARpqWC_SARSEM}, at most $a=\sum_{l=1}^p \alpha_l$ is identified, as discussed in the spatial econometrics literature.
Specifically, in the flow-case limit, $y_t$ is WN with variance $\gamma_y(0)=\Gamma\Sigma_*\Gamma'$ with $\Gamma=(I-aA)^{-1}$ and $\Sigma_* = q\Sigma$ (\cref{prop_NVARpq_SARSEM__limit_flow_y}).\footnote{
    Under the same variance-scaling, the stock-case limit is degenerate by \cref{prop_NVARpq_SARSEM__limit_stock}.
    %
}
Given $\Sigma_*$, $a$ is then identified if all eigenvalues of $A$ are real.\footnote{
    If a weakly stationary $\tilde a$ also satisfies $\tilde\Gamma\Sigma_*\tilde\Gamma' = \Gamma\Sigma_*\Gamma'$, 
    then $K = \tilde\Gamma^{-1}\Gamma = (I-\tilde aA)(I-aA)^{-1}$ satisfies $K\Sigma_*K' = \Sigma_*$, 
    so that $\Sigma_*^{-1/2}K\Sigma_*^{1/2}$ is orthogonal and every eigenvalue $(1-\tilde a\lambda_j)/(1-a\lambda_j)$ of $K$ has modulus one. 
    For real $\lambda_j$, weak
    stationarity makes numerator and denominator positive so that every eigenvalue equals one. 
    An orthogonal matrix with all eigenvalues equal to one is the identity, hence $K=I$ and $\tilde a=a$ for $A\neq 0$.
}
%
Joint identification with unrestricted $\Sigma_*$ fails since $(a,\Sigma_*) \mapsto \gamma_y(0)$ is not injective, though covariates may render $a$ identified:
in the SAR-limit of \cref{eq_NVARpqWC_specA}, the limit-loading on the $i$th covariate is $(I-aA)^{-1}(\beta_i I + \gamma_i A)$ by \cref{prop_NVARpqWC_SARSEM_specA} 
so that $a$ is identified provided $I,A,A^2$ are linearly independent and $\gamma + a \beta \neq 0$ \citep[Proposition 1]{BramoulleDjebbariFortin2009},
whereas in the SEM-limit of \cref{eq_NVARpqWC_specB}, the limit-loading $\beta_i I + \gamma_i A$ is independent of $a$
so that restrictions on $\Sigma_*$ are needed to identify $a$ \citep{KelejianPrucha1999,KelejianPrucha2010JoE,LeeYu2011}.

With covariates and finite $q$, the parameter of interest is $(\alpha,\beta,\gamma,\Sigma) \in \Theta_p^A \times \reals^{2k}$.
I define two points to be observationally equivalent if they induce the same conditional distribution of $\cbr{y_t}$ given $\cbr{Z_t}$,
and I write the identified sets as $\calI^{f_s,A}_{p,q}(\alpha,\beta,\gamma,\Sigma)$ for $s\in \{A,B\}$, with projections $\calI^{f_s,A}_{p,q}(\alpha,\beta,\gamma)$ and $\calI^{f_s,A}_{p,q}(\alpha)$.
It is convenient to write $m_t = Mz_t$ with $z_t = vec(Z_t) \in \reals^{nk}$ and $M=\beta' \otimes I_n + \gamma' \otimes A = [M_1,...,M_k]$, where $M_i = \beta_i I_n + \gamma_i A$.

\myassumptionNVARpqWC{}{ass_NVARpqWC_strictexo}{
    $\cbr{z_t}$ is independent of $\cbr{e_\tau}$, strictly stationary and ergodic with $\mean||z_t||^2 < \infty$, and has a spectral density $f_z(\omega)$ that is bounded and non-singular $\forall \; \omega \in [-\pi,\pi]$.
}

\myassumptionNVARpqWC{}{ass_NVARpqWC_contemploading}{
    $\Xi_0(\alpha) = q^{-1}\sum_{j=0}^{q-1}(q-j)\Theta_j$ from \cref{prop_NVARpqWC_observeddynamics} is non-singular.
}

\myassumptionNVARpqWC{}{ass_NVARpqWC_nonscalarA}{
    $\nexists c \in \reals$ s.t. $A = c I_n$.
}

Let $\mu_t = \mean[y_t|\cbr{Z_{t-s}}_{s=0}^\infty]$.
Under \cref{ass_NVARpqWC_strictexo}, we have
$\mu_t = \sum_{s=0}^\infty \Xi_s(\alpha)M(\beta,\gamma) z_{t-s}$ under \cref{eq_NVARpqWC_specA}
and
$\mu_t = M(\beta,\gamma) z_t$ under \cref{eq_NVARpqWC_specB},
while in either case $y_t - \mu_t=y_t^*$ equals the covariate-free NVAR($p,q$) (all by \cref{prop_NVARpqWC_observeddynamics}).
Under \cref{ass_NVARpq_ID_normality}, the conditional distribution of $\cbr{y_t}$ given $\cbr{Z_t}$ is Gaussian, so two points are observationally equivalent iff they induce the same $\mu_t$ and autocovariance function of $y_t^*$.
As a result, the identification problem splits into the problem for $(\alpha,\Sigma)$ treated above and a regression problem for $(\beta,\gamma)$ given $\alpha$.\footnote{
    \cref{ass_NVARpqWC_nonscalarA} makes $(\beta,\gamma) \mapsto M$ injective: if $\beta'\otimes I_n+\gamma'
    \otimes A=0$, then $\beta_iI_n+\gamma_iA=0$ for each $i$, which under this assumption forces $\gamma_i=\beta_i=0$.
    \cref{ass_NVARpqWC_contemploading} is only used under specification A and involves the contemporaneous loading of $y_t$ on $m_t$. Note that 
    $w'\Theta_j=\psi_j(\lambda\alpha)w'$ for a left eigenpair $(\lambda,w)$ of $A$, where $\psi_j(\lambda\alpha)$ are the impulse-responses of the scalar AR$(p)$ with polynomial $\phi_\lambda$, so that $\Xi_0$ has the eigenvalue $q^{-1}\sum_{j=0}^{q-1}(q-j)\psi_j(\lambda\alpha)$.
}
Using this insight, \cref{prop_ID_NVARpqWC} characterizes the identified set of the NVAR with covariates.
It shows that the reflection problem of \citet{Manski1993} only arises under contemporaneous network effects, while under any finite $q$, the condition of \citet{BramoulleDjebbariFortin2009} that $I, A, A^2$ must be linearly independent can be weakened to \cref{ass_NVARpqWC_nonscalarA}.

\myrepproposition{NVAR($p,q$) With Covariates: Identified Set}{prop_ID_NVARpqWC}{
    Let $y_t$ evolve as in \cref{eq_NVARpqWC_specA} or \cref{eq_NVARpqWC_specB} for some $q \in \naturals$.
    Suppose we observe $\cbr{y_t, Z_t}$ and \cref{ass_NVARpq_ID_stationarity,ass_NVARpq_ID_actualorderisp,ass_NVARpq_ID_distincteigvalqthpowers,ass_NVARpq_ID_normality,ass_NVARpqWC_strictexo,ass_NVARpqWC_contemploading,ass_NVARpqWC_nonscalarA} hold, and define $(\phi,v)$ as in \cref{prop_ID_NVARpq}.
    Then
    \begin{enumerate}
        \item[(i)] under \cref{eq_NVARpqWC_specB}, $\calI^{f_B,A}_{p,q}(\alpha,\beta,\gamma,\Sigma) = \calI^{f,A}_{p,q}(\alpha,\Sigma) \times \cbr{(\beta,\gamma)}$ so that \cref{prop_ID_NVARpq_flow_one,prop_ID_NVARpq_flow_two,prop_ID_NVARpq_flow_three} apply to $(\alpha,\Sigma)$ unchanged and $(\beta,\gamma)$ is point-identified;
        \item[(ii)] under \cref{eq_NVARpqWC_specA}, $\calI^{f_A,A}_{p,q}(\alpha) \subseteq \calI^{f,A}_{p,q}(\alpha)$, and if $(\tilde\alpha,\tilde\beta,\tilde\gamma,\tilde\Sigma) \in \calI^{f_A,A}_{p,q}(\alpha,\beta,\gamma,\Sigma)$ with $\tilde\alpha = \alpha$, then $(\tilde\beta,\tilde\gamma) = (\beta,\gamma)$;
        \item[(iii)] under \cref{eq_NVARpqWC_specA}, if $\calI^{f,1}_{p,q}(\phi,v) = \{(\phi,v)\}$ (conjecture in \citet{P017}), then $\calI^{f_A,A}_{p,q}(\alpha,\beta,\gamma) = \cbr{(\alpha,\beta,\gamma)}$, and if in addition \cref{ass_NVARpq_ID_Adiagonalizable,ass_NVARpq_ID_techassforSigma_flow} hold, then $\calI^{f_A,A}_{p,q}(\alpha,\beta,\gamma,\Sigma) = \cbr{(\alpha,\beta,\gamma,\Sigma)}$.
    \end{enumerate}
}




\subsection{Estimation}
\label{subsec_inference_est}

Under $q=1$, the NVAR without covariates from \cref{eq_NVARp1} can be written as the linear regression 
$y_{it} = x_{it}^{y\prime} \alpha + u_{it}$ or $y_t = X^y_t \alpha + u_t$,
where $x^y_{it}$ is the $i$th row of $X^y_t = [A y_{t-1}, ..., A y_{t-p}]$ and stacks $\sum_{j=1}^n a_{ij}y_{j,t-l}$ for $l=1:p$.
As usual, this leads to the conditional estimators
$\hat{\alpha}|\Sigma =  \sbr{ \sum_{t=1}^T X_t^{y\prime}\Sigma^{-1}X_t^y }^{-1} \sbr{  \sum_{t=1}^T X_t^{y\prime} \Sigma^{-1} y_t }$
and $\hat\Sigma|\alpha = T^{-1}\sum_{t=1}^T u_t u_t'$.
With covariates as in \cref{eq_NVARpqWC_specA}, we have $y_{it} = x_{it}'b + u_{it}$ with $b = (\alpha',\beta',\gamma')'$ and $x_{it} = (x_{it}^{y\prime},Z_{it}',x_{it}^{Z\prime})'$, where $Z_{it}'$ and $x_{it}^{Z\prime}$ are the $i$th rows of $Z_t$ and $AZ_t$, respectively, leading to analogous estimators.
With covariates as in \cref{eq_NVARpqWC_specB}, we have $F(L)(y_t - m_t) = u_t$, a regression of $y_t$ on $Z_t$ and $AZ_t$ with NVAR($p,1$)-errors that is linear in $\alpha$ given $(\beta,\gamma)$ and vice versa. 
This leads to three conditional estimators: 
regressing $y^*_t = y_t - m_t$ on $X^{y*}_t$ 
yields $\hat\alpha|(\Sigma,\beta,\gamma)$, 
regressing $F(L)y_t$ on $F(L)Z_t$ and $AF(L)Z_t$ yields $(\hat\beta,\hat\gamma)|(\Sigma,\alpha)$, and $\hat\Sigma|(\alpha,\beta,\gamma)$ is as above.
In each case, the joint estimator is obtained by iterating on these conditional estimators until convergence \citep{MengRubin1993}.



For $q>1$, $(\alpha,\beta,\gamma,\Sigma)$ can be estimated by Maximum Likelihood (ML), using the Kalman filter to numerically evaluate the likelihood, as I do in \cref{sec_app1}.\footnote{
    In the state space models defined by \cref{eq_NVARpq,eq_NVARpqWC_specA,eq_NVARpqWC_specB}, no data is observed in some periods $\tau$.
    In these periods, the updating-step of the Kalman filter is skipped and the updated distribution of states equals their predicted distribution. 
    See \citet{SchorfheideSong2015}, whose mixed-frequency VAR confronts the same issue. 
}
While the likelihood can be expressed analytically using the VARMA(X)-representation of the NVAR from \cref{prop_NVARpq_observeddynamics,prop_NVARpqWC_observeddynamics}, it does not yield analytical expressions for the estimators. 
Alternative estimation methods based on data augmentation are numerically less stable since there is little information to pin down $x_\tau$ in periods $\tau \neq tq$.



By \cref{prop_NVARpq_observeddynamics}, the observed process without covariates follows a VARMA whose autoregressive, moving-average- and innovation covariance-matrices are smooth functions of $(\alpha,\Sigma)$, given $A$ and $q$.
If 
(i) $x_\tau$ is w.s., 
(ii) the VARMA representation is invertible, 
(iii) $e_\tau$ is i.i.d. Gaussian -- or, more generally, the innovations $\nu_t$ of the VARMA representation are mean-zero i.i.d. with bounded fourth moments -- 
and (iv) $\Sigma_\nu = \var[\nu_t]$ is non-singular,
then \citet[Proposition 12.1 with its Remark 2]{Luetkepohl2005} applies to $\theta=(\alpha',vech(\Sigma)')'$: provided that $(\alpha,\Sigma)$ are point-identified, the Gaussian ML estimator satisfies 
$$ \sqrt{T}\br{\hat\theta-\theta} \convdist N\br{0, \sbr{G(\theta)'\calI(\vartheta)G(\theta)}^{-1}} \; , \quad G(\theta) = \partial\vartheta/\partial\theta' \; , $$
where $\vartheta$ collects the parameters of the unrestricted VARMA process and $\calI(\vartheta)$ is their asymptotic information matrix.
For $q=1$, $\vartheta$ collects $\{\alpha_lA\}_{l=1:p}$ and $\Sigma$, $G$ is available analytically, and the result reduces to the standard asymptotics of the restricted VAR estimator from \citet[Propositions 5.1 and 5.5]{Luetkepohl2005}.
For $q>1$, $G$ has to be computed numerically.

With covariates, the analogous limit-distribution is obtained for the Gaussian ML estimator of $\theta = (\alpha',\beta',\gamma',vech(\Sigma)')'$ conditional on $\cbr{Z_t}$ provided that in addition \cref{ass_NVARpqWC_strictexo} holds.
This follows from \citet[Theorems 4.2.1 and 4.3.1]{HannanDeistler1988}.
\cref{ass_NVARpqWC_strictexo} delivers their conditions on the observed input, 
and \cref{ass_NVARpq_ID_normality} delivers their conditions on the VARMAX-innovations $\nu_t$.



How precisely $\alpha$ can be estimated depends on $q$, and it does so non-monotonically.
%
\cref{appsec_inference} shows this analytically for an NVAR($1,q$) observed as a stock variable without covariates, exploiting again the link to \citet{P017}:
projecting on a real eigen-pair $(\lambda,w)$ of $A$ turns the observed process into an AR($1$) with autocorrelation coefficient $\psi = (\lambda\alpha)^q$.
Per-observation information about $\alpha$ then equals $\calI_{\alpha\alpha}(q) = (q^2/\alpha^2)\,\psi^2/(1-\psi^2)$.
For a persistent process, information about $\alpha$ rises with $q$ up to $q^{*} \approx -0.797/\log(|\lambda\alpha|)$.
Considering $q \in \naturals$, it is decreasing with $q$ throughout only if $|\lambda\alpha| < 0.58$.

%% file: part_App1.tex

\section{Lagged Input-Output Conversion \& Business Cycles}
\label{sec_app1}




\citet{LongPlosser1983} show that time lags between the production of goods and their subsequent use as intermediaries for producing other goods can generate endogenous business cycles.
In this application, I use the NVAR to empirically quantify the salience of this mechanism.
I generalize their RBC economy with one period-lagged input-output conversion (IOC) by assuming that firms' production requires inputs produced in the past $p$ periods.
At the model-frequency, sectoral output then evolves as an NVAR($p,1$), which translates into an NVAR($p,q$) for some $q\in \naturals$ at my monthly observational frequency. 
Thereby, $A$ is the input-output matrix, $\alpha$ shows how input-sourcing is spread out over the $p$ periods,
and $u_t$ is sectoral productivity. 
By estimating $\alpha$, $q$ and the variance and persistence of $u_t$, I quantify the extent to which lagged IOC can rationalize business cycles in this framework. 
\cref{appsec_app1} accompanies this section.





\subsection{Model}
\label{subsec_app1_theory}




As in \citet{Carvalho-TahbazSalehi2019}, suppose there are $n$ sectors,
in each of which a representative firm produces a differentiated good $i$ by combining labor $l_{i\tau}$ and goods produced by other sectors $j$, $\{x_{ij\tau}\}_{j=1}^n$, using a constant-returns-to-scale Cobb-Douglas production function. Firms maximize profits, taking prices as given. The profits of firm $i$ in period $\tau$ are
\begin{align*}
	\Pi_{i\tau} = p_{i\tau}y_{i\tau} - w_\tau l_{i\tau} - \sum_{j=1}^n p_{j\tau} x^{ij}_\tau \; , \quad \text{with} \; \; 	y_{i\tau} = z_{i\tau} l_{i\tau}^{b_i} \prod_{j=1}^n x_{ij\tau}^{a_{ij}} \; ,
\end{align*}
where $b_i > 0$, $a_{ij} \geq 0$ and $b_i + \sum_{j=1}^n a_{ij} = 1$. 
$z_{i\tau}$ denotes Total Factor Productivity (TFP) in sector $i$, $\{p_{i\tau}\}_{i=1}^n$ are the prices of the $n$ goods, and $w_\tau$ is the price of labor. 
$x_{ij\tau}$ is the amount of good $j$ used in the production at time $\tau$, which may differ from the amount of good $j$ purchased in period $\tau$, $x^{ij}_\tau$.
I define $\tilde{y}_\tau = ln(y_\tau)$, $y_\tau = (y_{1\tau}, ..., y_{n\tau})'$ and $\tilde{z}_\tau$ analogously.

Suppose further that there is a representative household who supplies one unit of labor inelastically and exhibits log-preferences over the $n$ goods:
\begin{align*}
	\underset{\{c_{i\tau}\}_{i=1}^n }{max} \; \sum_{i=1}^n \gamma_i \; ln (c_{i\tau}/\gamma_i) \; , \quad \text{s.t.} \; \sum_{i=1}^n p_{i\tau} c_{i\tau} = w_\tau \; , \quad \text{where} \; \sum_{i=1}^n \gamma_i = 1 \; .
\end{align*}
The first-order condition (FOC) yields $c_{i\tau} = \gamma_i w_\tau / p_{i\tau}$.\footnote{
	This holds even if households have access to a storage technology; market clearing under representative households in a closed economy implies that households spend their whole period $\tau$ income $w_\tau$ on consumption.
}

Different assumptions on the timing of IOC lead to different dynamics of sectoral output in this economy.
The literature predominantly assumes that inputs are converted into outputs in the same period when they are purchased, i.e. $x_{ij\tau} = x^{ij}_\tau$ (e.g. \citet{Carvalho2010,AcemogluEtAl2012}).
In the resulting static environment, sectoral output $\tilde{y}$ follows the SAR from \cref{subsec_model_furtherproperties}: dropping time subscripts,
\begin{align} \label{eq_app1_yTilde_CIOC}
	\tilde{y} = k^y + A \tilde{y} + \tilde{z} \; ,
\end{align}
where 
$a_{ij} = p_j x^{ij} / (p_i y_i)$ is the value of sector $i$'s purchases of good $j$ as a fraction of the value of its output.
In this setting, the input-output network $A$ amplifies sectoral TFP shocks and therefore affects the variance of $\tilde{y}$, but any autocorrelation in $\tilde{y}$ is inherited from that in $\tilde{z}$.

To analyze dynamics under lagged IOC, I assume that agents behave as under perfect foresight. This assumption is standard for analyzing dynamic spatial economies \citep{FanHongParro2023}, because accounting for expectations of endogenous variables would prevent analytical results.
If it takes one period to convert purchased inputs into output, as in \citet{LongPlosser1983}, then $x_{ij\tau} = x^{ij}_{\tau-1}$ and sectoral output follows an NVAR($1,1$):
\begin{align*}
	\tilde{y}_\tau = k^y + A \tilde{y}_{\tau-1} + \tilde{z}_\tau \; .
\end{align*}
Relative to the static economy, the interpretation of $a_{ij} = p_{j\tau} x^{ij\tau} / (p_{i\tau} y_{i\tau})$ and the steady state are unchanged,
yet $A$ drives transition dynamics under this process:
as discussed in \cref{sec_model},
despite equivalent long-term responses as in the static economy,
$\tilde{y}_{\tau}$ adjusts to any disturbance to $\tilde{z}_\tau$ \emph{dynamically} because the latter takes time to transmit along supply chains, hence producing the endogenous business cycles of \citet{LongPlosser1983}.

To take the economy with lagged IOC to the data, I generalize the lag length by assuming that firms require inputs produced in the past $p$ periods for production.
Let $x_{ij\tau}$ aggregate quantities of input $j$ purchased in the past $p$ periods using a CES-aggregator:
\begin{align*}
	x_{ij\tau} = \sbr{ \sum_{l=1}^p \alpha_l \br{ x^{ij}_{\tau,\tau-l} }^r }^{1/r} \; , \quad \alpha_l \geq 0 \; , \quad \sum_{l=1}^p \alpha_l = 1 \; , \quad r \in [0,1) \; ,
\end{align*}
where $x^{ij}_{\tau,\tau-l}$ denotes the use of good $j$ purchased at time $\tau-l$ in the production of good $i$ at time $\tau$.\footnote{
	This means that a good perishes after $p$ periods (with regard to its suitability as an input in production). 
	Therefore, the amount of good $j$ purchased at time $\tau$ can be used in production at periods $\tau+1, ..., \tau+p$: $x^{ij}_\tau = \sum_{l=1}^p x^{ij}_{\tau+l,\tau}$.
	Under a single lag in IOC -- $\alpha_1 = 1$ --, we have $x_{ij\tau} = x^{ij}_{\tau,\tau-1} = x^{ij}_{\tau-1}$.
}
This specification stands for frictions like delivery costs and -lags, storage capacity constraints and technological lengths of the production process itself, as discussed e.g. in \citefour{KhanThomas2007}{AlessandriaKaboskiMidrigan2010}{LiuTsyvinski2024}{AntrasTubdenov2025}.\footnote{
	As in the \citet{LongPlosser1983}-economy above, I presume the buyer stores the good. 
}
The parameters $\{\alpha_l\}_{l=1}^p$ show the shares of an input $j$ purchased at different periods in the past in the overall usage of input $j$ in the production of good $i$.
This time profile of input-sourcing is assumed to be constant over time $\tau$ and across sector-pairs $(i,j)$.
In the Cobb-Douglas case $r\rightarrow 0$, sectoral output follows an NVAR($p,1$):
\begin{align} \label{eq_app1_yTilde_LIOC}
	\tilde{y}_\tau = k^{y2} + \sum_{l=1}^p \alpha_l A \tilde{y}_{\tau-l} + \tilde{z}_\tau \; .
\end{align}
For other elasticities of substitution $r \in [0,1)$, an analogous, approximate relation is obtained by log-linearizing around the steady state.
In either case, the interpretation of $a_{ij} = p_{j} x^{ij} / (p_{i} y_{i})$ is preserved in steady state.
The restrictions $\alpha_l \geq 0$, $\sum_{l=1}^p \alpha_l = 1$ and $\sum_{j=1}^n a_{ij} < 1 \; \forall \; i$ imply that $\tilde{y}_\tau$ is stationary as long as $\tilde{z}_\tau$ is.\footnote{
	For an element-wise nonnegative matrix with row sums strictly smaller than 1, the largest eigenvalue is $<1$ in absolute value \citep[ch. 2]{BermanPlemmons1994}. Stationarity then follows by \cref{prop_WS_NVARpq}.
}

Suppose log sectoral TFP $\tilde{z}_{i\tau}$ is difference stationary with aggregate and idiosyncratic TFP $e^a_\tau$ and $e_{i\tau}$ evolving as AR(1) processes:
\begin{align*} 
	\Delta \tilde{z}_{i\tau} = \gamma_i + \delta_i e^a_\tau + e_{i\tau} \; , \quad \text{with} \quad (1- \rho_a L) e^a_{\tau} =  \varepsilon^a_{\tau} \sim WN \; \; \text{and} \; \; (1-\rho_i L) e_{i\tau} =  \varepsilon_{i\tau} \sim WN \; .
\end{align*}
%
%
Under contemporaneous IOC (\cref{eq_app1_yTilde_CIOC}), this leads to
\begin{align} \label{eq_app1_yTildeGrowth_CIOC}
	\Delta \tilde{y}_\tau = (I-A)^{-1} \sbr{ \gamma + \delta e^a_\tau + e_{\tau} } \; , 
\end{align}
and the IRFs $\partial \Delta \tilde{y}_{i,\tau+h}/ \partial \varepsilon_{j,\tau}$ and $\partial \Delta \tilde{y}_{i,\tau+h}/\partial \varepsilon^a_\tau$ are as in \cref{eq_GCGIRF_SAR_AR1innov} in \cref{subsec_model_furtherproperties}.
%
%
Under this specification, the persistence in output growth is exclusively due to persistence in the aggregate and sector-specific TFP processes $e^a_\tau$ and $\{e_{i\tau}\}_{i=1}^n$.
IOC provides an immediate amplification: it scales a sector $i$'s output growth response to a TFP shock in sector $j$ at every horizon $h$ by element $(i,j)$ of the Leontief-inverse $(I-A)^{-1}$ while preserving its shape. 

Under lagged IOC (\cref{eq_app1_yTilde_LIOC}), we get
\begin{align} \label{eq_app1_yTildeGrowth_LIOC}
	\Delta \tilde{y}_\tau = \gamma + \sum_{l=1}^p \alpha_l A \Delta \tilde{y}_{\tau-l} + \delta e^a_\tau + e_\tau
\end{align}
and IRFs as in \cref{eq_GCGIRF_NVARp1_AR1innov} in \cref{subsec_model_furtherproperties}.
As discussed there, the impulse-responses under \cref{eq_app1_yTildeGrowth_LIOC} are equivalent to those under \cref{eq_app1_yTildeGrowth_CIOC}, yet under the lagged-IOC specification, the amplification by IOC is not immediate but spread out over time and thus provides an additional source of persistence.
Thereby, lagged IOC and TFP-autocorrelations imply distinct dynamics: $\alpha$ relates the timing at which a sector's output is first impacted by a TFP shock in another sector and the strength of this response to network connections of different order,
whereas $\rho_a$ and $\{\rho_j\}_{j=1:n}$ induce an exponentially decaying response after every round of networked transmission.

To take the processes in \cref{eq_app1_yTildeGrowth_CIOC} and \cref{eq_app1_yTildeGrowth_LIOC} to the data, we must specify what a period in the theoretical models above signifies.
Writing observed output growth as $\{\Delta y_t \}_{t=1}^T$, I let
\begin{align} \label{eq_app1_yGrowth}
	\Delta y_t = \Delta \tilde{y}_{tq} + ... + \Delta \tilde{y}_{tq-q+1} \quad t = 1, ..., T 
\end{align}
for some $q \in \naturals$, allowing the model-frequency to differ from the observational frequency. 
%

Under either IOC-specification, TFP shocks propagate downstream, from suppliers to customers, because of the assumption of Cobb-Douglas production (see \citet{Carvalho-TahbazSalehi2019}, who also discuss the implications of alternative production functions).
\citet{AcemogluAkcigitKerr2016} show that this is in line with empirical evidence.

The economy abstracts from other mechanisms that delay the propagation of sectoral shocks: in particular, it has no capital, and its prices are flexible.
Since $\alpha$ captures lagged propagation along $A$ whatever its source, any other mechanism operating along the input-output network at monthly frequency would be absorbed by $\alpha$ and TFP autocorrelations, which is why my estimates are best read as an upper bound on the contribution of lagged IOC to business cycles.
As far as capital-based mechanisms are concerned, two considerations suggest that the bound is tight.
First, capital-based mechanisms have been found to generate little persistence in output growth \citep{Rouwenhorst1991,CogleyNason1995,Horvath2000}
because net investment is only about 0.4\% of the capital stock per quarter \citep[p. 506]{CogleyNason1995}, so that even long gestation lags leave the path of capital stock and thus of output nearly unchanged.
In contrast, lagged IOC delays the entire materials inputs whose shares in gross output are the row sums of $A$.
Second, as discussed above, $\alpha$ is identified from the timing at which a sector's output is first affected by a shock in another sector, as determined by their connections in $A$.
When capital goods are produced and traded across sectors, the delayed propagation follows the network of investment goods, which is much sparser than the production network,\footnote{
	In the capital-flow table used by \citet{FoersterSarteWatson2011}, roughly 85\% of sectors provide no capital to other sectors and six sectors account for 78\% of the off-diagonal entries (see their footnote 6).
}
and it unfolds at the time scale of years rather than the 40 days I estimate in \cref{subsec_app1_results}.\footnote{
	Sector-specific, non-traded capital adds a decay as captured by TFP-persistence \citep[p. 785--786]{Horvath1998}.
}




\subsection{Data \& Estimation}
\label{subsec_app1_dataest}


I construct $\{\Delta y_{it} \}_{i=1:n,t=1:T}$ for $n=23$ manufacturing sectors in the US economy January 2005 through August 2022 using the monthly industrial production (IP) indices provided by the Federal Reserve Board, the same series as used in \citethree{LongPlosser1987}{FoersterSarteWatson2011}{CarvalhoReischer2021}.
To eliminate seasonal patterns, I regress each series on month-dummies, take the residuals and re-scale the series before computing growth rates as log-differences.
Following \citet{AcemogluAkcigitKerr2016}, and in line with the theory in \cref{subsec_app1_theory}, I calibrate the links $a_{ij}$ in $A$ as
$a_{ij} = sales_{j \longrightarrow i} / sales_i$,
where $sales_{j \longrightarrow i}$ is the total value of goods and services purchased by sector $i$ from sector $j$ as determined by the corresponding entry in the Bureau of Economic Analysis' input-output \myquote{use} table, the same as used in \citethree{FoersterSarteWatson2011}{AcemogluEtAl2012}{KamepalliEtAl2025}.
I take the table from 2007.
$a_{ij}$ shows how many dollars worth of output of sector $j$ sector $i$ needs to purchase in order to produce one dollar's worth of its own output.
The calibration assumes that firms' input shares reported for the course of a year are equal to those at higher frequency intervals.
See \cref{appsubsec_app1_data} for more details on data construction.

I consider lagged-IOC specifications with $p=\{1,2,3,4,5,6\}$ and $q \in \{1,2,3,4\}$. 
The latter imply network interaction frequencies of a month, two weeks, 10 days and one week, respectively.
For the contemporaneous-IOC specification, I take $q=1$, in line with existing studies and my characterization of SAR-dynamics as the limit case of high-frequency of network interactions in \cref{subsec_model_furtherproperties}.
In both cases, I assume Normal TFP-innovations
$\varepsilon^a_{\tau} \sim N(0,\sigma^2_a)$ and $\varepsilon_{i\tau} \sim N(0,\sigma^2_i)$.
To identify the common TFP process, I normalize $\sigma^2_a = 1$ and $\delta_1 = 1$.
To accommodate the restrictions $\alpha_l \geq 0 \; \forall \; l$ and $\sum_{l=1}^p \alpha_l = 1$, I drop $\alpha_p$ from $\alpha$ and impose $\alpha_l \in [0,1]$ for $l=1:p-1$ and $\sum_{l=1}^{p-1} \alpha_l \leq 1$.
Under lagged IOC, this yields $p-1 + 4n$ unknown parameters: $\theta = (\alpha',\gamma',\delta',\rho_a,\rho', \sigma^{2\prime})'$, where $\rho$ and $\sigma^2$ stack $\{\rho_i\}_{i=1}^n$ and $\{\sigma^2_i\}_{i=1}^n$, respectively.
Under contemporaneous IOC, $\alpha$ is dropped from $\theta$, leading to $4n$ parameters.\footnote{
	With $n=23$ sectors and $T=211-p$ periods available for estimation, this yields an observations-to-parameters ratio of 52.5 under $p=1$ and 48.6 under $p=6$.
}

I conduct Bayesian inference on $\theta$ under Uniform priors on the respective domains.\footnote{
	Specifically, 
	$\rho_i, \rho_a \in [0,1)$ and
	$\sigma^2_i \in \reals_{++}$ for $i=1:n$, 
	$\delta \in \reals^{n-1}$, 
	and $\alpha \in [0,1]^{p-1} \cap \{\alpha: ||\alpha||_1 \leq 1\}$.
}
As a result, the posterior mode equals the Gaussian ML estimator.
I evaluate the likelihood using the Kalman filter and obtain the posterior numerically using the Sequential Monte Carlo (SMC) algorithm, which estimates the marginal data density (MDD) and therefore enables model selection.
In absence of TFP-autocorrelations ($\rho_a=0$, $\rho=0$), my analysis from \cref{sec_inference} implies that $(\alpha',\delta',\sigma^{2'})'$ is identified.
Because of the qualitatively distinct impulse-responses implied by $\alpha$ and $(\rho_a,\rho')'$, this identification result should extend to the entire $\theta$.\footnote{
	$\gamma$ is simply the unconditional mean of $\Delta \tilde{y}_\tau$ and, therefore, $\mean[\Delta y_t] = q\gamma$.
}
Any counterevidence to this claim should show itself in the form of a multi-model marginal posterior, which the SMC algorithm -- in contrast to alternative numerical sampling techniques -- is designed to capture \citep[Chapter 5]{HerbstSchorfheide2015}.




\subsection{Results}
\label{subsec_app1_results}


\cref{tab_app1_modelselection} reports the MDD for all lagged-IOC specifications considered.
All but a few specifications with $q=1$ yield a larger MDD than -11452, the value under contemporaneous IOC.
Hence, the data generally favors a temporally spread-out to an immediate amplification of TFP shocks;
it is informative for the transmission-timing and rejects the implicit assumption of the SAR model that the networked transmission is so high as to not be identified.
The preferred specification features $q=3$ and $p=4$, i.e. input-output conversion takes place at a frequency of 10 days (a third of a month) and is spread out over 40 days.
This is in line with 6.3 weeks of materials inventories for US durable-goods manufacturing reported in \citet[Table 2]{Ramey1989} and somewhat shorter than the cost-weighted average period of production of 69 days from \citet[p. 627]{AntrasTubdenov2025}.\footnote{
	Note that only \citet{Ramey1989}'s figure is the time between the purchase of materials and their use, the object $\alpha$ describes, whereas the number of \citet{AntrasTubdenov2025} measures purchase-to-sale times, thus also including work-in-process and finished-goods inventories.
	Also, these figures are \emph{average} times, whereas 40 days is the \emph{upper end of the support} of $\alpha$, and my implied mean lag $\sum_{l=1}^p l \alpha_l$ is shorter than 40 days.
}
My model selection corroborates the respective assertions of \citet[p. 428]{Shea2002} and \citet[p. 73--74]{Horvath2000} that the laggedness of IOC is not pronounced at yearly and quarterly frequency.

\begin{table}[h]
	\caption{Model Selection: Log MDD}
	\label{tab_app1_modelselection}
	\centering
	\begin{tabular}{C{0.5cm}C{0.5cm}C{1.75cm}C{1.75cm}C{1.75cm}C{1.75cm}C{1.75cm}C{1.75cm}}
		\hline \hline \tabspaceAfterRule
	& & \multicolumn{6}{c}{$p$} \tabspaceAfterHeaderRow \cline{3-8}
		\tabspaceAfterRule
	& & $1$ & $2$ & $3$ & $4$ & $5$ & $6$ \tabspaceAfterHeaderRow
	\multicolumn{1}{l|}{}  & 1 & -12015 & -11876 & -11386 & -11564 & -11190 & -11569 \tabspaceAfterBodyRow
	\multicolumn{1}{l|}{$q$}  &  2 & -10990 & -11158 & -11192 & -11273 & -11097 & -11218 \tabspaceAfterBodyRow
	\multicolumn{1}{l|}{}  &  3 & -10996 & -11238 & -10988 & -10936 & -11158 & -11017 \tabspaceAfterBodyRow
	\multicolumn{1}{l|}{}  & 4 & -11232 & -11056 & -11242 & -11237 & -11215 & -11014 \tabspaceAfterBodyRow
	\tabspaceAfterRule
		\bottomrule
		\multicolumn{8}{L{14cm}}{\scriptsize {\em Notes:} The table shows log MDDs  across lagged-IOC specifications. Rows (top to bottom) refer to IOC-frequencies of a month, two weeks, 10 days, and one week, respectively The log MDD under contemporaneous IOC is -11452.}\\
	\end{tabular}
	\vspace*{-15pt}
\end{table}

The subsequent analysis is based on this preferred specification.
In it, $(\alpha',\rho_a,\rho')'$ are estimated very precisely, as shown by the marginal posterior distributions in \cref{plot_app1_prpo_alphas,plot_app1_prpo_rhos}.
This suggests that lagged network-interactions and innovation-autocorrelations are indeed separately identified,
which is in line with the qualitatively different transitional dynamics they imply (see \cref{subsec_model_furtherproperties}). 

\begin{figure}[t]
	\begin{center}
		\figspaceBeforePlotArea
		\begin{subfigure}[b]{\plotSizeDouble\textwidth}
			\subcaption*{\scriptsize{Single Shock Network Transmission}}
				\centering
				\includegraphics[width=1\textwidth, clip]{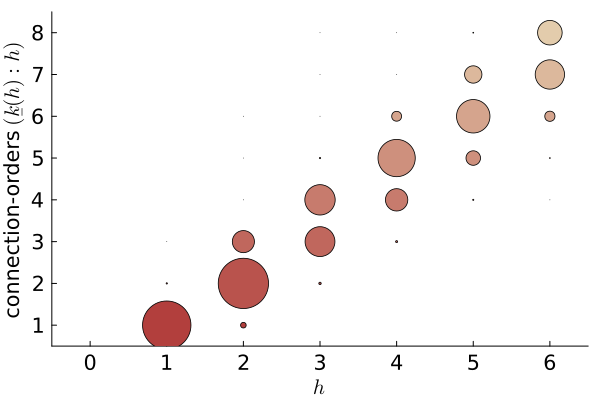}
		\end{subfigure}
		\begin{subfigure}[b]{\plotSizeDouble\textwidth}
			\subcaption*{\scriptsize{IRF: Fabr. Metals to Mining}}
				\centering
				\includegraphics[width=1\textwidth, clip]{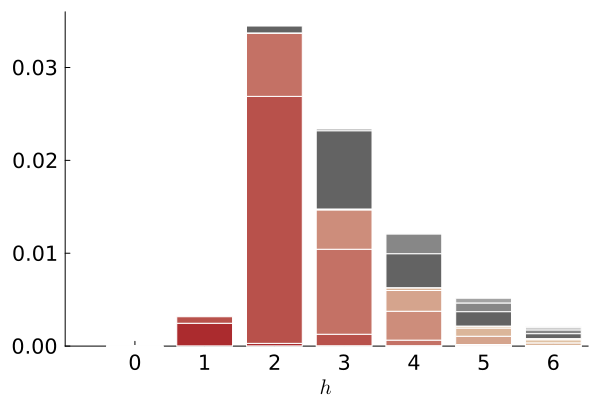}
		\end{subfigure}

		\figspaceBetweenPlotRows

		\begin{subfigure}[b]{\plotSizeDouble\textwidth}
			\subcaption*{\scriptsize{\myquote{Fabricated Metal Products}: Links}}
				\centering
				\includegraphics[width=1\textwidth, clip]{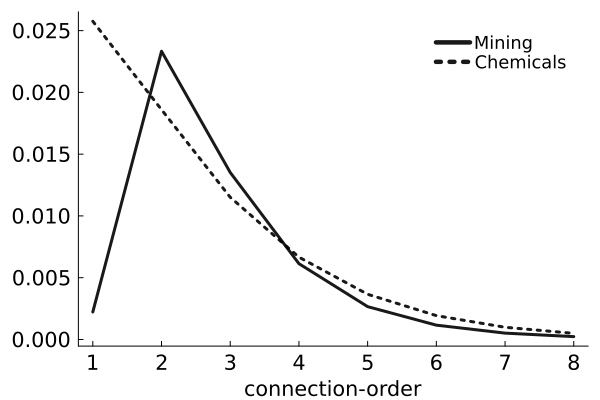}
		\end{subfigure}
		\begin{subfigure}[b]{\plotSizeDouble\textwidth}
			\subcaption*{\scriptsize{IRF: Fabr. Metals to Chemicals}}
				\centering
				\includegraphics[width=1\textwidth, clip]{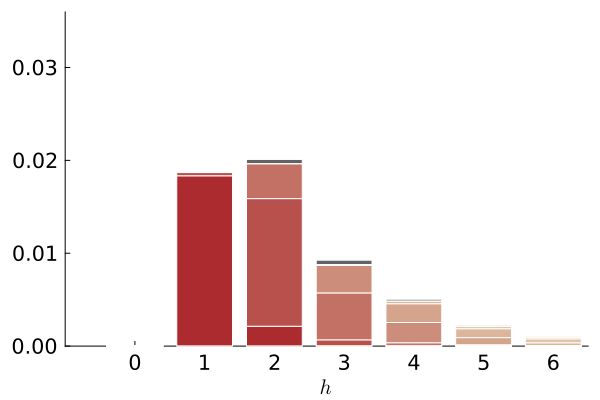}
		\end{subfigure}
		\figspaceAfterPlotArea
	\end{center}
\caption{Impulse-Responses of Sectoral Output Growth to Sectoral TFP Shocks\figspaceBetweenCaptionAndNotes
\scriptsize {\em Notes:} The top left panel depicts the coefficients $c^h_{\underline{k}(h)}:c^h_h$ from \cref{eq_GCGIRF_NVARp1_AR1innov} for $h=1:6$, illustrating how a single shock transmits along network connection-orders (y-axis) over time (x-axis; horizons). 
The dashed lines show the bounds on which connection-orders can matter in an NVAR($2,1$): $h\in \underline{k}(h):h$. 
The bottom left panel shows the supply chain connections of different order from \myquote{Fabricated Metal Products} to \myquote{Mining (except oil and gas)} and \myquote{Chemical Products}.
The right panels show the resulting IRFs to a TFP shock of one standard deviation.}
\label{plot_app1_sectorIRFs_HOs}

\end{figure}

\cref{plot_app1_sectorIRFs_HOs} illustrates the composition of impulse-responses. 
The top-left panel shows the estimated temporal propagation of TFP shocks along supply-chain links of different order.\footnote{
	The impulse-responses throughout this section refer to shocks at the end of an observational period -- in the last third of a month --,  implying that a sector reacts contemporaneously only to own TFP shocks. 
} 
Within the first two months after the shock, the impact is mostly limited to direct customer-sectors, after which it spreads to connections further downstream slightly faster than one connection-order per month. 
The bottom-left panel shows the strength of network connections of different order from the sector \myquote{Fabricated Metal Products} to the sectors \myquote{Mining (except oil and gas)} and \myquote{Chemical Products}, respectively.
Firms producing fabricated metals rely on chemical products directly as well as indirectly in their supply chain.
In contrast, they use mining products only indirectly, though this higher-order dependence is of a similar magnitude.
As a result, under the preferred lagged-IOC specification,
fabricated metals production reacts faster to a TFP shock in the chemicals sector than to a TFP shock in mining (see right panels of \cref{plot_app1_sectorIRFs_HOs}), 
yet its peak-magnitude is larger in the latter case, where production increases by about 0.035 percentage points.\footnote{
	Expectedly, responses of sectoral output to sector-specific TFP shocks in a single other sector are rather small, even when summing up the response of output growth to obtain the output-response in levels.
	Larger responses are obtained when considering impulses to common TFP shocks; see \cref{plot_app1_sectorIRFsToAgg}.
}
%
\todo{see here commented-out comment on Dufour \& Wang (2023)}


\begin{figure}[t]
	\begin{center}
		\figspaceBeforePlotArea
		\caption*{\footnotesize{IRFs of Aggregate Output Growth}}

		\begin{subfigure}[b]{\plotSizeTriple\textwidth}
			\subcaption*{\scriptsize{Aggregate Output: Links}}
				\centering
				\includegraphics[width=1\textwidth, clip]{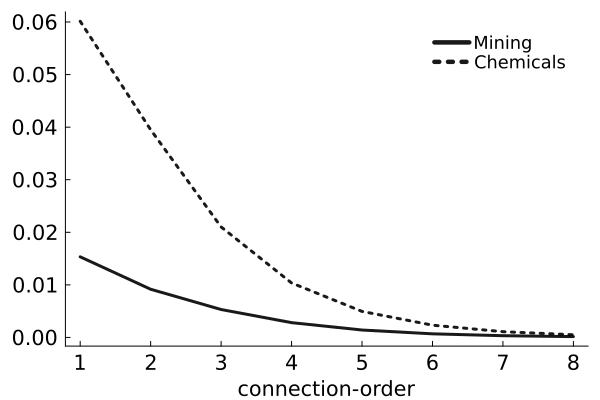}
		\end{subfigure}
		\begin{subfigure}[b]{\plotSizeTriple\textwidth}
			\subcaption*{\scriptsize{IRF to Mining}}
				\centering
				\includegraphics[width=1\textwidth, clip]{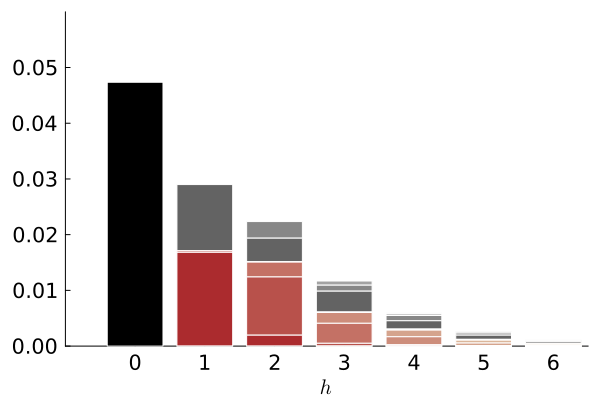}
		\end{subfigure}
		\begin{subfigure}[b]{\plotSizeTriple\textwidth}
			\subcaption*{\scriptsize{IRF to Chemicals}}
				\centering
				\includegraphics[width=1\textwidth, clip]{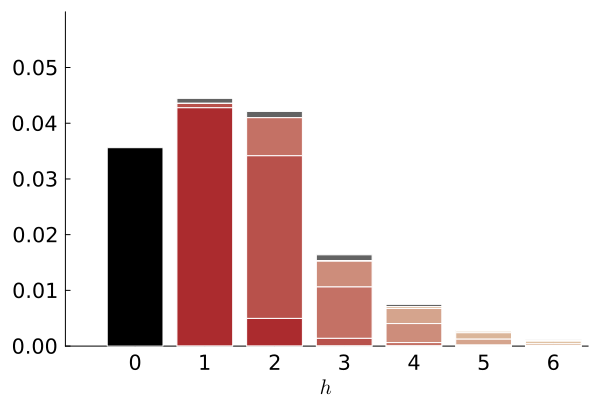}
		\end{subfigure}
		\figspaceAfterPlotArea
	\end{center}
	\caption{Impulse-Responses of Aggregate Output Growth to Sectoral TFP Shocks\figspaceBetweenCaptionAndNotes
	\scriptsize {\em Notes:} The left panel shows the links of aggregate output to the sectors \myquote{Mining (except oil and gas)} and \myquote{Chemical Products}, constructed as the weighted average of sectoral links to those two sectors, with weights given by contribution to aggregate output. The middle and right panels show the impulse-responses of aggregate IP growth to a one standard deviation shock to TFP in those two sectors, respectively.}
	\label{plot_app1_AggIRFsToSectors}

\end{figure}

Similar reasoning explains \emph{aggregate} output growth responses to sector-specific TFP shocks, illustrated in \cref{plot_app1_AggIRFsToSectors}.
The relevant network-quantity for aggregate responses is sectors' weighted reliance on the shocked sector as a supplier, with weights given by their contribution to aggregate output (left panel).
Despite a similar contemporaneous effect (black bars),\footnote{
	A sector's output growth reacts contemporaneously to own TFP shocks, hence prompting an aggregate response. 
	When considering, as here, TFP shocks at the end of an observational period, sectoral output growth reacts contemporaneously only to own TFP shocks. 
	The aggregate then only responds because of the shocked sector's own response, scaling it by the sector's weight in aggregate output. 
}  
due to chemicals' position at the top of supply chains, 
their productivity increases translate into more persistent output growth responses than do TFP shocks in mining (middle and right panel).


Using contemporaneous IOC, \citet{AcemogluEtAl2012} show that such aggregate output responses are larger after TFP shocks to sectors with more central positions in the supply-chain network.
The preferred lagged-IOC specification leads to the same output responses in the long run but implies that different network-positions affect not only the strength of sectoral TFP shocks' impact on aggregate output, but also its timing. 
This is illustrated in the left panel of \cref{plot_app1_aggIRFs_size_timing}, which shows the time profiles of the response of aggregate IP to TFP shocks in different sectors.
Sectors at the bottom of supply chains, such as furniture production, have a more immediate effect on aggregate output than sectors that act as important suppliers to other sectors in the economy, such as primary metals.
The right panel of \cref{plot_app1_aggIRFs_size_timing} shows that stronger effects tend to take more time to realize, but this relationship is not clear-cut; for example, a TFP shock in furniture production has a similar long-term effect on aggregate output as a shock to nonmetallic minerals, yet the latter materializes much slower. 


\begin{figure}
	\begin{center}
		\figspaceBeforePlotArea
		\begin{subfigure}[b]{\plotSizeDouble\textwidth}
				\centering
				\includegraphics[width=1\textwidth, clip]{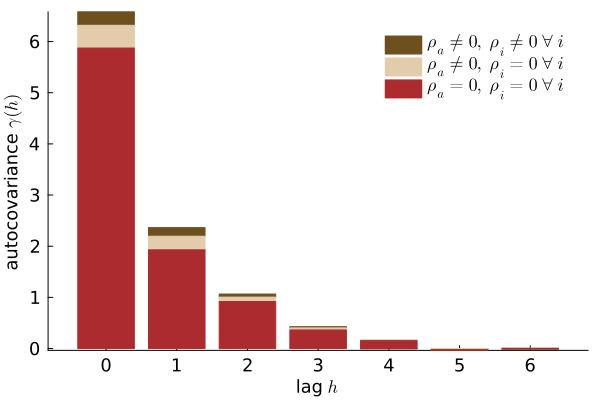}
		\end{subfigure}
		\begin{subfigure}[b]{\plotSizeDouble\textwidth}
				\centering
				\includegraphics[width=1\textwidth, clip]{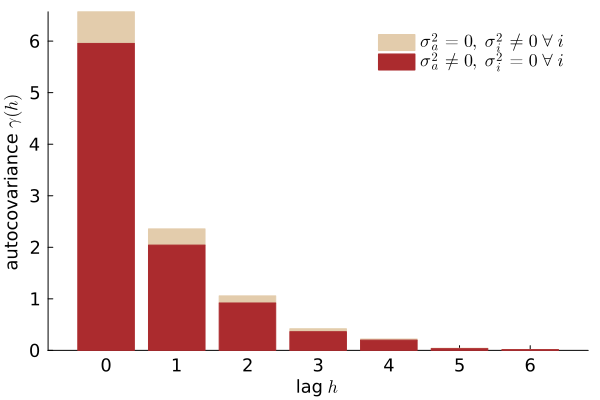}
		\end{subfigure}
		\figspaceAfterPlotArea
	\end{center}
    \caption{Decomposing the Autocovariance Function of Aggregate Output Growth\figspaceBetweenCaptionAndNotes
    \scriptsize {\em Notes:} The left panel illustrates the contributions of lagged IOC and the autocorrelations in TFP shocks to the autocovariance function of aggregate IP growth, while the right panel shows the contributions of the variation in those TFP shocks itself.}
    \label{plot_app1_autocovdecomp}
\end{figure}

\begin{figure}
	\begin{center}
		\figspaceBeforePlotArea
		\begin{subfigure}[b]{\plotSizeDouble\textwidth}
			\subcaption*{\scriptsize{TFP-Autocorrelations: $\rho_a$ and $\rho_i$, $i=1:n$}}
				\centering
				\includegraphics[width=1\textwidth, clip]{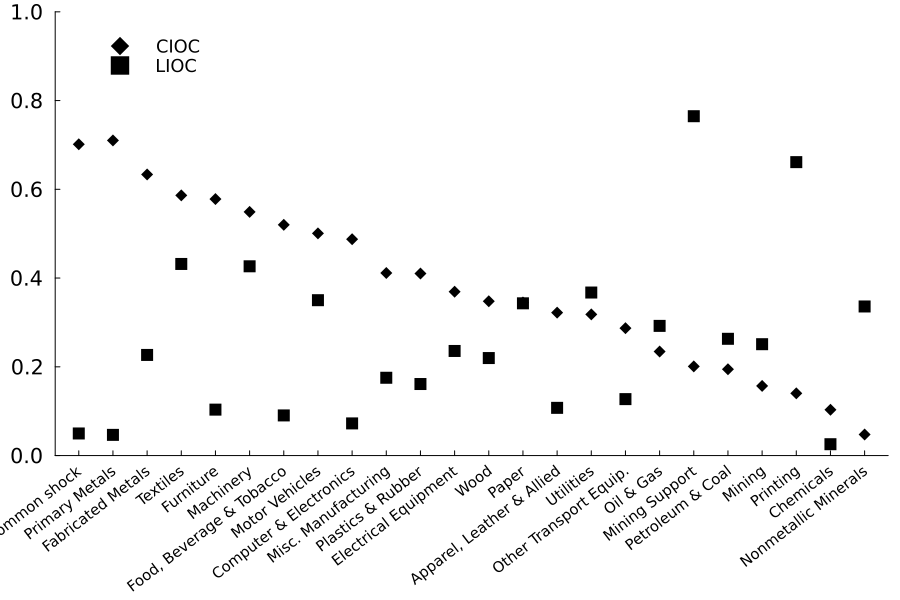}
		\end{subfigure}
		\begin{subfigure}[b]{\plotSizeDouble\textwidth}
			\subcaption*{\scriptsize{TFP-Variances: $\var[\delta_i e^a_\tau + e_{i,\tau}]$, $i=1:n$}}
				\centering
				\includegraphics[width=1\textwidth, clip]{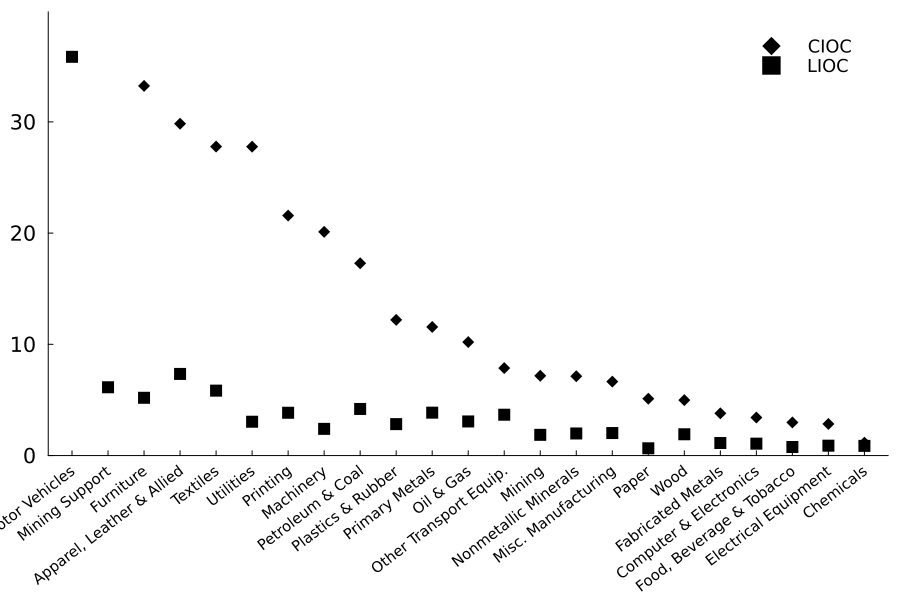}
		\end{subfigure}
		\figspaceAfterPlotArea
	\end{center}
    \caption{TFP-Autocorrelations and -Variances under Contemporaneous and Lagged IOC\figspaceBetweenCaptionAndNotes
    \scriptsize {\em Notes:} The left plot shows the posterior means of $\rho_a$ and $\{\rho_i\}_{i=1:n}$, i.e. the autocorrelations of the common TFP shock $e^a_\tau$ and the idiosyncratic sectoral TFP shocks $e_{i,\tau}$.
	Diamond-shaped markers refer to values obtained under contemporaneous IOC, while rectangular markers refer to values obtained under the preferred lagged-IOC specification. 
	The right plot analogously displays sectors' TFP-variances $\var[\delta_i e^a_\tau + e_{i,\tau}]$. 
	The y-axis in the right plot is trimmed to exclude outliers and render the illustration legible.}
    \label{plot_app1_CompareLIOCCIOC_VarRhoTFP}
\end{figure}

\cref{plot_app1_autocovdecomp} decomposes the autocovariance function of aggregate output growth.
The left panel gauges the contributions of the three sources of persistence: lagged IOC and autocorrelation in common and idiosyncratic TFP shocks.
Setting $\rho_a$ and $\rho$ to zero reveals that lagged IOC accounts for 82\%, 86\% and 85\% of the autocovariance in aggregate output growth at horizons $h=1:3$, respectively (red bars).
With autocorrelated common TFP, these numbers rise to 93\%, 94\% and 94\%, respectively (cream bars), leaving an even smaller role for autocorrelation in idiosyncratic sectoral TFP (brown bars).
Even the variance (horizon $h=0$) is almost perfectly replicated by lagged IOC, a single autocorrelated common TFP process and serially uncorrelated sector-specific TFP processes.
%
In line with that, the preferred lagged-IOC specification estimates most autocorrelations $\rho_a$ and $\{\rho_i\}_{i=1:n}$ as lower than does the contemporaneous-IOC specification (left panel in \cref{plot_app1_CompareLIOCCIOC_VarRhoTFP}).
The right panel of \cref{plot_app1_CompareLIOCCIOC_VarRhoTFP} shows an even clearer reduction of TFP variances $\var[\delta_i e^a_\tau + e_{i,\tau}]$, $i=1:n$, amounting to 72.78\% on average across sectors.
Their reduction is the intuitive explanation for why this lagged-IOC specification is preferred to the contemporaneous-IOC specification by model selection criteria in \cref{tab_app1_modelselection}.
Overall, these results lend strong support to the mechanism suggested by \citet{LongPlosser1983}:
lagged IOC substantially endogenizes business cycles, as it reduces both the variances and autocorrelations in exogenous TFP processes required to rationalize output fluctuations in the data.

Interestingly, in the preferred lagged-IOC specification, business cycles are mainly the result of common rather than sector-specific TFP shocks:
the right panel of \cref{plot_app1_autocovdecomp} calculates the autocovariance of aggregate output growth implied when setting $\sigma^2_i =0 \; \forall \; i$ (red bars) and shows that 90\% of the variance and similar fractions of the first three autocovariances of aggregate output growth (86\%, 86\% and 85\%, respectively) are explained by the single common TFP shock.
\citet{KamepalliEtAl2025} reach a qualitatively similar conclusion on annual data for 43 US sectors.\footnote{
	Using contemporaneous and one period-lagged network effects with factor-augmented errors, they estimate sector-specific shocks to directly contribute 2.4--4.0\% to the aggregate output growth variance, while they deal separately with the network-propagated part of idiosyncratic shocks.
}
Other decompositions 
assign a larger role to sector-specific shocks, whether they decompose the variance of aggregate output growth, of its innovations, or of its detrended level.
\citet{FoersterSarteWatson2011} attribute 19\% of the quarterly variance of aggregate IP growth to them over 1972--1983 and 50\% over 1984--2007.
The factor analysis of \citet{LongPlosser1987} leaves at least 53\% of the variance of aggregate output-growth innovations to them.
\citet{Bouakez-Cardia-RugeMurcia2014} obtain 28\% in an input-output economy 
with heterogeneous price rigidities, though their aggregate shocks are monetary and preference-related rather than technological.
My analysis suggests that abstracting from the laggedness of IOC may confound idiosyncratic sectoral shocks with sectoral heterogeneity in the temporal exposure to common shocks.
\citet{DeGraeve-Schneider2023}, for instance, identify sector-specific shocks by requiring the (grouped) sectoral output responses \emph{on impact} to be ranked as in the Leontief inverse, while imposing no restriction at later horizons, so that a common shock to which sectors are exposed at different times can satisfy their restrictions.
\citet[p. 335]{LongPlosser1987} themselves note that the comovement they measure is equally consistent with the propagation of industry-specific shocks and with a common shock that reaches different sectors at different points in time and that separating the two requires \myquote{put[ting] more structure on the problem}, which the NVAR does.

%% file: part_Conclusion.tex

\section{Conclusion}
\label{sec_conclusion}

In this paper, I develop the Network-Vector-Autoregression (NVAR), an econometric framework in which innovations transmit dynamically along bilateral links and drive the dynamics of a cross-sectional variable.
In turn, I estimate the potential of lagged input-output conversion to endogenize business cycles through the lens of a Real Business Cycle economy.

In my application, I abstract from alternative mechanisms that delay the propagation of sectoral shocks, such as capital accumulation over a network of investment goods and staggered price setting along production chains.
Embedding them in the model, with the investment network as a second adjacency matrix and prices alongside quantities, would allow us to estimate their relative roles in explaining business cycles.

Furthermore, the NVAR could be augmented to accommodate heterogeneous propagation patterns across units or over time. 
Rather than assuming time-invariant links, an important methodological step forward would be to jointly investigate network effect- and formation-dynamics. 
%
%
I leave these as suggestions for future research.

%% file: part_appendix_model.tex

\section{Lagged Network Effects: Theory}
\label{appsec_model}




\propositionmarker{prop_GCGIRF_NVARp1}

Let $\Phi_l = \alpha_l A$ for $l=1:p$, and define $F$ to be the companion-form matrix of \cref{eq_NVARp1}:
$$ F = \begin{bmatrix} 
			\Phi_1 & ... & \Phi_p  \\ 
			\multicolumn{2}{c}{I_{n(p-1)}} & 0_{n(p-1) \times n} \\
		\end{bmatrix} $$
s.t. $\partial y_{t+h}/\partial y_t = [I_n,0_{n \times n(p-1)}] F^h [I_n,0_{n \times n(p-1)}]' = (F^h)_{11}$. 
$(F^h)_{lm}$ is the $n\times n$ block in position $(l,m)$ of $F^h$.

\myproofof{prop_GCGIRF_NVARp1}{ 
	I prove by induction that $(F^h)_{1l}$ has powers of $A$ in the set $\text{ceil}((h+l-1)/p):h$. Note that the claim is true for $h=1$. Suppose it is true for $h$. For $h+1$ we have
	\begin{align*}
		F^{h+1} = 
		F^{h} F =
		\begin{bmatrix} (F^h)_{11}\Phi_1 + (F^h)_{12}  & ... & (F^h)_{11}\Phi_{p-1} + (F^h)_{1p} & (F^h)_{11}\Phi_p \\ \vdots  & \ddots & \vdots & \vdots \end{bmatrix}  \; .
	\end{align*}
	(Only the first row of blocks in $F^{h+1}$ are relevant to the argument.)
	Consider $m \in 0:(p-1)$ s.t $h+m$ is a multiple of $p$. 
	Then 
	$$ \text{ceil}\br{ \frac{h+l-1}{p} }= \begin{cases}
		\frac{h+m}{p} \quad &\text{for} \; l=1:(m+1) \\
		\frac{h+m}{p}+1 \quad &\text{for} \; l=(m+2):p \\
	\end{cases} \; . $$
	This means that for $l=1:(m+1)$, $(F^h)_{1l}$ has powers of $A$ in $\br{(h+m)/p}:h$, while for $l=(m+2):p$ it has powers in $\br{(h+m)/p+1}:h$. 
	Then by the equation above, for $l=1:m$, $(F^{h+1})_{1l}$ has powers of $A$ in $\br{(h+m)/p}:(h+1)$, while for $l=(m+1):p$ it has powers in $\br{(h+m)/p+1}:(h+1)$. These sets are both equal to 
	$\text{ceil}\br{(h+1+l-1)/p}:(h+1)$ and independent of $m$. Therefore, the claim holds in all possible cases. 
	$\blacksquare$
}

\propositionmarker{prop_GCGIRF_NVARpq_stock}

\myproofof{prop_GCGIRF_NVARpq_stock}{
	This follows from \cref{prop_GCGIRF_NVARp1} applied to 
	\begin{align*}
		\frac{\partial y_{i,t+h} } { \partial y_{j,t} } = \frac{ \partial x_{i,(t+h)q} }{\partial x_{j,tq}} = \frac{\partial x_{i,\tau+hq} }{ \partial x_{j,\tau} }  = \frac{\partial x_{i,\tau+hq} }{ \partial e_{j,\tau} }  \; . \quad \blacksquare
	\end{align*}
}

\myproposition{NVAR($p,q$), Flow Variable: Granger-Causality}{prop_GCGIRF_NVARpq_flow}{
	Let the flow variable $y_t$ evolve as in \cref{eq_NVARpq} for some $q \in \naturals \backslash\{1\}$.
	Then we can write
	$ \frac{\partial y_{i,t+h}}{\partial y_{j,t}}  = c^{h}_{\underline{k}} \sbr{A^{\underline{k}}}_{ij} + ... + c^{h}_{hq} \sbr{A^{hq}}_{ij} + R $,
	where $\underline{k} = \text{ceil}(hq/p)$, the coefficients $\{c^h_k\}_{k=\underline{k}:hq}$ are polynomials in $\{\alpha_l\}_{l=1:p}$,
	and $R$ is a linear combination of $A^k$ for some $k \notin \underline{k}:hq$.
}

\myproof{
	We have $y_t = x_{tq} + ... + x_{tq -q +1}$.
	Suppose the first (latest) term, $x_{tq}$ is responsible for the change in $y_t$. Then
	$$ \frac{ \partial y_{t+h} }{ \partial x_{tq} }   =  \frac{ \partial (x_{(t+h)q} + ... + x_{(t+h)q - q +1}) }{ \partial x_{tq} }  = \frac{ \partial x_{\tau+hq} }{ \partial x_{\tau} }  + ... +  \frac{ \partial x_{\tau+hq-q+1} }{ \partial x_{\tau} } \; ,$$ 
	and, by \cref{prop_GCGIRF_NVARp1}, connection-orders $k\in \{\text{ceil} ((q(h-1)+1)/p), ..., hq\}$ matter. Analogous calculations show that if the last (earliest) term, $x_{tq -q +1}$, is responsible for the change in $y_t$, connection-orders $k\in \{\text{ceil} (hq/p), ..., hq+q-1\}$ matter, while the cases in-between lead to sets contained in the union of these two sets.
	Therefore, regardless of which term is responsible for the change in $y_t$, $\partial y_{i,t+h} / \partial u_{j,t}$ is a linear combination of $A^k$ for connection-orders $k$ in the intersection of these two sets, 
	$ \{\text{ceil} (hq/p), ..., hq\} = \{\text{ceil} ((q(h-1)+1)/p), ..., hq\} \cap \{\text{ceil} (hq/p), ..., hq+q-1\} $. $\blacksquare$
}


Define $g(z) = \sum_{k=1}^p \alpha_k z^k$ and $F(z) = I_n - g(z)A$ s.t. $F(L)x_\tau = e_\tau$.
For some $q \in \naturals$, let $\wq = e^{2\pi \ii/q}$.
Note that $\wq^q = 1$.
%
%
Define
$$ \calF_q(z) = \prod_{m=0}^{q-1} F(\omega_q^m z) \;, \quad \calG_q(z^q) = \calF_q(z) \; , \quad C_q(z) = \prod_{m=1}^{q-1} F(\omega_q^m z) = F(z)^{-1} \calG_q(z^q) \; . $$
\cref{lemma_calFqRewriting} establishes that such a polynomial $\calG_q(z) = I_n - \sum_{k=1}^{p} \calG_{q,k}z^k$ exists and has degree $\leq p$ and real coefficient-matrices $\calG_{q,k}$.
Given this, $C_q(z)$ is a polynomial of degree $p(q-1)$ with real coefficients $C_{q,k}$.
For the flow case, define also
$$ s_q(z) = 1 + z + \dots + z^{q-1} \; , \quad S_q(z) = s_q(z)s_q(z^{-1}) \quad \text{and} \quad D_q(z) = C_q(z)s_q(z) = \sum_{k=0}^{(p+1)(q-1)} D_{q,k}z^k \; , $$
and let $X_\tau = s_q(L)x_\tau$ s.t. $y_t = X_{tq}$. 
$D_q(z)$ is a polynomial of degree $(p+1)(q-1)$ whose coefficient-matrices $D_{q,k}$ are polynomials in $A$, because the $C_{q,k}$ are (\cref{lemma_calFqRewriting}) and this property is preserved under multiplication by $s_q(z)$.
Note that $s_q(1)=q$ and $s_q(z) = (1-z^q)/(1-z)$ for $z \neq 1$, so the zeros of
$s_q$ are exactly the $q$th roots of unity other than $1$,
while $s_q(\eta) \neq 0$ and $s_q(1/\eta) \neq 0$ for $0<|\eta|<1$. 
Also, note that $\gamma_X(h)=\sum_{|m|\leq q-1}(q-|m|)\gamma_x(h+m)$,\footnote{
    $\gamma_X(h) = \sum_{l=0}^{q-1}\sum_{k=0}^{q-1} \gamma_x(h+k-l)$ since $X_\tau = s_q(L)x_\tau = \sum_{l=0}^{q-1} x_{\tau-l}$, and there are $q-|m|$ pairs with $k-l=m$.
} 
and by similar reasoning we can see that $S_q(z) = \sum_{|m| \leq q-1}(q-|m|)z^m$. 
If the flow is observed as an average, $y_t = (x_{tq}+\dots+x_{tq-q+1})/q$,
then $S_q(z)/q^2$ replaces $S_q(z)$ throughout and all results hold unchanged.

\mylemma{}{lemma_calFqRewriting}{
    Let $\alpha \in \reals^p$, $q \in \naturals$ and $A$ be a real $n \times n$ matrix with eigenvalues $\lambda_1, \ldots, \lambda_n$. Then
    \begin{itemize}
    \item[(i)] for each $z$, the matrices $F(\omega_q^m z)$, $m = 0, \ldots, q-1$, commute with one another so that $\calF_q(z)$ and $C_q(z)$ do not depend on the order of their factors;
    \item[(ii)] $\calF_q(z) = \calG_q(z^q)$, where $\calG_q(z) = I_n - \sum_{k=1}^{p} \calG_{q,k} z^k$ is a real polynomial of degree $\leq p$ whose coefficient matrices $\calG_{q,k}$ are polynomials in $A$;
    \item[(iii)] $C_q(z) = F(z)^{-1} \calG_q(z^q)$ is a real polynomial of degree $\leq p(q-1)$ whose coefficient matrices $C_{q,k}$ are polynomials in $A$.
    \end{itemize}
}

\myproof{
    (i) We have $F(\omega_q^m z) = I_n - g(\omega_q^m z) A$, and any two matrices of the form $I_n - cA$ and $I_n - c'A$ with scalars $c$ and $c'$ commute, since $(I_n - cA)(I_n - c'A)$ is symmetric in $(c,c')$.

    (ii) Write $\calF_q(z) = \sum_{l=0}^{pq} B_l z^l$ with matrix coefficients $B_l$.
    Note that $\calF_q(\omega_q z) = \calF_q(z)$; replacing $z$ by $\omega_q z$ replaces the factor $F(\omega_q^m z)$ by $F(\omega_q^{m+1} z)$, and since $\omega_q^q = 1$, this permutes the factors cyclically, without changing the product (by (i)).
    Comparing coefficients, $B_l(\omega_q^l - 1) = 0$, and since $\omega_q^l \neq 1$ for $l$ not divisible by $q$, we have $B_l = 0$ for all such $l$. 
    Hence $\calG_q(z) = \sum_{k=0}^{p} B_{qk} z^k$ satisfies $\calG_q(z^q) = \calF_q(z)$, with degree $\leq p$ and constant term $B_0 = \calF_q(0) = \prod_m F(0) = I_n$. 
    Each $B_l$ is a polynomial in $A$, because each factor is and this property is preserved under multiplication. 

    For a matrix polynomial $P(z) = \sum_l P_l z^l$, write $\bar P(z) = \sum_l \bar P_l z^l$ for the polynomial with entrywise conjugated coefficients.
    Note that $\overline{PQ} = \bar P \bar Q$ for two polynomials $P$ and $Q$. 
    Also, $\overline{P(c \,\cdot)} = \bar P(\bar c \,\cdot)$ for a scalar $c$. 
    Since $\alpha$ and $A$ are real, $\bar F = F$. 
    Hence
    $$ \overline{\calF_q}(z) \;=\; \prod_{m=0}^{q-1} F\!\left(\overline{\omega_q^m}\, z\right) \;=\; \prod_{m=0}^{q-1} F\!\left(\omega_q^{q-m} z\right) \;=\; \calF_q(z) \; , $$
    where the last equality uses (i). 
    Thus $\bar B_l = B_l$; all coefficient matrices are real.
	Analogous arguments apply to $C_q(z)$, proving statement (iii).
    $\blacksquare$
}

\propositionmarker{prop_NVARpq_observeddynamics}

\myproofof{prop_NVARpq_observeddynamics}{
    Let $y_t$ first be a stock variable.
	Multiplying $F(L)x_\tau = e_\tau$ by $C_q(L)$ gives $\calF_q(L)x_\tau = C_q(L)e_\tau$. 
    Since $\calF_q(L) = \calG_q(L^q)$ and $L^{qk}x_{tq} = y_{t-k}$, evaluating at $\tau = tq$ gives $\calG_q(L)y_t = \nu_t$ with $\nu_t = C_q(L)e_{tq}$, i.e. $B_{q,k} = C_{q,k}$ for $k = 0:d$.
    The innovations $e_{tq-k}$ in $\nu_t$ and $e_{tq-qh-k'}$ in $\nu_{t-h}$ coincide iff $k' = k - qh$, yielding the stated formula for the autocovariances, which vanish iff $qh > p(q-1)$, i.e. for $h > q^* = \text{floor}(p(q-1)/q)$.
	Since $p(q-1)/q < p$, we have $q^* \leq p-1$.
	The stated properties of $\calG_q(z)$ and $C_q(z)$ carry over from \cref{lemma_calFqRewriting}.

	Now suppose $y_t$ is a flow variable.
	Apply $s_q(L)I_n$ to $F(L)x_\tau = e_\tau$ and proceed as above with $D_q(z)$ in place of $C_q(z)$, which gives $\calG_q(L)y_t = \nu_t$ with $\nu_t = D_q(L)e_{tq}$, i.e. $B_{q,k} = D_{q,k}$ for $k = 0:d$, and autocovariances that vanish iff $qh > (p+1)(q-1)$, i.e. for $h > q^* = \text{floor}((p+1)(q-1)/q)$.
	Since $(p+1)(q-1)/q < p+1$, we have $q^* \leq p$.
    $\blacksquare$
}


\propositionmarker{prop_WS_NVARpq}

For $\zeta \in \complexnumbers$, let $\phi_\zeta(z) = 1 - \zeta g(z)$, which for an eigenvalue $\lambda$ of $A$ is the polynomial from \cref{prop_WS_NVARpq}, 
and write $\phi_\zeta(z) = \prod_{i=1}^p (1-\eta_{\zeta,i}z)$, so that $\{\eta_{\zeta,i}\}_{i=1:p}$ are the roots of the monic polynomial $w^p\phi_\zeta(1/w) = w^p - \zeta\sum_{k=1}^p \alpha_k w^{p-k}$.
Let $\bar\eta = \max_{j=1:n,i=1:p}|\eta_{\lambda_j,i}|$.
Throughout, $x_\tau$ is defined to be w.s. if $\det F(z) \neq 0$ for all $|z| \le 1$.\footnote{
	Equivalently, $|l_i|<1$ for all eigenvalues $l_i$ of the companion-form matrix $F$ \citep[Section 2.1]{Luetkepohl2005}.
}
Since $\phi_\lambda(z) \neq 0$ for all $|z| \leq 1$ iff $|\eta_{\lambda,i}|<1$ for all $i=1:p$, \cref{prop_WS_NVARpq} implies that $x_\tau$ is w.s. iff $\bar\eta<1$.

\myproofof{prop_WS_NVARpq}{
	The eigenvalues of $g(z)A$ are $g(z)\lambda_j$, so those of $F(z) = I_n - g(z)A$ are $1 - g(z)\lambda_j = \phi_{\lambda_j}(z)$, and the determinant is their product.
	Therefore, $\det F(z) = \prod_{j=1}^{n} \phi_{\lambda_j}(z)$, and the statement for $x_\tau$ follows trivially.

	By \cref{prop_NVARpq_observeddynamics}, we know $\calG_q(L) y_t = \nu_t$.
	Consequently, $\det \calG_q(z^q) = \det \calF_q(z) = \prod_{m} \prod_j \phi_{\lambda_j}(\omega_q^m z)$.
	Note that $\prod_{m=0}^{q-1} \phi_\lambda(\wq^m z) = \prod_{m=0}^{q-1} \prod_{i=1}^p (1-\eta_{\lambda,i}\wq^m z) = \prod_{i=1}^p (1-\eta_{\lambda,i}^q z^q)$.\footnote{
        It suffices to show $\prod_{m=0}^{q-1}(1-\eta \wq^m z) = 1 - \eta^q z^q$ for any $\eta \in \complexnumbers$.
        For $\eta = 0$, this trivially holds. 
        For $\eta \neq 0$,
        both sides are polynomials in $z$ of degree $q$ with the same $q$ distinct roots $\{\wq^{-m}\eta^{-1}\}_{m=0}^{q-1}$, and are therefore equal up to a multiplicative constant.
        Since both have constant term $1$, that constant is one: the two polynomials are equal. 
    }
	Hence, we have $\det \calG_q(z) = \prod_{j} \prod_{i=1}^p (1-\eta_{\lambda_j,i}^q z)$.
	If $x_\tau$ is stationary, then $|\eta_{\lambda_j,i}|<1$ for all $i=1:p$ and $j=1:n$, 
	which implies $\det \calG_q(z) \neq 0$ for all $|z|\leq 1$, i.e. $y_t$ is stationary.
	$\blacksquare$ 
}


\mylemma{}{lemma_LTresp_NVARp1}{%
	Let $y_t$ evolve as in \cref{eq_NVARp1}, and let $\bar y_t = aA \bar y_t + v_t$ with $a=\sum_{l=1}^p \alpha_l$.
	Assume $y_t$ is w.s..
	Then
	$\underset{H\rightarrow \infty}{\lim} \sum_{h=0}^H \partial y_{i,t+H} / \partial u_{j,t+h} =  \partial \bar y_{it} / \partial v_{jt}  = \sbr{ (I-aA)^{-1} }_{ij}$.
}

\myproof{
	First, note that $\bar y_t = (I-aA)^{-1} v_t$ and therefore $\partial \bar y_t / \partial v_t = (I-aA)^{-1}$.
	Turning to $y_t$, under weak stationarity,
	\begin{align*}
		R \equiv \underset{H\rightarrow \infty}{\lim} \sum_{h=0}^{H} \frac{\partial y_{t+H}}{\partial u_{t+h}} = \underset{H\rightarrow \infty}{\lim}  \sum_{h=0}^{H} \frac{\partial y_{t+h}}{\partial u_{t}} = \sum_{h=0}^\infty \frac{\partial y_{t+h}}{\partial u_{t}}  \; .
	\end{align*}
	To find $R$, note that $\partial y_{t+h} / \partial u_t  = (F^h)_{11}$ for the companion-form matrix $F$.
	Since $\sum_{h=0}^\infty F^h = (I-F)^{-1}$, $R = \br{ (I-F)^{-1} }_{11}$ is the $n \times n$ matrix in position (1,1) in the $np \times np$ matrix $ (I-F)^{-1}$.
	
	Let $M = (I-F)^{-1}$ and partition it into $p^2$ blocks of dimension $n \times n$, denoted by $\{M_{lm}\}_{l,m=1:p}$. Then $R = M_{11}$. We have $I = M(I-F)$ and therefore
	\begin{align*}
		I = \begin{bmatrix} 
				M_{11} & M_{12} & ... & M_{1p} \\
				\vdots & \vdots & \ddots & \vdots \\
				M_{p1} & M_{p2} & ... & M_{pp}
			\end{bmatrix} 
		\begin{bmatrix} 
			I-\alpha_1 A & -\alpha_2 A  & ... & -\alpha_{p-1} A  & -\alpha_p A  \\
			-I_n & I_n & & & 0_n \\ 
			0_n & -I_n & \ddots &  & \vdots \\ 
			\vdots & & \ddots & \ddots &  \\ 
			0_n & \dots & & -I_n & I_n 
		\end{bmatrix}
	\end{align*}
	Comparing the left- and right-hand sides for block $(1,p)$, we get $0_n = -M_{11}\alpha_p A + M_{1p}$,
	which implies $M_{1p} = M_{11}\alpha_p A$. 
	For block $(1,l)$, $l \in 2:(p-1)$, we get $0_n = -M_{11} \alpha_l A + M_{1l} - M_{1,l+1}$ so that
	\begin{align*}
		M_{12} = M_{11}\alpha_2 A + M_{13} = M_{11}\alpha_2 A + M_{11}\alpha_3 A + M_{14} = ... = M_{11}(\alpha_2 + ... + \alpha_p)A \; .
	\end{align*}
	Finally, $I_n = M_{11}(I-\alpha_1 A) - M_{12} = M_{11}(I-aA)$ 
	so that $M_{11} = \br{ (I-F)^{-1} }_{11} = (I-aA)^{-1}$. 
	$\blacksquare$
}


\mylemma{}{lemma_LTresp_NVARp1_AR1innov}{
	Let $y_t$ evolve as in \cref{eq_NVARp1}, and let $\bar y_t = a A \bar y_t + v_t$ with $a =\sum_{l=1}^p \alpha_l$.
	Assume $y_t$ and $\bar y_t$ are w.s., and assume $(1-\rho_j L) u_{jt} = \varepsilon_{jt} \sim WN$ and $(1-\rho_j L) v_{jt} = \epsilon_{jt} \sim WN$, with $\rho_j \in [0,1)$.
	Then
	$\underset{H\rightarrow \infty}{\lim} \sum_{h=0}^H \partial y_{i,t+H} / \partial \varepsilon_{j,t+h}  = \underset{H\rightarrow \infty}{\lim} \sum_{h=0}^H \partial \bar y_{i,t+H} / \partial \epsilon_{j,t+h} = (1-\rho_j)^{-1}\sbr{(I- a A)^{-1}}_{ij}$.
}

\myproof{
	It holds that
	\begin{align*}
		\frac{\partial \bar y_{t+h}}{\partial \epsilon_{t}} = \sum_{l=0}^h \frac{\partial \bar y_{t+h}}{\partial v_{t+h-l}} \frac{\partial v_{t+h-l}}{\partial \epsilon_{t}} \quad \text{and} \quad \frac{\partial y_{t+h}}{\partial \varepsilon_{t}} = \sum_{l=0}^h \frac{\partial y_{t+h}}{\partial u_{t+h-l}} \frac{\partial u_{t+h-l}}{\partial \varepsilon_{t}} \; .
	\end{align*}
	By \cref{lemma_LTresp_NVARp1}, under stationarity of $\bar y_t$ and $y_t$, we also know that
	$$ \frac{\partial \bar y_{t}}{\partial v_{t}} = \underset{H\rightarrow \infty}{\lim} \sum_{l=0}^H \frac{\partial y_{t+l}}{\partial u_{t}} = (I- a A)^{-1} \; . $$
	For $\bar y_t$, $\partial \bar y_{t+h}/\partial v_{t+h-l} = 0$ for $l \geq 1$. We then have
	\begin{align*}
		\underset{H\rightarrow \infty}{\lim} \sum_{h=0}^{H} \frac{\partial \bar y_{t+H}}{\partial \epsilon_{t+h}} = \sum_{h=0}^\infty \frac{\partial \bar y_{i,t+h}}{\partial \epsilon_{j,t}}
		= \sum_{h=0}^\infty \frac{\partial \bar y_{i,t+h}}{\partial v_{j,t+h}} \frac{\partial v_{j,t+h}}{\partial \epsilon_{j,t}}
		= \sum_{h=0}^\infty \sbr{(I-aA)^{-1}}_{ij} \rho_j^h = (1-\rho_j)^{-1}\sbr{(I- a A)^{-1}}_{ij} \; . 
	\end{align*}
	For $y_t$, we have 
	\begin{align*}
		\underset{H\rightarrow \infty}{\lim} \sum_{h=0}^H \frac{\partial y_{i,t+h}}{\partial \varepsilon_{j,t}} 
		&= \underset{H\rightarrow \infty}{\lim} \sum_{h=0}^H \sum_{l=0}^h \frac{\partial y_{i,t+l}}{\partial u_{j,t}} \rho_j^{h-l} 
		= \underset{H\rightarrow \infty}{\lim} \sum_{l=0}^H \frac{\partial y_{i,t+l}}{\partial u_{j,t}} \sum_{h=l}^{H} \rho_j^{h-l}  \\
		&= \underset{H\rightarrow \infty}{\lim} \sum_{l=0}^H \frac{\partial y_{i,t+l}}{\partial u_{j,t}} \frac{1-\rho_j^{H-l+1}}{1-\rho_j} 
		=\frac{1}{1-\rho_j} \underset{H\rightarrow \infty}{\lim} \sum_{l=0}^H \frac{\partial y_{i,t+l}}{\partial u_{j,t}}  - \underset{H\rightarrow \infty}{\lim} \sum_{l=0}^H  \frac{\rho_j^{H-l+1}}{1-\rho_j}  \frac{\partial y_{i,t+l}}{\partial u_{j,t}} \; .
	\end{align*}
	By \cref{lemma_LTresp_NVARp1}, the first term equals $(1-\rho_j)^{-1} \sbr{(I- a A)^{-1}}_{ij}$. 

	The second term equals zero. To see this, break it up as follows:
	$$ \underset{H\rightarrow \infty}{\lim} \sum_{l=0}^{H-1}  \frac{\rho_j^{H-l+1}}{1-\rho_j}  \frac{\partial y_{i,t+l}}{\partial u_{j,t}} + \underset{H\rightarrow \infty}{\lim} \frac{\rho_j}{1-\rho_j}  \frac{\partial y_{i,t+H}}{\partial u_{j,t}} \; .$$
	The latter term equals zero since $y_t$ is stationary.
	The former term can be written as 
	$$ \frac{1}{1-\rho_j}  \underset{H\rightarrow \infty}{\lim} \rho_j^{H+1-c} \sum_{l=0}^{H-1} \rho_j^{c-l}  \frac{\partial y_{i,t+l}}{\partial u_{j,t}} \; \forall \; c > 0 \; .$$
	Let $c = H(1-\delta)$ for $\delta > 0$ small.
	Since $\underset{H\rightarrow \infty}{\lim} \rho_j^{H+1-c} = 0$, it remains to be shown that the limt of the second term in this product is finite, in which case the limit of products is equal to the product of their limits, which is zero.

	Let $\tilde{\rho}_j = \rho_j^{c/(H-1)-1}$. For $\delta$ small enough, $\tilde{\rho}_j \in [0,1)$. We then have
	\begin{align*}
		\underset{H\rightarrow \infty}{\lim} \sum_{l=0}^{H-1}  \rho_j^{c-l} \frac{\partial y_{i,t+l}}{\partial u_{j,t}} 
		= \underset{H\rightarrow \infty}{\lim} \sum_{l=0}^{H-1}  \br{\rho_j^{c/l-1}}^l \frac{\partial y_{i,t+l}}{\partial u_{j,t}} 
		&\leq \underset{H\rightarrow \infty}{\lim} \sum_{l=0}^{H-1}  \tilde{\rho}_j^l \frac{\partial y_{i,t+l}}{\partial u_{j,t}} \\
		&\leq \underset{H\rightarrow \infty}{\lim} \sum_{l=0}^{H-1} \frac{\partial y_{i,t+l}}{\partial u_{j,t}} = \sbr{(I- a A)^{-1}}_{ij} < \infty \quad \blacksquare 
	\end{align*}
}





\mylemma{}{lemma_LTresp_NVARpq_AR1innov}{%
	Let $y_t$ evolve as in \cref{eq_NVARpq} for some $q \in \naturals \backslash\{1\}$. 
	Assume $x_\tau$ and $y_t$ are w.s. and $(1-\rho_jL)e_{j\tau} = \tilde{\varepsilon}_{j\tau} \sim WN$, with $\rho_j \in [0,1)$,
	and let $\kappa_q = 1$ if $y_t$ is a stock variable and $\kappa_q = q$ if $y_t$ is a flow variable.
	Then
	$\underset{H\rightarrow \infty}{\lim}  \sum_{h=0}^{Hq} \partial y_{t+H} / \partial \tilde{\varepsilon}_{tq+h}
	= 
	\kappa_q \underset{H\rightarrow \infty}{\lim}  \sum_{h=0}^{H} \partial x_{\tau + H} / \partial \tilde{\varepsilon}_{\tau+h}$.
}

\myproof{
	If $y_t$ is a stock, we have $y_{t+H} = x_{(t+H)q}$, and so
	\begin{align*}
		\underset{H\rightarrow \infty}{\lim}  \sum_{h=0}^{Hq} \frac{\partial y_{t+H}}{\partial \tilde{\varepsilon}_{tq+h}} = 
		\underset{H\rightarrow \infty}{\lim}  \sum_{h=0}^{Hq} \frac{\partial x_{(t+H)q}}{\partial \tilde{\varepsilon}_{tq+h}} =
		\underset{H\rightarrow \infty}{\lim}  \sum_{h=0}^{Hq} \frac{\partial x_{\tau + Hq}}{\partial \tilde{\varepsilon}_{\tau+h}} = \underset{H\rightarrow \infty}{\lim}  \sum_{h=0}^{H} \frac{\partial x_{\tau + H}}{\partial \tilde{\varepsilon}_{\tau+h}} \; .
	\end{align*}
	If $y_t$ is a flow, we have $y_{t+H} = x_{(t+H)q} + ... + x_{(t+H)q-q+1}$, and each of its $q$ terms contributes the same limit:
	\begin{align*}
		\underset{H\rightarrow \infty}{\lim}  \sum_{h=0}^{Hq} \frac{\partial y_{t+H}}{\partial \tilde{\varepsilon}_{tq+h}} &= \underset{H\rightarrow \infty}{\lim}  \sum_{h=0}^{Hq} \sum_{k=0}^{q-1} \frac{\partial x_{(t+H)q-k}}{\partial \tilde{\varepsilon}_{tq+h}} \one{h \leq Hq -k }  \\
		&= \underset{H\rightarrow \infty}{\lim}  \sum_{k=0}^{q-1} \sum_{h=0}^{Hq-k}  \frac{\partial x_{(t+H)q-k}}{\partial \tilde{\varepsilon}_{tq+h}} 
		= q \underset{H\rightarrow \infty}{\lim}  \sum_{h=0}^{H}  \frac{\partial x_{\tau + H}}{\partial \tilde{\varepsilon}_{\tau+h}} \; .\footnotemark \quad \blacksquare
	\end{align*}
	\footnotetext{The indicator function ensures that only responses to contemporaneous or past impulses are summed-up.}
}

\mylemma{}{lemma_contempResp_timeagg}{%
	Let the flow variable $y_t$ evolve as in \cref{eq_NVARpq} for some $q \in \naturals \backslash\{1\}$.
	Assume $x_\tau$ and $y_t$ are w.s. and $(1-\rho_jL)e_{j\tau} = \tilde{\varepsilon}_{j\tau} \sim WN$, with $\rho_j \in [0,1)$. 
	Let $\tilde{\varepsilon}_\tau = (\tilde{\varepsilon}_{1\tau}, ..., \tilde{\varepsilon}_{n\tau})'$.
	Then for $g \in 0:(q-1)$,
	$\underset{g\rightarrow \infty}{\lim}  \partial y_{t} / \partial \tilde{\varepsilon}_{\tau-g}
	= 
	\underset{g\rightarrow \infty}{\lim}  \sum_{h=0}^{g} \partial x_{\tau + g} / \partial \tilde{\varepsilon}_{\tau+h}$.
}

\myproof{
	We consider the response of $y_{t} = \sum_{k=0}^{q-1} x_{tq-k}$ to a (high-frequency) innovation $\tilde{\varepsilon}_{tq-g}$, $g \in 0:(q-1)$ that occurs within observational period $t$.
	We have
	$$ \frac{\partial y_{t}}{\partial \tilde{\varepsilon}_{tq-g}} 
	= \sum_{k=0}^{q-1} \frac{\partial x_{tq-k}}{\partial \tilde{\varepsilon}_{tq-g}} \one{k \leq g}
	= \sum_{k=0}^{q-1} \frac{\partial x_{\tau+g-k}}{\partial \tilde{\varepsilon}_{\tau}} \one{k \leq g}
	= \sum_{k=0}^{g} \frac{\partial x_{\tau+g-k}}{\partial \tilde{\varepsilon}_{\tau}}
	= \sum_{h=0}^{g} \frac{\partial x_{\tau+h}}{\partial \tilde{\varepsilon}_{\tau}} \; . \quad \blacksquare $$
}

\propositionmarker{prop_NVARpq_SARSEM}


\propositionpartmarker{prop_NVARpq_SARSEM__LTresp}

\propositionpartmarker{prop_NVARpq_SARSEM__contempresp}

\propositionpartmarker{prop_NVARpq_SARSEM__varobsinnov}

\propositionpartmarker{prop_NVARpq_SARSEM__limit_AutoregPol}

\propositionpartmarker{prop_NVARpq_SARSEM__limit_stock}

\propositionpartmarker{prop_NVARpq_SARSEM__limit_flow_obsinnov}

\propositionpartmarker{prop_NVARpq_SARSEM__limit_flow_y}

Recall that $\Theta_h = \partial x_{\tau+h}/\partial x_\tau$ and $\Gamma_h = \sum_{m=0}^h \Theta_m$, and define $\Gamma^{(q)}_h = \sum_{m=\max(0,h-q+1)}^{h}\Theta_m$, which equals $\Gamma_h$ for $h \leq q-1$.
Since $\Theta_0 = I_n$ and $\Theta_h = \sum_{l=1}^{\min(h,p)}\alpha_l A\Theta_{h-l}$, the $\Theta_h$, and hence the $\Gamma_h$ and $\Gamma^{(q)}_h$, are polynomials in $A$.
Under weak stationarity of $x_\tau$, $F(L)$ is invertible with $\Psi(z) = F(z)^{-1} = \sum_{h=0}^\infty \Theta_h z^h$, 
so that $x_\tau = \Psi(L)e_\tau = \sum_{h=0}^\infty \Theta_h e_{\tau-h}$ and, in the flow case, $y_t = s_q(L)\Psi(L)e_{tq} = \sum_{h=0}^\infty \Gamma^{(q)}_h e_{tq-h}$.
Note that $\Psi(1) = (I_n - g(1)A)^{-1} = \Gamma$ since $g(1) = a$.

\myproofof{prop_NVARpq_SARSEM__LTresp}{It follows from \cref{lemma_LTresp_NVARp1,lemma_LTresp_NVARpq_AR1innov}, applied for $\rho_j=0$. $\blacksquare$}

\myproofof{prop_NVARpq_SARSEM__contempresp}{It follows from \cref{lemma_LTresp_NVARp1,lemma_contempResp_timeagg}, applied for $\rho_j=0$. $\blacksquare$}

\myproofof{prop_NVARpq_SARSEM__varobsinnov}{
	Every $e_{tq-h}$ with $h \geq q$ is a function of $x_{tq-h}, x_{tq-h-1}, ...$ and hence contained in $\calF_{t-1}$, while $e_{tq}, ..., e_{tq-q+1}$ are orthogonal to $\calF_{t-1}$.
	In the stock case, $y_t = x_{tq} = \sum_{h=0}^\infty \Theta_h e_{tq-h}$, so that $u_t \equiv y_t - \mean[y_t|\calF_{t-1}] = \sum_{h=0}^{q-1}\Theta_h e_{tq-h}$
	and $\var[u_t] = \Sigma + \sum_{h=1}^{q-1} \Theta_h \Sigma \Theta_h'$ as $e_\tau \sim WN(0,\Sigma)$. 
	In the flow case, $y_t = \sum_{h=0}^\infty \Gamma^{(q)}_h e_{tq-h}$, so that $u_t = \sum_{h=0}^{q-1}\Gamma_h e_{tq-h}$ 
	and $\var[u_t] = \Sigma + \sum_{h=1}^{q-1} \Gamma_h \Sigma \Gamma_h'$.
	$\blacksquare$
}

To illustrate \cref{prop_NVARpq_SARSEM__varobsinnov}, consider an NVAR($p,q$) for $q=2$ and suppose $\Sigma =\text{diag}(\sigma^2_1, ..., \sigma^2_n)$. 
If $y_t = x_\tau$ for $t=\tau/q \in \naturals$, then 
$u_t = e_\tau + \alpha_1 A e_{\tau-1}$, with
$$ \var[u_t] = \Sigma + \alpha_1^2 A \Sigma A' \; . $$
This reveals that $\covar(u_{it},u_{jt}) = \one{i=j}\sigma_i^2 + \alpha_1^2 \sum_{k=1}^n a_{ik}a_{jk} \sigma^2_k$. 
If instead $y_t = x_\tau + ... + x_{\tau-q+1}$ for $t=\tau/q \in \naturals$, then
$u_t = e_\tau + (I+\alpha_1 A) e_{\tau-1}$, with
$$ \var[u_t] = 2\Sigma + \alpha_1 \br{ A \Sigma + (A \Sigma)' } + \alpha_1^2 A \Sigma A' \; , $$
which leads to $\covar(u_{it},u_{jt}) = \one{i=j} 2 \sigma_i^2 + \alpha_1 (a_{ij}\sigma_j^2 + a_{ji}\sigma_i^2) + \alpha_1^2 \sum_{k=1}^n a_{ik}a_{jk} \sigma^2_k$.
In the former case, $\covar(u_{it},u_{jt})$ is determined by common exposure to third units, while in the latter case bilateral exposure matters as well. In either case, exposure means first-order connections.\footnote{
	To understand the different results for stock and flow variables, assume for simplicity that there are no self-links: $a_{ii}=0 \; \forall\; i$. 
	For a stock variable, $y_{it}=x_{i,\tau}$, which implies that $u_{it}$ only depends on $e_{i,\tau}$, not on $e_{i,\tau-1}$: we have $u_{it} = e_{i,\tau} + \alpha_1 \sum_{j=1}^n a_{ij}e_{j,\tau-1}$. As a result, the only (possibly) common terms in $u_{it}$ and $u_{jt}$ are the one-period lagged high-frequency innovations of third units, $e_{k,\tau-1}$, and any comovement between $u_{it}$ and $u_{jt}$ is due to common exposure to these units: $\covar\br{u_{it},u_{jt}} \neq 0$ iff $\exists \; k$ s.t. $a_{ik},a_{jk} \neq 0$.
	In contrast, for a flow variable, $y_{it}=x_{i,\tau}+ x_{i,\tau-1}$, which implies that $u_{it}$ also depends on $e_{i,\tau-1}$: we have $u_{it} = e_{i,\tau} + e_{i,\tau-1} + \alpha_1 \sum_{j=1}^n a_{ij}e_{j,\tau-1}$. Besides common exposure to third units, comovement is also due to bilateral exposure: $a_{ij} \neq 0 \; | \; a_{ji}\neq 0 \; \Rightarrow \; \covar\br{u_{it},u_{jt}} \neq 0$. 
	} 
Under $q=3$, exposure is determined by first- and second-order connections: e.g. for a stock variable, we have $u_t = e_\tau + \alpha_1 A e_{\tau-1} + (\alpha_2 A + \alpha_1^2A^2)e_{\tau-2}$, with 
$$ \var[u_t] = \Sigma + (\alpha_1^2 + \alpha_2^2) A \Sigma A' + \alpha_1^2 \alpha_2 \br{ A \Sigma (A^2)' + \sbr{ A \Sigma (A^2)' }' } + \alpha_1^4 A^2 \Sigma (A^2)' \; . $$

\myproofof{prop_NVARpq_SARSEM__limit_flow_obsinnov}{
	By the proof of \cref{prop_NVARpq_SARSEM__varobsinnov}, $u_t = \sum_{h=0}^{q-1}\Gamma_h e_{tq-h}$.
	Define $e_h \equiv \Gamma_h e_{tq-h}$ with $\mean[e_h]=0$ and $V_h \equiv \var[e_h] = q^{-1}\Gamma_h \Sigma_* \Gamma_h'$.
	Fix $d \in \reals^n$, $d \neq 0$, and consider $d'u_t = \sum_{h=0}^{q-1} d'e_h$, a sum of independent, mean-zero terms.
	Let $s^2_q = \sum_{h=0}^{q-1} d'V_hd$ and $\sigma_* = d'V_*d$ with $V_* = \Gamma\Sigma_*\Gamma'$.
	Below, I show $s^2_q \rightarrow \sigma_*$; note that $\sigma_*>0$ since $\Sigma_*$ is positive definite and $\Gamma$ is invertible, the latter because $\det F(1) = \det(I-aA) \neq 0$ under weak stationarity of $x_\tau$.
	Lyapunov's condition holds: since $\Gamma_h \rightarrow \Gamma$, we have $\sup_h\|\Gamma_h\|<\infty$ and hence $\mean|d'e_h|^{2+\delta} \leq \|\Gamma_h'd\|^{2+\delta}\, q^{-1-\delta/2}\, \mean\|\sqrt{q}\,e_\tau\|^{2+\delta} \leq \bar{c}\, q^{-1-\delta/2}$ uniformly in $h$ and $q$ (because the distribution of $\sqrt{q}\,e_\tau$ does not depend on $q$), so that
	$$ \frac{1}{s^{2+\delta}_q}\sum_{h=0}^{q-1}\mean|d'e_h|^{2+\delta} \leq \frac{\bar{c} \, q^{-\delta/2}}{(s^2_q)^{1+\delta/2}} = O\br{q^{-\delta/2}} \rightarrow 0 \; .$$
	Therefore, by Lyapunov's Central Limit Theorem,
	$ s_q^{-1}d'u_t = s_q^{-1} \sum_{h=0}^{q-1} d'e_h \convdist N(0,1) $,
	and, by Slutsky's theorem,
	$ d'u_t = s_q \cdot s_q^{-1}d'u_t \convdist N(0,\sigma_*) $.
	As this argument applies for arbitrary $d \neq 0$, by the Cramer-Wold theorem, $u_t \convdist N(0,V_*)$.

	It remains to show that $s^2_q = q^{-1}\sum_{h=0}^{q-1} \sigma_h \rightarrow d'V_*d \equiv \sigma_*$ for $\sigma_h = q\,d'V_hd = d'\Gamma_h\Sigma_*\Gamma_h'd$.
	By \cref{lemma_LTresp_NVARp1}, as $h\rightarrow \infty$, $\Gamma_h \rightarrow \Gamma$. 
	Therefore, $qV_h \rightarrow \Gamma \Sigma_* \Gamma'$,
	and $\sigma_h \rightarrow \sigma_*$.
	In other words, $\forall \; \epsilon > 0$, $\exists \; H$ s.t. $|\sigma_h - \sigma_*|< \epsilon \; \forall \; h>H$.
	Also, for $H \leq q-1$, 
	\begin{align*}
		|s^2_q - \sigma_*| &= \bigabslines{\frac{1}{q}\sum_{h=0}^{q-1} (\sigma_h - \sigma_*)} 
		\leq \frac{1}{q} \sum_{h=0}^{H-1} |\sigma_h - \sigma_*| + \frac{1}{q} \sum_{h=H}^{q-1} |\sigma_h - \sigma_*| 
		\leq \frac{1}{q} \sum_{h=0}^{H-1} |\sigma_h - \sigma_*| + \epsilon \; .
	\end{align*}
	The first term vanishes as $q \rightarrow \infty$.
	Hence, $\forall \; \epsilon > 0$, $\underset{q\rightarrow \infty}{\lim} |s^2_q - \sigma_*| \leq \epsilon$, i.e. $s^2_q \rightarrow \sigma_*$. $\blacksquare$
}


I define a few objects to prove \cref{prop_NVARpq_SARSEM__limit_AutoregPol,prop_NVARpq_SARSEM__limit_flow_y,prop_NVARpq_SARSEM__limit_stock,prop_NVARpqWC_observeddynamics} more easily.
Throughout the proofs, fix $\varrho \in (\bar\eta,1)$. 
Since weak stationarity of $x_\tau$ means $\bar\eta<1$, there is some $c<\infty$ s.t.
$\|\Theta_h\| \leq c \varrho^{h}$ and $\|\gamma_x(h)\| \leq c\,\|\Sigma\|\,\varrho^{|h|}$, where $\gamma_x(h) = \mean[x_\tau x_{\tau-h}']$.
Neither $\varrho$ nor $c$ depend on $q$, as both are determined by $\alpha$ and $A$ alone.

\myproofof{prop_NVARpq_SARSEM__limit_AutoregPol}{
	I show that $\forall \; \varsigma \in (\bar\eta,1) \; \exists \; K_\varsigma < \infty$, independent of $q$, s.t. $\|\calG_{q,k}\| \leq K_\varsigma \varsigma^{qk}$ for $k=1:p$.
	As in the proof of \cref{prop_WS_NVARpq},
	$$ \prod_{m=0}^{q-1}\phi_\zeta(\wq^m z) = \prod_{i=1}^p \br{1-\eta_{\zeta,i}^q z^q} = 1 - \sum_{k=1}^p (-1)^{k+1}E_k(\zeta) z^{qk} \; , \quad E_k(\zeta) = e_k\br{\eta_{\zeta,1}^q, ..., \eta_{\zeta,p}^q} \; , $$
	where $e_k$ is the $k$th elementary symmetric polynomial. 
	Being symmetric in the $\eta_{\zeta,i}$, $E_k$ is a polynomial in the elementary symmetric functions of the $\eta_{\zeta,i}$, i.e.\ in the coefficients of $\phi_\zeta$.
	Since these are linear in $\zeta$, $E_k$ is a polynomial in $\zeta$.
	Note that $|E_k(\zeta)| \leq {p \choose k}\br{\max_i |\eta_{\zeta,i}|}^{qk}$.
	Since $\calG_q(z^q) = \calF_q(z) = \prod_{m=0}^{q-1}F(\wq^m z)$ and all factors are polynomials in $A$ that commute (\cref{lemma_calFqRewriting}(i)), the same algebra applies with $\zeta$ replaced by $A$, giving $\calG_{q,k} = (-1)^{k+1}E_k(A)$.
	If $A$ is diagonalizable, then $A = W \Lambda W^{-1}$, so that $\calG_{q,k} = (-1)^{k+1}W \text{diag}\br{E_k(\lambda_1), ..., E_k(\lambda_n)}W^{-1}$ and $\|\calG_{q,k}\| \leq \|W\|\|W^{-1}\|{p \choose k}\bar\eta^{qk}$.\footnote{
		The $\eta_{\zeta,i}$ are continuous in $\zeta$, being the roots of a monic polynomial whose coefficients are linear in $\zeta$, so there is an $r>0$ s.t. 
		$\max_i|\eta_{\zeta,i}| \leq \varsigma$, and therefore $|E_k(\zeta)| \leq {p \choose k}\varsigma^{qk}$, for every $\zeta$ within distance $r$ of an eigenvalue of $A$.
		By the Jordan normal form, $A = SJS^{-1}$ for some invertible $S$, where $J$ is block-diagonal with square blocks $J_1, ..., J_B$ of sizes $m_1, ..., m_B$ summing to $n$, and where block $b$ carries one eigenvalue $\lambda_b$ of $A$ along its diagonal, ones directly above it and zeros elsewhere:
		$ J_b = \lambda_b I_{m_b} + N_b $.
		$A$ is diagonalizable exactly when every block has size one, i.e. when every $N_b=0$.
		We have $\|N_b\| \leq 1$, since $N_b$ maps each standard basis vector to another one or to zero.
		Also, $N_b^{m_b}=0$, since each multiplication by $N_b$ shifts its ones one place further up, so that after $m_b$ steps none is left.
		Let $m = \max_b m_b$, so that $N_b^m = 0$ for every $b$.
		The Taylor expansion of the polynomial $E_k$ around $\lambda_b$ terminates after $m$ terms, $E_k(J_b) = \sum_{l=0}^{m-1} E_k^{(l)}(\lambda_b)N_b^l/l!$, 
		and Cauchy's estimate on the circle of radius $r$ around $\lambda_b$ bounds each $|E_k^{(l)}(\lambda_b)|/l! \leq r^{-l}\max_{|\zeta-\lambda_b|=r}|E_k(\zeta)| \leq r^{-l}{p \choose k}\varsigma^{qk}$.
		Since $E_k(J)$ is block-diagonal with blocks $E_k(J_b)$, we have $\|E_k(A)\| \leq \|S\|\|S^{-1}\|\max_b\|E_k(J_b)\|$, and hence $\|\calG_{q,k}\| \leq K_\varsigma \varsigma^{qk}$ for $K_\varsigma = \|S\|\|S^{-1}\|{p \choose k}\sum_{l=0}^{m-1}r^{-l} < \infty$, which does not depend on $q$.
	}
	Since $\bar\eta<1$ and $\varsigma<1$, both bounds vanish as $q \rightarrow \infty$, giving $\calG_{q,k} \rightarrow 0$ for $k=1:p$ and hence $\calG_q(z) \rightarrow I_n$. $\blacksquare$
}

\myproofof{prop_NVARpq_SARSEM__limit_flow_y}{
	We have $y_t = \sum_{l=0}^{q-1}x_{tq-l}$, so that $\gamma_y(h) = \sum_{l,m=0}^{q-1}\gamma_x(hq-l+m)$.
	For $h=0$ this equals $\sum_{|d|\leq q-1}(1-|d|/q)\,q\gamma_x(d)$, which converges to $\sum_{d=-\infty}^{\infty}q\gamma_x(d)$ by dominated convergence, since $q\gamma_x(d)$ does not depend on $q$ and is absolutely summable, as shown next.
	Since $x_\tau = \sum_{j=0}^\infty \Theta_j e_{\tau-j}$ and $\mean[e_se_t'] = q^{-1}\Sigma_*\one{s=t}$, only the terms with $j = h+d$ survive in $\gamma_x(d) = \sum_{j,h \geq 0}\Theta_j \mean[e_{\tau-j}e_{\tau-d-h}']\Theta_h'$, so that $q\gamma_x(d) = \sum_{h=0}^\infty \Theta_{h+d}\Sigma_*\Theta_h'$ for every $d \in \integers$, where $\Theta_j = 0$ for $j<0$.
	Since $\|\Theta_j\| \leq c\varrho^j$, the resulting double sum over $d$ and $h$ converges absolutely, so we may sum over $d$ first, and for fixed $h$, as $d$ runs through $\integers$, so does $h+d$, so that $\sum_{d \in \integers} \Theta_{h+d} = \sum_{j=0}^\infty \Theta_j = \Psi(1)$ irrespective of $h$. 
	Hence
	$$ \sum_{d=-\infty}^{\infty} q\gamma_x(d) = \sum_{h=0}^\infty \br{\sum_{d=-\infty}^{\infty}\Theta_{h+d}} \Sigma_* \Theta_h' = \Psi(1)\Sigma_*\sum_{h=0}^\infty \Theta_h' = \Psi(1)\Sigma_*\Psi(1)' = \Gamma\Sigma_*\Gamma' \; , $$
	i.e.\ $q$ times the long-run variance of $x_\tau$, as for any linear process $\Psi(L)e_\tau$.
	For $h \geq 1$, all indices satisfy $hq-l+m \geq (h-1)q+1 > 0$, so
	$$ \|\gamma_y(h)\| \leq \frac{c\|\Sigma_*\|}{q}\varrho^{hq}\br{\sum_{l=0}^{q-1}\varrho^{-l}}\br{\sum_{m=0}^{q-1}\varrho^{m}} \leq \frac{c\|\Sigma_*\|}{q}\frac{\varrho^{(h-1)q+1}}{(1-\varrho)^2} \leq \frac{c\|\Sigma_*\| \, \varrho}{q(1-\varrho)^2} \rightarrow 0 \; . $$

	Since $y_t = \sum_{k=0}^{\infty}\Gamma^{(q)}_k e_{tq-k}$, we have $y_t = u_t + r_t$ with
	$$ u_t = \sum_{k=0}^{q-1}\Gamma_k e_{tq-k} \; , \quad r_t = \sum_{k=q}^{\infty}\Gamma_k^{(q)} e_{tq-k} \; , $$
	where $u_t$ is the innovation from the proof of \cref{prop_NVARpq_SARSEM__varobsinnov}.
	Since $\|\Gamma_k^{(q)}\| \leq c \varrho^{k-q+1}/(1-\varrho)$, we have $\|\var[r_t]\| = \| q^{-1} \sum_{k \geq q} \Gamma_k^{(q)}\Sigma_* \Gamma_k^{(q)\prime} \| = O(q^{-1})$, so $r_t \rightarrow 0$ in mean square.
	\cref{prop_NVARpq_SARSEM__limit_flow_obsinnov} gives $u_t \convdist N(0,\Gamma\Sigma_*\Gamma')$.
	Moreover, $u_t, ..., u_{t+S}$ are functions of the disjoint innovation-blocks $\{e_{(t+s)q}, ..., e_{(t+s)q-q+1}\}_{s=0:S}$ and therefore independent, so their joint limit is the limit of independent Normals.
	The claim follows then by Slutsky's theorem.
	Finally, $\bar y_t = (I-aA)^{-1}v_t = \Gamma v_t$ in the covariate-free SAR, so $v_t \sim N(0,\Sigma_*)$ implies $\bar y_t \sim N(0,\Gamma\Sigma_*\Gamma')$. $\blacksquare$
}

\myproofof{prop_NVARpq_SARSEM__limit_stock}{
	If $y_t$ is a stock variable, $y_t = x_{tq}$ and thus $\gamma_y(h) = \gamma_x(hq)$.
	Hence $\gamma_y(0) = \var[x_\tau] = \sum_{h=0}^\infty \Theta_h \Sigma \Theta_h'$ for every $q$, and $\|\gamma_y(h)\| \leq c\|\Sigma\|\varrho^{hq} \rightarrow 0$ for $h \geq 1$.
	For fixed $\Sigma$ and Normal $e_\tau$, $\br{y_t', ..., y_{t+S}'}'$ is jointly Normal for each $q$ with each $y_{t+s}$ distributed as $x_\tau$, so vanishing cross-covariances imply convergence to independent Normals distributed as $x_\tau$.
	For $\Sigma = q^{-1}\Sigma_*$, $\mean\|y_t\|^2 = \text{tr}\,\gamma_y(0) = q^{-1}\,\text{tr}\sum_{h=0}^\infty \Theta_h \Sigma_* \Theta_h' \rightarrow 0$.
	$\blacksquare$
}

\citet[Proposition 1]{BreitungSwanson2002} discuss convergence for skip-sampled stocks and scaled flows under a general stationary linear process, thereby providing the machinery for \cref{prop_NVARpq_SARSEM__limit_stock,prop_NVARpq_SARSEM__limit_flow_y}.


\todo{see here in comment Lemma linking $u_t$ and $\nu_t$}

\propositionmarker{prop_NVARpqWC_observeddynamics}

\propositionpartmarker{prop_NVARpqWC_observeddynamics_specA}

\propositionpartmarker{prop_NVARpqWC_observeddynamics_specB}

For $s \geq 0$, define
$$ \Xi_s = \frac1q\sum_{i=sq}^{(s+1)q-1}\Gamma^{(q)}_i \quad \text{and} \quad
   \Pi_s = \frac1q\sum_{k=sq}^{(s+1)q-1}D_{q,k} \; , $$
where $D_{q,k} = 0$ for $k > (p+1)(q-1)$, 
and let $m^*_\tau = m_{t(\tau)}$ denote the covariate-term at interaction frequency, so that $m^*_{tq-i} = m_{t-s}$ iff $sq \leq i \leq (s+1)q-1$.

\myproofof{prop_NVARpqWC_observeddynamics_specA}{
	Consider first the VARMAX-representation.
	Under \cref{eq_NVARpqWC_specA}, $F(L)x_\tau = q^{-1}m^*_\tau + e_\tau$.
	Applying $s_q(L)C_q(L)$ to both sides and using $C_q(L)F(L) = \calG_q(L^q)$ (\cref{lemma_calFqRewriting}) gives 
	$\calG_q(L^q)X_\tau = D_q(L)\br{q^{-1}m^*_\tau + e_\tau}$, and evaluating at $\tau=tq$,
	$$ \calG_q(L)y_t = \frac1q\sum_{k=0}^{(p+1)(q-1)}D_{q,k}\,m^*_{tq-k} + \nu_t \; , $$
	where $\nu_t = D_q(L)e_{tq}$ is the moving-average term from \cref{prop_NVARpq_observeddynamics}.
	Collecting the terms with $m^*_{tq-k} = m_{t-s}$, i.e. $k \in sq:(s+1)q-1$, gives $\sum_{s=0}^{q^*}\Pi_s m_{t-s}$, 
	since $\Pi_s = 0$ whenever $sq > (p+1)(q-1)$, i.e. for $s > \text{floor}((p+1)(q-1)/q) = q^*$.
	The $\Pi_s$ are polynomials in $A$ because the $D_{q,k}$ are.

	Now consider the reduced form.
	Under weak stationarity of $x_\tau$ in \cref{eq_NVARpq}, $x_\tau = \Psi(L)\br{q^{-1}m^*_\tau + e_\tau}$, 
	where the series $\sum_{j \geq 0}\Theta_j m^*_{\tau-j}$ converges absolutely with probability one, because $\|\Theta_j\| \leq c\varrho^j$ and $\mean\|m_t\| \leq \br{\|\beta\| + \|A\|\|\gamma\|}\mean\|Z_t\|$ is bounded under weak stationarity of $Z_t$.
	Hence
	$$ y_t = \sum_{l=0}^{q-1}x_{tq-l} = \frac1q\sum_{l=0}^{q-1}\sum_{j \geq 0}\Theta_j\,m^*_{tq-l-j} + \sum_{l=0}^{q-1}\sum_{j \geq 0}\Theta_j\,e_{tq-l-j} \; , $$
	where the second term is $y^*_t$.
	In the first term, the coefficient of $m^*_{tq-i}$ with $i = l+j$ is $\sum_{l=0}^{\min(i,q-1)}\Theta_{i-l} = \Gamma^{(q)}_i$, 
	and collecting the terms with $m^*_{tq-i} = m_{t-s}$ gives $\Xi_s$, which is a polynomial in $A$ because the $\Gamma^{(q)}_i$ are.
	For $s=0$, $\Gamma^{(q)}_i = \Gamma_i$, so that $\Xi_0 = q^{-1}\sum_{i=0}^{q-1}\sum_{j=0}^{i}\Theta_j = q^{-1}\sum_{j=0}^{q-1}(q-j)\Theta_j$.
	For $s \geq 1$, every $i \geq sq \geq q$, so that $\|\Gamma^{(q)}_i\| \leq \sum_{m=i-q+1}^{i}c\varrho^m \leq c\varrho^{i-q+1}/(1-\varrho)$, 
	and summing over $i \geq sq$ gives $\|\Xi_s\| \leq c\varrho^{(s-1)q+1}/\br{q(1-\varrho)^2}$.
	$\blacksquare$
}

\myproofof{prop_NVARpqWC_observeddynamics_specB}{
	The first statement is immediate from \cref{eq_NVARpqWC_specB}. 
	To get the second, apply $\calG_q(L)$ to $y_t=m_t+y^*_t$ and use $\calG_q(L)y^*_t=\nu_t$ (\cref{prop_NVARpq_observeddynamics}). $\blacksquare$
}


\mylemma{}{lemma_NVARpqWC_limit}{%
	Let $y_t$ evolve as in \cref{eq_NVARpqWC_specA} or \cref{eq_NVARpqWC_specB} for some $q\in\naturals$, with $e_\tau\sim WN(0,\Sigma)$, $Z_t$ w.s. and independent of $\{e_\tau\}$, and $(\alpha,A)$ s.t. $x_\tau$ in \cref{eq_NVARpq} is w.s..
	Then, as $q \rightarrow \infty$,
	\begin{enumerate}
		\item[(i)] $\Xi_0 \rightarrow \Gamma$ and $\sum_{s \geq 1}\|\Xi_s\| \rightarrow 0$;
		\item[(ii)] $\mean\big\|\sum_{s \geq 0}\Xi_s m_{t-s} - \Gamma m_t\big\| \rightarrow 0$;
		\item[(iii)] $y^*_t$ is independent of $\{Z_t\}$ for every $q$, and if $e_\tau \overset{i.i.d.}{\sim} (0,\Sigma)$ with $\Sigma = q^{-1}\Sigma_*$ as in \cref{prop_NVARpq_SARSEM__limit_flow_obsinnov}, then $\forall \; S \in \naturals$,
		$\br{y^{*\prime}_t, ..., y^{*\prime}_{t+S}}' \convdist N\br{0, I_{S+1}\otimes\Gamma\Sigma_*\Gamma'}$ jointly with $\{Z_t\}$, where the limit is independent of $\{Z_t\}$.
	\end{enumerate}
}

\myproof{
	(i) By \cref{prop_NVARpqWC_observeddynamics}, $\Xi_0 = \sum_{j=0}^{q-1}\Theta_j - q^{-1}\sum_{j=0}^{q-1}j\Theta_j$.
	The first term converges to $\Psi(1) = \Gamma$ because $\sum_{j \geq 0}\|\Theta_j\| \leq c\sum_{j \geq 0}\varrho^j < \infty$, and the second is bounded in norm by $q^{-1}c\sum_{j \geq 0}j\varrho^j \rightarrow 0$.
	Also, by the bound in \cref{prop_NVARpqWC_observeddynamics}, 
	$$ \sum_{s \geq 1}\|\Xi_s\| \leq \frac{c}{q(1-\varrho)^2}\sum_{s \geq 1}\varrho^{(s-1)q+1} = \frac{c\,\varrho}{q(1-\varrho)^2(1-\varrho^q)} \rightarrow 0 \; . $$

	(ii) We have $\|\sum_{s \geq 0}\Xi_s m_{t-s} - \Gamma m_t\| \leq \|\Xi_0 - \Gamma\|\|m_t\| + \sum_{s \geq 1}\|\Xi_s\|\|m_{t-s}\|$, where $\mean\|m_t\|$ is bounded under weak stationarity of $Z_t$.
	Taking expectations and using (i) gives the statement.

	(iii) $y^*_t = \sum_{h \geq 0}\Gamma^{(q)}_h e_{tq-h}$ is a function of $\{e_\tau\}$ alone and therefore independent of $\{Z_t\}$.
	Under the stated conditions on $e_\tau$, \cref{prop_NVARpq_SARSEM__limit_flow_y} gives $\xi_q \equiv \br{y^{*\prime}_t, ..., y^{*\prime}_{t+S}}' \convdist \xi \sim N\br{0, I_{S+1}\otimes\Gamma\Sigma_*\Gamma'}$.
	Fix any finite set of dates $t_1, ..., t_k$ and let $\zeta = \br{\vecop(Z_{t_1})', ..., \vecop(Z_{t_k})'}'$.
	By independence of $\xi_q$ and $\zeta$, the joint characteristic function factorizes, $\mean\exp\{\ii(r_1'\xi_q + r_2'\zeta)\} = \mean\exp\{\ii r_1'\xi_q\}\mean\exp\{\ii r_2'\zeta\}$, which converges to $\mean\exp\{\ii r_1'\xi\}\mean\exp\{\ii r_2'\zeta\}$, the characteristic function of $(\xi',\zeta')'$ with $\xi$ independent of $\zeta$.
	L\'evy's continuity theorem then gives joint convergence in distribution. $\blacksquare$
}

\propositionmarker{prop_NVARpqWC_SARSEM}

\propositionpartmarker{prop_NVARpqWC_SARSEM_specA}

\propositionpartmarker{prop_NVARpqWC_SARSEM_specB}

Under the assumptions of \cref{prop_NVARpqWC_SARSEM}, \cref{lemma_NVARpqWC_limit}(iii) gives, $\forall \; S \in \naturals$, $\br{y^{*\prime}_t, ..., y^{*\prime}_{t+S}}' \convdist N\br{0, I_{S+1}\otimes\Gamma\Sigma_*\Gamma'}$ jointly with $\{Z_t\}$, the limit being independent of $\{Z_t\}$.
This is the distribution of $\br{(\Gamma v_t)', ..., (\Gamma v_{t+S})'}'$ for $v_t \overset{i.i.d.}{\sim} N(0,\Sigma_*)$ independent of $\{Z_t\}$.

\myproofof{prop_NVARpqWC_SARSEM_specA}{
	By \cref{prop_NVARpqWC_observeddynamics}, $y_t - \Gamma m_t = y^*_t + \br{\sum_{s \geq 0}\Xi_s m_{t-s} - \Gamma m_t}$, where the second term converges to zero in $L^1$ by \cref{lemma_NVARpqWC_limit}(ii), and hence in probability.
	By Slutsky's theorem, $\br{(y_t - \Gamma m_t)', ..., (y_{t+S} - \Gamma m_{t+S})'}'$ has the same limit as $\br{y^{*\prime}_t, ..., y^{*\prime}_{t+S}}'$, jointly with $\{Z_t\}$.
	Finally, \cref{eq_SAR} is equivalent to $\bar y_t = \Gamma m_t + \Gamma v_t$. $\blacksquare$
}

\myproofof{prop_NVARpqWC_SARSEM_specB}{
	By \cref{prop_NVARpqWC_observeddynamics}, $y_t - m_t = y^*_t$, so the claim follows from the above, and \cref{eq_SEM} is equivalent to $\bar y_t = m_t + \Gamma v_t$. $\blacksquare$
}


\propositionmarker{prop_LTresp_NVARpq_AR1innov}

\myproofof{prop_LTresp_NVARpq_AR1innov}{It follows from \cref{lemma_LTresp_NVARp1_AR1innov,lemma_LTresp_NVARpq_AR1innov}. $\blacksquare$}


\paragraph*{Generality of $q \in \naturals$}

Throughout the paper, I discuss the NVAR($p,q$) for $q\in\naturals$.
In \cref{subsubsec_model_laggednetworkeffects_NVARpq}, I mention that in the stock case we can accommodate $q\in \rationals_{++}$.

If $q^{-1} \in \naturals \backslash \{1\}$, observational frequency is an integer-multiple of the network interaction frequency.
For example, under monthly observations, $q=1/3$ shall indicate quarterly network interactions, and $p$ shall signify over how many past quarters transmission is spread out.
For this, we can let $y_t$ follow a restricted NVAR($p^*,1$) with $p^*=p/q$:
\begin{align*}
	y_t = \gamma_1 A y_{t-1} + ... + \gamma_{p^*} A y_{t-p^*} + u_t \; , \quad \gamma_l = \begin{cases}
		\alpha_{lq} \; \quad &\text{if $l$ is multiple of $q^{-1}$} \\
		0 \; \quad &\text{otherwise}
	\end{cases} \; . 
\end{align*}
In my example, the observed monthly series depends on its value in the past $p$ quarters, i.e. on its value three ($1/q$) months ago, six ($2/q$) months ago, etc., up to $3p$ ($3/q$) months ago.\footnote{
	This assumes that transmission is instantaneous at the end of each (network interaction-) period. Alternative definitions with smoother transmission inevitably lead towards abandoning the paradigm of discrete time.
}

If $q \in \rationals_{++}$, we can write $q=q_1q_2$ with $q_1^{-1} \in \naturals$ and $q_2 \in \naturals$.\footnote{
	Note that $q_2$ is the least common multiple of $q$ and $1$, and $q_1$ is their greatest common denominator.} 
For example, under monthly observations, $q=4/3$ means that network interactions occur every three weeks, $q=4/5$ that they occur every five weeks, and $q=30/4$ that they occur every four days. 
We can interpret this as observing every $q_2 \in \naturals$ periods a snapshot of an NVAR($p^*,1)$ process with $p^* = p/q_1 \in \naturals$ and with parameters restricted as in the preceding paragraph.
In the example with monthly observations and $q=4/3$, this amounts to observing every fourth period a snapshot of a weekly process that depends on its value three weeks ago, six weeks ago, etc.


%% file: part_appendix_inference.tex


\section{Lagged Network Effects: Large $T$-Inference}
\label{appsec_inference}


\subsection{Identification}
\label{appsubsec_inference_ID}

In addition to the definitions in \cref{appsec_model},
let $\Theta_p = \cbr{ (\phi,v) : \phi(z) \neq 0 \; \text{for all} \; |z| \leq 1 \; , \; v > 0 }$ be the parameter space of the AR($p,q$), 
let $(\phi,v) = (\lambda_*\alpha, w_*'\Sigma w_*)$ as in \cref{prop_ID_NVARpq} and, for any $(\tilde\alpha,\tilde\Sigma) \in \Theta_p^A$, $(\tilde\phi,\tilde v) = (\lambda_*\tilde\alpha, w_*'\tilde\Sigma w_*)$, 
and let $x_{*\tau} = w_*'x_\tau$ and $y_{*t} = w_*'y_t$, s.t. $y_{*t} = x_{*tq}$ in the stock case and $y_{*t} = s_q(L)x_{*tq}$ in the flow case.
Recall that $\alpha^- = (-\alpha_1,\alpha_2,-\alpha_3,\dots,(-1)^p\alpha_p)'$, s.t. $\phi_\lambda^{\alpha^-}(z) = \phi_\lambda(-z)$ and $F^{\alpha^-}(z) = F(-z)$.
For a stationary process $w_t$ with autocovariances $\gamma_w(h)$, define the Autocovariance Generating Function (ACGF) $\Gamma_w(z) = \sum_{h=-\infty}^{\infty} \gamma_w(h) z^h$.
Note that the ACGF of $x_\tau$ is $\Gamma_x(z) = F(z)^{-1}\Sigma F(z^{-1})^{-1\prime}$.
%
For a Laurent polynomial (or Laurent series) $Q$, define 
$[Q]_q(z) = q^{-1}\sum_{j=0}^{q-1} Q(\wq^j z)$,
which keeps only the powers of $z$ divisible by $q$ (by the same argument as in \cref{lemma_calFqRewriting}).
Finally, note the assumptions from \citet{P017}:

\myassumptionARpq{}{ass_ARpq_ID_stationarity}{
    $x_\tau$ is w.s., i.e. $\phi(z) = \prod_{i=1}^p(1-\eta_i z)$ with $|\eta_i| < 1$ for $i=1:p$, or, equivalently, $\phi(z) \neq 0$ for all $|z| \leq 1$.
}

\myassumptionARpq{}{ass_ARpq_ID_actualorderisp}{
    $\phi_p \neq 0$, or, equivalently, $\eta_i \neq 0$ for $i=1:p$ (since $\phi_p = (-1)^{p+1}\prod_i \eta_i$).
}

\myassumptionARpq{}{ass_ARpq_ID_distincteigvalqthpowers}{
    $\eta_i^q \neq \eta_j^q$ for all $i \neq j$, $i,j = 1:p$.
}

\myassumptionARpq{}{ass_ARpq_ID_normality}{
    $e_\tau \overset{i.i.d.}{\sim} N(0,v)$.
}

\mylemma{}{lemma_NVARtoARlink}{
    Let $(\alpha,\Sigma) \in \Theta_p^A$ and suppose \cref{ass_NVARpq_ID_stationarity,ass_NVARpq_ID_actualorderisp,ass_NVARpq_ID_distincteigvalqthpowers} hold. Then
    \begin{enumerate}
        \item[(i)] $(\tilde\phi,\tilde v) \in \Theta_p$ for any $(\tilde\alpha,\tilde\Sigma) \in \Theta_p^A$;
        \item[(ii)] $(\phi, v) \in \Theta_p$ satisfies
        \cref{ass_ARpq_ID_stationarity,ass_ARpq_ID_actualorderisp,ass_ARpq_ID_distincteigvalqthpowers};
        \item[(iii)] if also \cref{ass_NVARpq_ID_normality} holds, then \cref{ass_ARpq_ID_normality} holds for $w_*'e_\tau$;
        \item[(iv)] the map $\alpha \mapsto \lambda_*\alpha$ is a bijection $\reals^p \to \reals^p$ with $\lambda_*(\alpha^-) = (\lambda_*\alpha)^-$, and $\alpha^- \neq \alpha$ for $q$ even.
    \end{enumerate}
}

\myproof{
    $(\tilde\phi,\tilde v) \in \Theta_p$ because $\tilde\phi$ satisfies \cref{ass_ARpq_ID_stationarity} since \cref{ass_NVARpq_ID_stationarity} -- implied by writing $(\tilde\alpha,\tilde\Sigma) \in \Theta_p^A$ -- implies $\tilde\phi_{\lambda_*}(z) \neq 0$ for $|z| \leq 1$ by \cref{prop_WS_NVARpq}, and because $\tilde v = w_*'\tilde\Sigma w_* > 0$ for $\tilde\Sigma$ positive definite and $w_* \neq 0$ real.
    The same holds for $(\phi,v)$.
    In addition, 
    \cref{ass_ARpq_ID_actualorderisp} holds because $\phi_p = \lambda_*\alpha_p \neq 0$,
    \cref{ass_ARpq_ID_distincteigvalqthpowers} is the second part of \cref{ass_NVARpq_ID_distincteigvalqthpowers},
    and \cref{ass_NVARpq_ID_normality} implies $w_*'e_\tau \overset{i.i.d.}{\sim} N(0, v)$.
    Bijectivity of $\alpha \mapsto \lambda_*\alpha$ and $\lambda_*(\alpha^-) = (\lambda_*\alpha)^-$ are immediate from $\lambda_* \neq 0$.
    Then, by Lemma 2 from \citet{P017} applied to
    $\phi = \lambda_*\alpha$, we have $(\lambda_*\alpha)^- \neq \lambda_*\alpha$ for $q$ even, hence $\alpha^- \neq \alpha$. \todo{keep verifying hardcoded reference to Lemma from P017 until P017 is published}
    $\blacksquare$
}

\mylemma{}{lemma_NVAR_eigenproj}{
    Let $(\alpha,\Sigma) \in \Theta_p^A$ and let $x_\tau$ evolve as in \cref{eq_NVARp1} with $e_\tau \sim WN(0,\Sigma)$.
    \begin{enumerate}
        \item[(i)] Let $(\lambda, w)$ be a real left eigenpair of $A$, i.e.\ $w'A = \lambda w'$ with $\lambda \in \reals$, $w \in \reals^n$, $w \neq 0$. 
        Then $x^*_\tau = w'x_\tau$ follows an AR($p$) with coefficients $\phi_k =
        \lambda\alpha_k$ and innovations $e^*_\tau = w'e_\tau \sim WN(0, v)$, $v = w'\Sigma w > 0$.
        If also $e_\tau \overset{i.i.d.}{\sim} N(0,\Sigma)$, then $e^*_\tau \overset{i.i.d.}{\sim} N(0,v)$.
        \item[(ii)] Let $(\lambda_j, w_j)$ and $(\lambda_k, w_k)$ be (possibly
        complex) left eigenpairs of $A$. 
        Then
        $$ w_j' \Gamma_x(z)\, w_k \;=\; \sigma_{jk}\, H_{jk}(z)\;, \quad \sigma_{jk} = w_j'\Sigma w_k\;, \quad H_{jk}(z) = \sbr{ \phi_{\lambda_j}(z)\phi_{\lambda_k}(z^{-1}) }^{-1}\;. $$
    \end{enumerate}
}

\myproof{
    Multiplying \cref{eq_NVARp1} from the left by $w'$ and using $w'A = \lambda w'$ yields $w'x_\tau = \sum_k \alpha_k \lambda\, w'x_{\tau-k} + w'e_\tau$.
    Trivially, $e^*_\tau \sim WN(0, w'\Sigma w)$ with $w'\Sigma w > 0$, and $e^*_\tau \overset{i.i.d.}{\sim} N(0,v)$ if $e_\tau \overset{i.i.d.}{\sim} N(0,\Sigma)$.
    From $w_j'F(z) = w_j' - g(z)w_j'A = \phi_{\lambda_j}(z)w_j'$ we get $w_j'F(z)^{-1} = \phi_{\lambda_j}(z)^{-1}w_j'$ and likewise $F(z^{-1})^{-1\prime}w_k = \phi_{\lambda_k}(z^{-1})^{-1}w_k$. 
    Hence, $w_j'\Gamma_x(z)w_k = w_j'F(z)^{-1}\Sigma F(z^{-1})^{-1\prime}w_k = \sigma_{jk}H_{jk}(z)$. 
    $\blacksquare$
}

\propositionmarker{prop_ID_NVARpq}

\propositionpartmarker{prop_ID_NVARpq_stock_one}

\propositionpartmarker{prop_ID_NVARpq_stock_two}

\propositionpartmarker{prop_ID_NVARpq_stock_three}

\propositionpartmarker{prop_ID_NVARpq_stock_four}

\propositionpartmarker{prop_ID_NVARpq_flow_one}

\propositionpartmarker{prop_ID_NVARpq_flow_two}

\propositionpartmarker{prop_ID_NVARpq_flow_three}

\myproofof{prop_ID_NVARpq_stock_one,prop_ID_NVARpq_stock_two,prop_ID_NVARpq_stock_three,prop_ID_NVARpq_stock_four}{
    To prove (i) -- that $\calI^{s,A}_{p,q}(\alpha)$ is contained in the stated set --, I show that if $(\tilde\alpha,\tilde\Sigma) \in \Theta_p^A$ is observationally equivalent to $(\alpha,\Sigma)$, then $(\tilde\phi,\tilde v)$ is observationally equivalent to $(\phi,v)$ in the stock case-AR($p,q$).
    By \cref{lemma_NVAR_eigenproj}, $x_{*\tau}$ and $y_{*t}$ evolve as a stock case-AR($p,q$) with parameter $(\phi,v) \in \Theta_p$ that satisfies \cref{ass_ARpq_ID_stationarity,ass_ARpq_ID_actualorderisp,ass_ARpq_ID_distincteigvalqthpowers,ass_ARpq_ID_normality} by \cref{lemma_NVARtoARlink}.
    Under $(\tilde\alpha,\tilde\Sigma)$, they evolve likewise with parameter $(\tilde\phi,\tilde v) \in \Theta_p$. 
    Since $y_{*t} = w_*'y_t$ is a known, fixed and linear function of $y_t$, observational equivalence of $(\tilde\alpha,\tilde\Sigma)$ and $(\alpha,\Sigma)$ implies that $(\tilde\phi,\tilde v)$ and $(\phi, v)$ are observationally equivalent in the scalar model. 
    Hence $(\tilde\phi,\tilde v) \in \calI^{s,1}_{p,q}(\phi,v)$, and $\tilde\alpha = \lambda_*^{-1}\tilde\phi$ by \cref{lemma_NVARtoARlink}.

    For (ii), note that $(\alpha^-,\Sigma) \in \Theta_p^A$ since $\det F(-z) \neq 0$ for $|z| \leq 1$ by \cref{ass_NVARpq_ID_stationarity}.
    Under $(\alpha^-,\Sigma)$, the latent ACGF is $F(-z)^{-1}\Sigma F(-z^{-1})^{-1\prime} = \Gamma_x(-z)$, so $\gamma_{x^-}(h) = (-1)^h\gamma_x(h)$ and $\gamma_{y^-}(h) =(-1)^{qh}\gamma_x(qh) = \gamma_y(h)$ for $q$ even.
    By Normality (\cref{ass_NVARpq_ID_normality}) and mean zero, this is observational equivalence.
    By \cref{lemma_NVARtoARlink}, $\alpha^- \neq \alpha$.

    To prove (iii), note that by (i) and if $\calI^{s,1}_{p,q}(\phi,v)$ is as stated, then $\tilde\phi = \lambda_*\tilde\alpha \in \{\phi, \phi^-\}=\{\alpha,\alpha^-\}$ for $q$ even and $\tilde\phi = \lambda_*\tilde\alpha=\phi=\lambda_*\alpha$ for $q$ odd.
    Conversely, $\alpha^- \in \calI^{s,A}_{p,q}(\alpha)$ by (ii).

    Finally, for (iv), let $(\tilde\alpha,\tilde\Sigma)$ be observationally equivalent to $(\alpha,\Sigma)$.
    By (iii), $\tilde\alpha = \alpha$ for $q$ odd and $\tilde\alpha \in \{\alpha,\alpha^-\}$ for $q$ even.
    Fix a pair $(j,k)$ of indices in $1:n$.
    By \cref{lemma_NVAR_eigenproj} and $\gamma_y(h) = \gamma_x(qh)$, 
    $w_j'\gamma_y(h)w_k$ is the coefficient of $z^{qh}$ in $\sigma_{jk}H_{jk}(z)$, i.e.
    $$ w_j'\,\Gamma_y(z^q)\,w_k \;=\; \sigma_{jk}\,[H_{jk}]_q(z)\;, $$
    and the same identity holds with tildes. 
    If $\tilde\alpha = \alpha$, then $\tilde H_{jk} = H_{jk}$.
    For $q$ even, if $\tilde\alpha = \alpha^-$, then $\tilde H_{jk}(z) = H_{jk}(-z)$ and $[\tilde H_{jk}]_q = [H_{jk}]_q$ since $\{-\omega_q^m\}_m = \{\omega_q^{m+q/2}\}_m = \{\omega_q^m\}_m$ for $q$ even.
    Observational equivalence implies $\Gamma_{\tilde y} = \Gamma_y$.
    Hence, $(\tilde\sigma_{jk} - \sigma_{jk})[H_{jk}]_q(z) = 0$ for all $z$, and so $\tilde\sigma_{jk} = \sigma_{jk}$ using $[H_{jk}]_q \neq 0$ (\cref{ass_NVARpq_ID_techassforSigma_stock}).
    Since this holds for all $j,k = 1:n$, we have $W'\tilde \Sigma W = W'\Sigma W$, and invertibility of $W$ (\cref{ass_NVARpq_ID_Adiagonalizable}) gives $\tilde\Sigma = \Sigma$. 
    Conversely, $(\alpha^-,\Sigma) \in \calI^{s,A}_{p,q}(\alpha,\Sigma)$ for $q$ even by (ii). 
    $\blacksquare$
}

\myproofof{prop_ID_NVARpq_flow_one,prop_ID_NVARpq_flow_two,prop_ID_NVARpq_flow_three}{
    For (v), note that $x_{*\tau}$ and $y_{*t}$ evolve as a flow case-AR($p,q$). The argument from the proof of \cref{prop_ID_NVARpq_stock_one} applies one-to-one.
    Statement (vi) then holds trivially: any observationally equivalent $\tilde\alpha = \lambda_*^{-1}\tilde\phi = \lambda_*^{-1}\phi = \alpha$.
    Finally, to prove (vii), let $(\tilde\alpha,\tilde\Sigma)$ be observationally equivalent to $(\alpha,\Sigma)$.
    By (vi), $\tilde\alpha = \alpha$. 
    Fix a pair $(j,k)$ of indices in $1:n$.
    Note that $\gamma_y(h) = \gamma_X(qh)$ and $\Gamma_X(z) = S_q(z)\Gamma_x(z)$, so \cref{lemma_NVAR_eigenproj} gives $w_j'\Gamma_X(z)w_k = \sigma_{jk}S_q(z)H_{jk}(z)$, and, therefore,
    $$ w_j'\,\Gamma_y(z^q)\,w_k = \sigma_{jk} [S_qH_{jk}]_q(z)\;, $$
    and the same with tildes.
    We have $\tilde H_{jk} = H_{jk}$ by $\tilde\alpha = \alpha$,
    and observational equivalence implies $(\tilde\sigma_{jk}-\sigma_{jk})[S_qH_{jk}]_q(z) = 0 \; \forall \; z$, so that $\tilde\sigma_{jk} = \sigma_{jk}$ using $[S_qH_{jk}]_q\neq 0$ (\cref{ass_NVARpq_ID_techassforSigma_flow}).
    This analysis holds for all $j,k$. Thus, $\tilde\Sigma = \Sigma$ using \cref{ass_NVARpq_ID_Adiagonalizable} as in the proof of \cref{prop_ID_NVARpq_stock_four}.
    $\blacksquare$
}


\mylemma{}{lemma_ID_NVARpqWC}{
    Let $y_t$ evolve as in \cref{eq_NVARpqWC_specA} or \cref{eq_NVARpqWC_specB} for some $q \in \naturals$, and suppose \cref{ass_NVARpq_ID_stationarity,ass_NVARpq_ID_normality,ass_NVARpqWC_strictexo} hold.
    Then $(\tilde\alpha,\tilde\beta,\tilde\gamma,\tilde\Sigma)$ is observationally equivalent to $(\alpha,\beta,\gamma,\Sigma)$ iff
    \begin{enumerate}
        \item[(i)] $(\tilde\alpha,\tilde\Sigma) \in \calI^{f,A}_{p,q}(\alpha,\Sigma)$, and
        \item[(ii)] $\Xi_s(\tilde\alpha)M(\tilde\beta,\tilde\gamma) = \Xi_s(\alpha)M(\beta,\gamma) \; \forall \; s \geq 0$ under \cref{eq_NVARpqWC_specA}, and $M(\tilde\beta,\tilde\gamma) = M(\beta,\gamma)$ under \cref{eq_NVARpqWC_specB}.
    \end{enumerate}
}

\myproof{
    The conditions are sufficient for observational equivalence under \cref{ass_NVARpq_ID_normality} since
    (ii) gives equal $\mu_t$ and (i) gives equal autocovariances of $y^*_t$.
    For necessity under \cref{eq_NVARpqWC_specA}, equal conditional distributions give equal conditional means, so $\sum_{s \geq 0}\Delta_s z_{t-s} = 0 \; \forall \; t$, where $\Delta_s = \Xi_s(\tilde\alpha)M(\tilde\beta,\tilde\gamma) - \Xi_s(\alpha)M(\beta,\gamma)$ is absolutely summable by the bound in \cref{prop_NVARpqWC_observeddynamics}.
    Hence, with $\Delta(z) = \sum_{s \geq 0}\Delta_s z^s$,
    $$ 0 = \mean\Big\|\sum_{s \geq 0}\Delta_s z_{t-s}\Big\|^2 = \int_{-\pi}^{\pi} \text{tr}\sbr{ \Delta(e^{-\ii\omega}) f_z(\omega) \Delta(e^{-\ii\omega})^* } d\omega \; . $$
    The integrand is the trace of a p.s.d. matrix, so $\Delta(e^{-\ii\omega})f_z(\omega)\Delta(e^{-\ii\omega})^* = 0$ for almost every $\omega$, and non-singularity of $f_z(\omega)$ gives $\Delta(e^{-\ii\omega}) = 0$ for almost every $\omega$.
    Since $\Delta(e^{-\ii\omega})$ is continuous, it vanishes everywhere, and so do its coefficients $\Delta_s$.
    This proves (ii).
    Given (ii), the conditional law of $y^*_t = y_t - \mu_t$ is the same under both points, which is (i).
    Under \cref{eq_NVARpqWC_specB}, the same argument applies with $\Delta_0 = M(\tilde\beta,\tilde\gamma) - M(\beta,\gamma)$ and $\Delta_s = 0$ for $s \geq 1$.
    $\blacksquare$
}

The necessity-part of \cref{lemma_ID_NVARpqWC} is the classical result that the transfer function from an observed input to the output is determined by the second moments of input and output; see \citet[Section 1.3]{HannanDeistler1988} and \citet[Eq. 2.83 and Theorem 8.4]{Ljung1999}.

\propositionmarker{prop_ID_NVARpqWC}

\propositionpartmarker{prop_ID_NVARpqWC_specB}

\propositionpartmarker{prop_ID_NVARpqWC_specA_containment}

\propositionpartmarker{prop_ID_NVARpqWC_specA_pointID}

\myproofof{prop_ID_NVARpqWC_specB}{
    Under \cref{eq_NVARpqWC_specB}, condition (ii) of \cref{lemma_ID_NVARpqWC} reads $M(\tilde\beta,\tilde\gamma) = M(\beta,\gamma)$, which is $(\tilde\beta,\tilde\gamma) = (\beta,\gamma)$ by \cref{ass_NVARpqWC_nonscalarA}, and condition (i) means $(\tilde\alpha,\tilde\Sigma) \in \calI^{f,A}_{p,q}(\alpha,\Sigma)$.
    %
    $\blacksquare$
}

\myproofof{prop_ID_NVARpqWC_specA_containment}{
    The containment follows by condition (i) of \cref{lemma_ID_NVARpqWC}.
    If $\tilde\alpha = \alpha$, condition (ii) for $s=0$ reads $\Xi_0(\alpha)\sbr{M(\tilde\beta,\tilde\gamma) - M(\beta,\gamma)} = 0$ so that $M(\tilde\beta,\tilde\gamma) = M(\beta,\gamma)$ by \cref{ass_NVARpqWC_contemploading} and $(\tilde\beta,\tilde\gamma) = (\beta,\gamma)$ by \cref{ass_NVARpqWC_nonscalarA}. $\blacksquare$
}

\myproofof{prop_ID_NVARpqWC_specA_pointID}{
    By \cref{prop_ID_NVARpq_flow_two} and \cref{prop_ID_NVARpqWC_specA_containment}, the conjecture gives $\tilde\alpha = \alpha$, and then \cref{prop_ID_NVARpqWC_specA_containment} gives $(\tilde\beta,\tilde\gamma) = (\beta,\gamma)$.
    $\tilde\Sigma = \Sigma$ follows as in the proof of \cref{prop_ID_NVARpq_flow_three}, applied to $y^*_t$. $\blacksquare$
}


\todo{see here commented-out details on \cref{ass_NVARpq_ID_techassforSigma_stock,ass_NVARpq_ID_techassforSigma_flow}}

\subsection{Estimation}
\label{appsubsec_inference_est}

\paragraph*{Estimation Precision}

By \cref{lemma_NVAR_eigenproj}, a real left eigenpair $(\lambda,w)$ of $A$ turns the NVAR($p,q$) into an AR($p,q$) from \citet{P017}, with 
$$ x_\tau = \phi_1 x_{\tau-1} + ... + \phi_p x_{\tau-p} + e_\tau \; , \quad e_\tau \sim WN(0,v) \; , $$
where $\phi_l = \lambda \alpha_l$ and $v=w'\Sigma w$.
Since $\lambda$ is known, information about $\phi$ translates into information about $\alpha$ by the factor $\lambda^2$.
%
For the stock case with $p=1$, the observed process is an AR($1$), and \citet{P017} shows that $\sqrt{T}(\hat\phi-\phi) \convdist N\br{0,(\partial\psi/\partial\phi)^{-2}(1-\psi^2)}$, 
and,
writing $\calI_{\phi\phi}(q)$ for the inverse of this asymptotic variance, derives the per-observation information about $\phi$ as
\begin{align*} 
  \calI_{\phi\phi}(q) = 
  \Bigl(\frac{\partial\psi}{\partial\phi}\Bigr)^{\!2}\frac{1}{1-\psi^{2}} = 
  \frac{q^{2}\phi^{2q-2}}{1-\phi^{2q}} \; .
\end{align*}
Using $\phi=\lambda\alpha$, we have $\calI_{\alpha\alpha}(q) = \lambda^2 \calI_{\phi\phi}(q) = ( q^{2} / \alpha^{2} ) \psi^{2} / (1-\psi^{2}) $, and differentiating this expression shows -- as in \citet{P017} -- that $\calI_{\alpha\alpha}(q)$ does not necessarily always decrease with $q$ but is maximized at $q^{*} \approx - 0.797 / \log(|\lambda\alpha|)$.
For $|\lambda\alpha|=0.3,0.5,0.7,0.9,0.95$, this yields $q^*=0.66,1.15,2.23,7.56,15.53$, respectively.


%% file: part_appendix_app1.tex

\section{Lagged Input-Output Conversion \& Business Cycles}
\label{appsec_app1}

\subsection{Theory}
\label{appsubsec_app1_theory}

The notation, the period $\tau$-profits of firm $i$ and the respresentative household's problem are as stated in the main text. 
The latter gives the FOC $c_{i\tau} = \gamma_i w_\tau / p_{i\tau}$.
%
%
Define $\lambda_{i\tau} = p_{i\tau}y_{i\tau}/w_\tau$ and $G^w_{\tau,\tau-l}=w_\tau/w_{\tau-l}$.
Let a tilde denote logs for $\tilde{y}_{i\tau}$, $\tilde{z}_{i\tau}$, $\tilde{\lambda}_{i\tau}$, etc., except $\tilde{p}_{i\tau} = ln(p_{i\tau}/w_\tau)$.
By construction, $\tilde{y}_{i\tau}=\tilde{\lambda}_{i\tau} - \tilde{p}_{i\tau}$.
Note that $b = (I-A)\iota$, where $\iota$ is an $n \times 1$ vector of ones.
Then, for $\lambda$ solving $\lambda = \gamma + A'\lambda$, we have $b'\lambda = \iota'(I-A)'(I-A')^{-1}\gamma = \iota'\gamma = 1$.

\paragraph*{Contemporaneous IOC} 

In this case, the amount of good $j$ purchased at $\tau$ and used in the production at $\tau$ coincide: $x_{ij\tau} = x_\tau^{ij} \equiv x^{ij}$.
Dropping time subscripts, firm $i$ solves
\begin{align*}
	\underset{l_i,\{x^{ij}\}_{j=1}^n }{max} \; p_i z_i l_i^{b_i} \prod_{j=1}^n \br{x^{ij}}^{a_{ij}} - w l_i - \sum_{j=1}^n p_j x^{ij} \; .
\end{align*}
The first-order conditions (FOCs) w.r.t. $l_i$ and $x^{ij}$ give
$l_i = b_i p_i y_i / w$
and 
$x^{ij} = a_{ij} p_i y_i /p_j$.
Plugging these expressions into the production function and taking logs yields
\begin{align*}
	\tilde{p}_{i} = k^p_i + \sum_{j=1}^n a_{ij} \tilde{p}_{j}- \tilde{z}_i \; , \quad k^p_i = - \sbr{ b_i ln(b_i) + \sum_{j=1}^n a_{ij}ln(a_{ij}) } \; .
\end{align*}
Stacking over $i$, we have $\tilde{p} = k^{p} + A \tilde{p} - \tilde{z}$.
%
The second FOC implies $a_{ij} = p_{j}x^{ij} / p_{i}y_{i}$, i.e. $a_{ij}$ is the value of good $j$ purchased by sector $i$ divided by the value of sector $i$'s output.

Market clearing for good $j$ requires $y_j = c_j + \sum_{i=1}^n x^{ij}$. Plugging in the expressions for $c_j$ and $x^{ij}$ and multiplying by $p_j/w$ yields
\begin{align*}
	\lambda_j = \gamma_j + \sum_{i=1}^n a_{ij} \lambda_i \; , \quad \text{i.e.} \quad \lambda = \gamma + A' \lambda \; .
\end{align*}
Thus, $\lambda = (I-A')^{-1} \gamma$.\footnote{
	The Domar weights of sector $i$, $\lambda_i$, reflects its importance as a supplier to other sectors, where relevance is defined by households' expenditure shares $\{\gamma_j\}_{j=1}^n$: $\lambda_i = \sum_{j=1}^n \gamma_j \sbr{(I-A)^{-1}}_{ji}$.
	%
}
Inserting $\tilde{p} = \tilde{\lambda} - \tilde{y}$ in the equation for $\tilde{p}$ and solving for $\tilde{y}$ yields
\begin{align*}
	\tilde{y} = k^y + A \tilde{y} + \tilde{z} \; , \quad k^y = (I-A)\tilde{\lambda} - k^p \; .
\end{align*}
Given $\lambda = \gamma + A' \lambda$, the labor market clears -- $\sum_{i=1}^n l_i = \sum_{i=1}^n b_i \lambda_i = b'\lambda=1$ -- and $w = \sum_{i=1}^n b_i p_{i} y_{i}$.

\paragraph*{Lagged IOC}

In the lagged-IOC economies from \cref{subsec_app1_theory} -- discussed in detail below --, we have
$x^{ij}_\tau=\sum_{l=1}^p x^{ij}_{\tau+l,\tau}$, i.e. the amount of good $j$ purchased by firm $i$ in period $\tau$, $x^{ij}_\tau$, can be used in production in periods $\tau+1$, $\tau+2$, ..., $\tau+p$.
Under a CES-aggregator, the amount of good $j$ used in production at time $\tau$ is then 
$$ x_{ij\tau} = \sbr{ \sum_{l=1}^p \alpha_l \br{ x^{ij}_{\tau,\tau-l} }^r }^{1/r} \; , \quad \alpha_l \geq 0 \; , \quad \sum_{l=1}^p \alpha_l = 1 \; , \quad r \in [0,1) \; . $$
Under a single lag in IOC -- $p=1$ --, we have $x_{ij\tau} = x^{ij}_{\tau,\tau-1} = x^{ij}_{\tau-1}$.

\mylemma{}{lemma_macromodelnumeraire}{
	Consider any of the lagged-IOC economies from \cref{subsec_app1_theory}. 
	Suppose firms' FOCs imply 
	$w_\tau l_{i\tau} = b_i p_{i\tau}y_{i\tau}$ 
	as well as 
	$p_{j\tau}\,x^{ij}_{\tau+l,\tau}=\beta^l\,\theta_l\,a_{ij}p_{i,\tau+l}y_{i,\tau+l}$ for $l=1:p$ and some weights $\theta_l \geq 0$, $\sum_{l=1}^p\theta_l=1$. 
	Then, in any equilibrium with zero profits and a constant rate of nominal growth, 
	$G^w_{\tau+1,\tau}=1/\beta$, $a_{ij}=p_{j\tau}x^{ij}_\tau/(p_{i\tau}y_{i\tau})$, and $\lambda_{i\tau}=\lambda_i \; \forall \; \tau$.
}

\myproof{
	Given the labor-FOC, zero profits imply
	$\sum_{j=1}^n p_{j\tau}x^{ij}_\tau = (1-b_i)p_{i\tau}y_{i\tau}$.
	Summing the input-FOC over $l$ shows that $a_{ij} \propto p_{j\tau}x^{ij}_\tau$ for all $j$.
	Hence, $a_{ij} = p_{j\tau}x^{ij}_\tau / p_{i\tau}y_{i\tau}$, since $1-b_i = \sum_{j=1}^n a_{ij}$.

	Summing the input-FOC over both $j$ and $l$ and using $\sum_ja_{ij}=1-b_i$ and
	$x^{ij}_\tau=\sum_lx^{ij}_{\tau+l,\tau}$,
	\[
	\sum_j p_{j\tau}x^{ij}_\tau=(1-b_i)\sum_{l=1}^p\beta^l\theta_l V_{i,\tau+l} \; , \quad V_{i,\tau+l} \equiv p_{i,\tau+l}y_{i,\tau+l} \; .
	\]
	Zero profits then imply $V_{i\tau}=\sum_l\beta^l\theta_lV_{i,\tau+l}$.
	Under constant nominal growth $V_{i\tau}=V_ig^\tau$, this reads $\sum_{l}\theta_l (\beta g)^l=1$.
	Hence, we must have $\beta g=1$, i.e. $g=1/\beta$.
	Then $\lambda_{i,\tau+1}/\lambda_{i\tau}=g/G^w_{\tau+1,\tau}=1/(\beta G^w_{\tau+1,\tau})$ for all $i$.
	Hence, $b'\lambda_{\tau+1}=b'\lambda_\tau/(\beta G^w_{\tau+1,\tau})$. 
	Labor market clearing in every period requires $b'\lambda_\tau=1$.
	Therefore, $G^w_{\tau+1,\tau}=1/\beta$ and $\lambda_\tau$ is constant. 
	$\blacksquare$
}

\paragraph*{One-Period Lagged IOC}

If good $j$ used in production at time $\tau$ is purchased at time $\tau-1$, then $x_{ij\tau} = x^{ij}_{\tau-1}$ and firm $i$ solves 
\begin{align*}
	\underset{\substack{\{l_{i\tau}, \{x^{ij}_\tau\}_{j=1}^n \}_{\tau=0}^\infty}}{max} \; \sum_{\tau=0}^\infty \beta^\tau &\Pi_{i\tau} \; , \quad \Pi_{i\tau} = p_{i\tau} z_{i\tau} l_{i\tau}^{b_i} \prod_{j=1}^n \br{ x_{\tau-1}^{ij} }^{a_{ij}} - w_\tau l_{i\tau} - \sum_{j=1}^n p_{j\tau} x^{ij}_{\tau}  \; .
\end{align*}
%
The FOC w.r.t. $l_{i\tau}$ and $x^{ij}_\tau$ give
$l_{i\tau} = b_i p_{i\tau}y_{i\tau} / w_\tau$
and 
$x^{ij}_\tau = \beta a_{ij} p_{i,\tau+1}y_{i,\tau+1} / p_{j\tau}$.
Plugging these expressions into the production function and taking logs gives
\begin{align*}
	\tilde{p}_{i\tau} = k^{p1}_{i\tau} + \sum_{j=1}^n a_{ij} \tilde{p}_{j,\tau-1} - \tilde{z}_{i\tau} \; , \quad \text{i.e.} \quad \tilde{p}_\tau = k^{p1}_\tau + A \tilde{p}_{\tau-1} - \tilde{z}_\tau \; ,
\end{align*}
where $k^{p1}_{i\tau} = k^{p1}_i - (1-b_i)\tilde{G}^w_{\tau,\tau-1}$ with $k^{p1}_i = - \sbr{ b_i ln(b_i) + \sum_{j=1}^n a_{ij}ln(\beta a_{ij}) } = k^p_i - (1-b_i)\ln \beta$.
The above FOCs allow us to apply \cref{lemma_macromodelnumeraire}: any equilibrium with zero profits and a constant rate of nominal growth must have $G^w_{\tau+1,\tau}=1/\beta$.
Hence, $\tilde{G}^w_{\tau,\tau-1} = -\ln \beta$, $k^{p1}_{i\tau}$ is constant and equals $k^p_i$ from the static economy, and 
$$ \tilde{p}_\tau = k^{p} + A \tilde{p}_{\tau-1} - \tilde{z}_\tau \; . $$
Thereby, the interpretation of $a_{ij}$ is preserved by \cref{lemma_macromodelnumeraire}.
By market clearing of good $j$,
\begin{align*}
	y_{j\tau} = c_{j\tau} + \sum_{i=1}^n x_{\tau}^{ij} = \gamma_j \frac{w_\tau}{p_{j\tau}} + \sum_{i=1}^n \beta a_{ij} \frac{p_{i,\tau+1}y_{i,\tau+1}}{p_{j\tau}} \; .
\end{align*}
Multiplying again by $p_{j\tau}/w_\tau$ yields
\begin{align*}
	\lambda_{j\tau}  = \gamma_j + \sum_{i=1}^n \beta a_{ij}  G^w_{\tau+1,\tau} \lambda_{i,\tau+1} \; , \quad \text{i.e.} \quad \lambda_\tau = \gamma + \beta G^w_{\tau+1,\tau} A'\lambda_{\tau+1} \; .
\end{align*}
By \cref{lemma_macromodelnumeraire}, this collapses to $\lambda = \gamma + A'\lambda$, the same expression as in the static economy.
Combining $\tilde{y}_\tau = \tilde{\lambda}_\tau - \tilde{p}_\tau$ with the expression for $\tilde{p}_\tau$ and using $\tilde{\lambda}_\tau = \tilde{\lambda}$ yields
\begin{align*}
	\tilde{y}_\tau = k^{y} + A \tilde{y}_{\tau-1} + \tilde{z}_\tau \; .
\end{align*}
In a steady state (SS) with $\tilde{z}_\tau = \tilde{z} \; \forall \; \tau$, the expressions for $\tilde{p}_\tau$ and $\tilde{y}_\tau$ reveal that their values are as in the static economy.

\paragraph*{General Lags in IOC}

Firm $i$'s problem is then
\begin{align*}
	\underset{\substack{\{l_{i\tau}, \{x^{ij}_\tau, \{x^{ij}_{\tau,\tau-l}\}_{l=1}^p \}_{j=1}^n \}_{\tau=0}^\infty}}{max}& \; \sum_{\tau=0}^\infty \beta^\tau \Pi_{i\tau} 
	\quad \text{s.t.} \quad x^{ij}_\tau = \sum_{l=1}^p x^{ij}_{\tau+l,\tau}  \; , \\
	&\Pi_{i\tau} = p_{i\tau}z_{i\tau}l_{i\tau}^{b_i} \prod_{j=1}^n \sbr{ \sum_{l=1}^p \alpha_l \br{ x^{ij}_{\tau,\tau-l} }^r }^{\frac{a_{ij}}{r}} - w_\tau l_{i\tau} - \sum_{j=1}^n p_{j\tau}x^{ij}_\tau  \; .
\end{align*}
For each input $j$, the firm chooses how much to buy in period $\tau$, $x^{ij}_\tau$, and how to distribute the purchased amount for production over periods $\tau+1, ..., \tau+p$. 
Abstracting from perfect substitutability -- i.e. $r=1$ -- allows me to ignore the boundary constraints $l_{i\tau}, x^{ij}_{\tau+l,\tau} \geq 0 \; \forall \; \tau, i, j, l$.
Differentiating w.r.t. $l_{i\tau}$ and $x^{ij}_{\tau+l,\tau}$ yields the FOCs
\begin{align*}
	l_{i\tau} = b_i \frac{p_{i\tau} y_{i\tau} }{w_\tau} \; , \quad x^{ij}_{\tau,\tau-l} = \sbr{ a_{ij}\alpha_l \beta^l \frac{p_{i\tau}y_{i\tau}}{p_{j\tau-l} x_{ij\tau}^r} }^{\frac{1}{1-r}} \; , \; \; l=1:p \; .
\end{align*}

\paragraph*{Cobb-Douglas Aggregation}


Under $r\rightarrow 0$, we have $x_{ij\tau} = \prod_{l=1}^p \br{ x^{ij}_{\tau,\tau-l} }^{\alpha_l}$ and the optimality conditions yield
\begin{align*}
	l_{i\tau} = b_i \frac{p_{i\tau}y_{i\tau}}{w_\tau} \; , \quad
	x^{ij}_{\tau,\tau-l} = \beta^l \alpha_l a_{ij} \frac{p_{i\tau}y_{i\tau}}{p_{j,\tau-l}} \; , \; \; l=1:p \; .
\end{align*}
Inserting these expressions into the production function, taking logs and using $\sum_{l=1}^p \alpha_l =1$ and $\sum_{j=1}^n a_{ij} = 1-b_i$ leads to
\begin{align*}
	\tilde{p}_{i\tau} = k^{p2}_{i\tau} + \sum_{j=1}^n a_{ij} \sum_{l=1}^p  \alpha_l \tilde{p}_{j,\tau-l} - \tilde{z}_{i\tau} \; , \quad \text{i.e.} \quad \tilde{p}_\tau = k^{p2}_\tau + \sum_{l=1}^p \alpha_l A \tilde{p}_{\tau-l} - \tilde{z}_\tau \; ,
\end{align*}
where 
$k^{p2}_{i\tau} = k^{p2}_i - (1 - b_i)\sum_{l=1}^p \alpha_l \tilde{G}^w_{\tau,\tau-l} $ and $k^{p2}_i = - b_i ln(b_i) - \sum_{j=1}^n a_{ij} \sum_{l=1}^p \alpha_l ln(\beta^l \alpha_l a_{ij}) $. 
The above FOCs allow us to apply \cref{lemma_macromodelnumeraire}: any equilibrium with zero profits and a constant rate of nominal growth must have $G^w_{\tau+1,\tau}=1/\beta$.
Hence, $\tilde{G}^w_{\tau,\tau-l} = -l \ln \beta$, $k^{p2}_{i\tau}$ is constant and equals 
$k^p_i + (1-b_i)H(\alpha)$ with $H(\alpha) = -\sum_{l=1}^p \alpha_l \ln \alpha_l \geq 0$.
As a result, 
\begin{align*}
	\tilde{p}_\tau = k^{p2} + \sum_{l=1}^p \alpha_l A \tilde{p}_{\tau-l} - \tilde{z}_\tau \; , \quad k^{p2} = k^p +  H(\alpha) A\iota \; .
\end{align*}
By \cref{lemma_macromodelnumeraire}, the interpretation of $a_{ij}$ is preserved.
Market clearing for good $j$ now yields
\begin{align*}
	y_{j\tau} = c_{j\tau} + \sum_{i=1}^n x^{ij}_\tau = c_{j\tau} + \sum_{i=1}^n \sum_{l=1}^p x^{ij}_{\tau+l,\tau} \; .
\end{align*}
Plugging in the optimality conditions and multiplying by $p_{j\tau}/w_\tau$ to solve for $\lambda_{j\tau}$ gives
\begin{align*}
	\lambda_{j\tau} = \gamma_j + \sum_{l=1}^p \beta^l \alpha_l  G^w_{\tau+l,\tau} \sum_{i=1}^n a_{ij} \lambda_{i,\tau+l} \; ,
\end{align*}
which by \cref{lemma_macromodelnumeraire} again collapses to $\lambda = (I-A')^{-1} \gamma$.
For output we then get
\begin{align*}
	\tilde{y}_\tau = k^{y2} + \sum_{l=1}^p \alpha_l A \tilde{y}_{\tau-l} + \tilde{z}_\tau \; ,
\end{align*}
where $k^{y2} = (I-A)\tilde{\lambda} - k^{p2} = k^y - H(\alpha) A\iota$.

These expressions for $\tilde{p}_\tau$ and $\tilde{y}_\tau$ reveal that in a SS with $\tilde{z}_\tau = \tilde{z} \; \forall \; \tau$, prices are higher and output is lower by the amount $H(\alpha)(I-A)^{-1} A\iota \geq 0$.
This is the entropy of the sourcing profile scaled by each sector's total intermediate intensity $\sbr{(I-A)^{-1} A\iota }_i = \sbr{(I-A)^{-1} \iota }_i -1$.
This difference in SS vanishes iff $\alpha_l =1$ for some $l$, i.e. when IOC occurs at a single lag.

\paragraph*{CES Aggregation}

For $r>0$, the FOCs for vintage demands depend on $x_{ij\tau}$, for which we need to solve first.
Write $m = r/(1-r)$, $\eta_l = \alpha_l (\alpha_l \beta^l)^m$ and $S_{j\tau} = \sum_{l=1}^p \eta_l p_{j,\tau-l}^{-m}$.
Raising the FOCs to the power $r$, multiplying by $\alpha_l$ and summing over $l$ gives $x_{ij\tau}^r = (a_{ij}p_{i\tau}y_{i\tau}/x_{ij\tau}^r)^m S_{j\tau}$, and since $r+r^2/(1-r)=m$, we get
$x_{ij\tau} = a_{ij} p_{i\tau}y_{i\tau} S_{j\tau}^{\frac{1-r}{r}}$.
Inserting this back for $x_{ij\tau}$ into the FOCs yields
\begin{align} \label{eq_GeneralLIOC_vintagedemand}
	x^{ij}_{\tau,\tau-l} = \beta^l \vartheta^j_{l\tau} a_{ij} \frac{p_{i\tau}y_{i\tau}}{p_{j,\tau-l}}  \; , \; \; \vartheta^j_{l\tau} = \frac{\eta_l}{\sum_{k=1}^p \eta_k (p_{j,\tau-k}/p_{j,\tau-l})^{-m}} \; , \; \; l=1:p \; ,
\end{align}
where I used $(\alpha_l\beta^l)^{1/(1-r)} = \beta^l\eta_l$.\footnote{
	This holds because $\alpha_l^{1/(1-r)} = \alpha_l^{1+m} = \eta_l \beta^{-lm}$ and $\beta^{l/(1-r)-lm} = \beta^l$.
}
Let $p^*_{i\tau} = p_{i\tau}/w_\tau$.
Since $p_{j,\tau-l} = p^*_{j,\tau-l}w_\tau/G^w_{\tau,\tau-l}$ and $S_{j\tau} = w^{-m}_\tau S^*_{j\tau}$ with $S^*_{j\tau} = \sum_{l=1}^p \eta_l (G^w_{\tau,\tau-l}/p^*_{j,\tau-l})^m$, we can then write $x_{ij\tau} = a_{ij} p^*_{i\tau}y_{i\tau} \br{S^*_{j\tau}}^{(1-r)/r}$.
Substituting this and $l_{i\tau} = b_i p^*_{i\tau}y_{i\tau}$ into the production function and dividing by $y_{i\tau}$ gives an exact relation among prices alone:
\begin{align} \label{eq_GeneralLIOC_relationamongprices}
	 1 = z_{i\tau} b_i^{b_i} p_{i\tau}^* \prod_{j=1}^n a_{ij}^{a_{ij}}\sbr{ \sum_{l=1}^p \eta_l \br{ \frac{G^w_{\tau,\tau-l}}{p^*_{j,\tau-l}} }^{\frac{r}{1-r}} }^{a_{ij}\frac{1-r}{r}} \; .
\end{align}
An exact expression for the dynamics of $\lambda_\tau$ is again obtained by inserting the optimality conditions for $c_{j\tau}$ and $x^{ij}_{\tau+l,\tau}$ into the market clearing condition for good $j$:
\begin{align*}
	\lambda_{j\tau} = \gamma_j + \sum_{i=1}^n a_{ij} \sum_{l=1}^p \beta^l \vartheta^j_{l,\tau+l}G^w_{\tau+l,\tau} \lambda_{i,\tau+l} \; .
\end{align*}

I now show a SS exists and \cref{lemma_macromodelnumeraire} applies in it.
Consider a candidate SS with $p^*_{j\tau}=p^*_j \; \forall \; j,\tau$ and $G^w_{\tau,\tau-1}=G^w \; \forall \; \tau$.
Then $p_{j,\tau-k}=p^*_j w_\tau (G^w)^{-k}$, so $p_{j,\tau-k}/p_{j,\tau-l}= (G^w)^{l-k}$ and the weights in the vintage demands become non-negative constants that sum to one: $\vartheta^j_{l\tau}=\vartheta_l = \alpha_l^{1/(1-r)} / \sum_{k=1}^p \alpha_k^{1/(1-r)}$.
Hence, \cref{lemma_macromodelnumeraire} applies: we have $G^w = 1/\beta$, the interpretation of $a_{ij}$ is preserved in steady state.
Evaluating the expression for $\lambda_{j\tau}$ at SS then shows that $\lambda = (I-A')^{-1}\gamma$.
Evaluating \cref{eq_GeneralLIOC_relationamongprices} at SS and taking logs gives 
$\tilde{p} = k^{p3} + A \tilde{p} - \tilde{z}$, 
where $k^{p3}$ stacks $k^{p3}_i = k^p_i + (1-b_i)H_r(\alpha)$ with $H_r(\alpha) = - (1-r)/r \ln \br{ \sum_{l=1}^p \alpha_l^{1/(1-r)}} \geq 0$.
Hence, $k^{p3} = k^p + H_r(\alpha)A\iota$.
For output $\tilde{y} = \tilde{\lambda}-\tilde{p}$, we then get 
$\tilde{y} = k^{y3} + A \tilde{y} + \tilde{z}$ with $k^{y3} = k^y - H_r(\alpha)A\iota$.
In SS, prices are again higher and output is lower than in the contemporaneous-IOC economy by $H_r(\alpha)(I-A)^{-1}A\iota$.

Linearizing \cref{eq_GeneralLIOC_relationamongprices} around this SS, we get 
\begin{align*}
	\hat{p}^*_{i\tau} = c^{p}_{i\tau} + \sum_{j=1}^n a_{ij} \sum_{l=1}^p \vartheta_l \hat{p}^*_{j,\tau-l} - \hat{z}_{i\tau} \; , \quad \text{i.e.} \quad \hat{p}^*_\tau = c^{p}_\tau + \sum_{l=1}^p \vartheta_l A \hat{p}^*_{\tau-l} - \hat{z}_\tau \; ,
\end{align*}
where $c^{p}_{i\tau} = - (1-b_i) \sum_{l=1}^p \vartheta_l \hat{G}^w_{\tau,\tau-l}$.
Combining this with $\hat{y}_\tau = \hat{\lambda}_\tau - \hat{p}^*_\tau$ yields 
$$ \hat{y}_\tau = c^{y}_\tau + \sum_{l=1}^p \vartheta_l A \hat{y}_{\tau-l} + \hat{z}_\tau\; , $$
where $c^{y}_\tau = \hat{\lambda}_\tau - \sum_{l=1}^p \vartheta_l A \hat{\lambda}_{\tau-l} - c^{p}_\tau $.
Note that $c^{y}_\tau$ and $c^{p}_\tau$ vanish at the SS.




\subsection{Data}
\label{appsubsec_app1_data}

\paragraph*{Sources}

The IP indices are the monthly, non-seasonally
adjusted series published by the Federal Reserve Board in its G.17 release. 
I perform the seasonal adjustment as described in \cref{subsec_app1_dataest}. 
The IO tables are the BEA's annual \myquote{use} tables at the summary level, available for 1997 through 2020; I use the table for 2007.
The output weights mentioned below are the monthly, seasonally adjusted industry
output series from the same G.17 release.

\paragraph*{Sector-Selection}

The summary-level use table distinguishes industries at a level of aggregation
that is finer than what the IP data supports. I first drop the two farm-related
industries and all government industries, which leaves 64 industries. Of these, I
retain the $n=23$ for which the Federal Reserve publishes IP indices. 
The resulting sample runs from January 2005 through August 2022, giving $T=212$ monthly observations before differencing.

\paragraph*{Matching IO- and IP-Sector-Classifications}

The BEA and Federal Reserve classifications do not correspond one-to-one. For 17
of the 23 sectors the two coincide and the IP index is used directly. The
remaining six BEA sectors each span several Federal Reserve series, listed in
\cref{tab_app1_concordance}. For these I construct a single index as the
output-weighted average of its components, 
$IP_{it} = \sum_{k \in \mathcal{K}_i} w_{kt} \, IP_{kt} / \sum_{k \in \mathcal{K}_i} w_{kt}$,
where $\mathcal{K}_i$ collects the Federal Reserve series making up BEA sector $i$
and $w_{kt}$ is series $k$'s output in month $t$. The weights therefore vary over
time, so that the composite index tracks shifts in the relative size of its
components.

\begin{table}[h]
	\caption{BEA Sectors Spanning Several Federal Reserve IP Series}
	\label{tab_app1_concordance}
	\centering
	\begin{tabular}{L{9.0cm}L{6.0cm}}
		\hline \hline \tabspaceAfterRule
		BEA sector & Federal Reserve IP series \tabspaceAfterHeaderRow \cline{1-2}
		\tabspaceAfterRule
		Utilities & G2211, G2212 \tabspaceAfterBodyRow
		Motor vehicles, bodies and trailers, and parts & G3361, G3362, G3363 \tabspaceAfterBodyRow
		Other transportation equipment & G3364, G3365T9 \tabspaceAfterBodyRow
		Food and beverage and tobacco products & G311, G312 \tabspaceAfterBodyRow
		Textile mills and textile product mills & G313, G314 \tabspaceAfterBodyRow
		Apparel and leather and allied products & G315, G316 \tabspaceAfterBodyRow
		\tabspaceAfterRule
		\bottomrule
		\multicolumn{2}{L{15cm}}{\scriptsize {\em Notes:} The table lists the six sectors from the BEA data were matched with IP-data available at finer disaggregation. 
		The IP-column refers to NAICS codes.
		The IP-sub-sectors exhaust all six BEA-sectors except Utilities, for which G2213 (water, sewage) is missing.}\\
	\end{tabular}
	\vspace*{-15pt}
\end{table}




\subsection{Estimation}
\label{appsubsec_app1_estimation}

\paragraph*{Likelihood Evaluation}

Under $q=1$, the model with contemporaneous IOC (\cref{eq_app1_yTildeGrowth_CIOC,eq_app1_yGrowth}) yields the following state space form:
\begin{align*}
	s_t = \Phi_0 + \Phi_1 s_{t-1} + v_t \; , \quad \Delta y_t = \Psi s_t \; ,
\end{align*}
where $s_t = (\Delta y_t', e_t', e^a_t)'$ and $v_t = (0, \varepsilon_t', \varepsilon^a_t)' \sim N(0,\Sigma_v)$ are $(2n+1) \times 1$, $\Psi = \bmat{I_n, 0}$, and
\begin{align*}
	\Phi_0 = \bmat{L\gamma \\ 0 \\ 0} \; , \quad \Phi_1 = \bmat{0 & L & L \delta \\ 0 & P & 0 \\ 0 & 0 & \rho_a} \; , \quad \Sigma_v = \bmat{0 & 0 & 0 \\ 0 & \Sigma & 0 \\ 0 & 0 & \sigma^2_a} \; ,
\end{align*}
and we further define $L = (I-A)^{-1}$ to be the Leontief-inverse and $P$ and $\Sigma$ to be diagonal matrices containing $(\rho_1, ..., \rho_n)$ and $(\sigma^2_1, ..., \sigma^2_n)$, respectively.

To write the model with lagged IOC (\cref{eq_app1_yTildeGrowth_LIOC,eq_app1_yGrowth}) in state space form, let $\tilde{p} = \max\{p,q\}$ and $\alpha_l = 0$ for $l=(p+1):\tilde{p}$. Define the $(n\tilde{p}+n+1)\times 1$ vectors $s_\tau = (\Delta \tilde{y}_\tau', ..., \Delta \tilde{y}_{\tau-\tilde{p}+1}',e_\tau', e^a_\tau)'$ and $v_\tau = (0, ..., 0, \varepsilon_\tau', \varepsilon^a_\tau)'$. We have the analogous state space form as above, with
\begin{align*}
	\Phi_0 = \bmat{\gamma \\ 0 \\ \vdots \\ 0} \; , \quad \Phi_1 = \bmat{\alpha_1 A & \dots & \alpha_{\tilde{p}-1}A & \alpha_{\tilde{p}}A & I & \delta \\
	& I_{np-n} & & 0 & 0 & 0 \\
	& 0 & & 0 & P & 0 \\ 
	& 0 & & 0 & 0 & \rho_a} \; , \quad \Sigma_v = \bmat{0 & 0 & 0 & 0 \\
	0 & 0 & 0 & 0 \\
	0 & 0 & \Sigma & 0 \\
	0 & 0 & 0 & \sigma^2_a } \; .
\end{align*}
Under $q=1$, we observe $\Delta y_t = \Delta \tilde{y}_\tau$, meaning that $\Psi = [I_n, 0]$. Under $q>1$, we have observe $\Delta y_t = \sum_{l=0}^{q-1}  \Delta \tilde{y}_{\tau-l}$ -- i.e. $\Psi = [I_n, ..., I_n, 0]$ -- only for periods $t = \tau/q \in \naturals$ (see \cref{appsec_inference}).

\paragraph*{Prior-Construction \& -Drawing}


Under a Uniform prior for $\alpha \in [0,1]^{p-1} \cap \{\alpha: \sum_{l=1}^{p-1} \alpha_l \leq 1\}$, 
\begin{align*}
	p(\alpha_1,...,\alpha_{p-1}) = p(\alpha_1|\alpha_2,...,\alpha_{p-1}) p(\alpha_2|\alpha_3,...,\alpha_{p-1}) ... p(\alpha_{p-2}|\alpha_{p-1}) p(\alpha_{p-1}) \; ,
\end{align*}
where 
$p(\alpha_l|\alpha_{l+1},...,\alpha_{p-1}) = l \br{ 1- \sum_{m=l}^{p-1} \alpha_m }^{l-1} / \br{ 1- \sum_{m=l+1}^{p-1} \alpha_m }^{l-1} $ if $\alpha_l \in \sbr{ 0, 1-\sum_{m=l+1}^{p-1} \alpha_m }$ for $l=1:p-2$
and 
$p(\alpha_{p-1}) = (p-1) \br{ 1 -\alpha_{p-1}}^{p-2}$ if $\alpha_{p-1} \in \sbr{ 0, 1 }$, 
and these expressions equal zero otherwise.\todo{derivation in P009diary3,p.39}
To draw from $p(\alpha_1,...,\alpha_{p-1})$, I draw $\alpha_{p-1}$ from its marginal distribution and iteratively draw $\alpha_{p-2}, ..., \alpha_1$ from the conditionals, using the inverse-cdf method in each step: 
for $l=1:p-1$, given $x_l \sim U(0,1)$, I compute
\begin{align*}
	\alpha_{p-1} = 1-(1-x_{p-1})^{1/(p-1)} \; \; \text{and} \; \; \alpha_l|(\alpha_{l+1}, ..., \alpha_{p-1}) = \br{ 1-\sum_{m=l+1}^{p-1}\alpha_m }\sbr{ 1- (1-x_l)^{1/l} } \; , \; \; l=1:(p-2) \; .
\end{align*}
%
The Uniform distributions for $\sigma^{2} \in \reals_{++}^n$ need upper bounds since draws from them are needed to initialize the SMC sampler.
A high value avoids restricting parts of the domain with non-trivial likelihood values, but a lower value leads to more efficient sampling.
I choose a low value to initialize the sampler, but allow the algorithm to ignore it in search for parameter-values in high likelihood-regions. 
Ex-post, I redefine the upper bound as the largest posterior draw among 
$\sigma^{2}$ 
across all models.
The resulting MDDs could be corrected by rescaling, but model selection is unaffected, so I ignore this.\footnote{
	Let $\sigma^{2}\in [0,\bar{s}]^{n}$.  
	The MDD is
	$ p(Y) = \int p(Y|\theta_{-s},s)p(\theta_{-s})p(s)d(\theta_{-s},s) = \bar{s}^{-(n+1)} \int p(Y|\theta_{-s},s)p(\theta_{-s})d(\theta_{-s},s) $.
	Re-scaling the upper bound ex-post to $s^*$ requires multiplying the MDD by
	$(\bar{s}/s^*)^{n}$. 
}
I set $\bar{\sigma}^2_i = 0.5\var[\Delta \tilde{y}_{i\tau}]$.

\paragraph*{Numerical Posterior Sampling}

I use the adaptive tempering variant of the SMC algorithm (see \citet{CaiEtAl2021}).
It ensures a precise estimation of the posterior even as the distance between the prior and posterior distributions is difficult to assess. 

\subsection{Results}
\label{appsubsec_app1_results}



\begin{figure}[H]
	\captionsetup[subfigure]{justification=centering}
	\begin{center}
		\figspaceBeforePlotArea
		\begin{subfigure}[b]{\plotSizeQuadruple\textwidth}
				\centering
				\includegraphics[width=1\textwidth, clip]{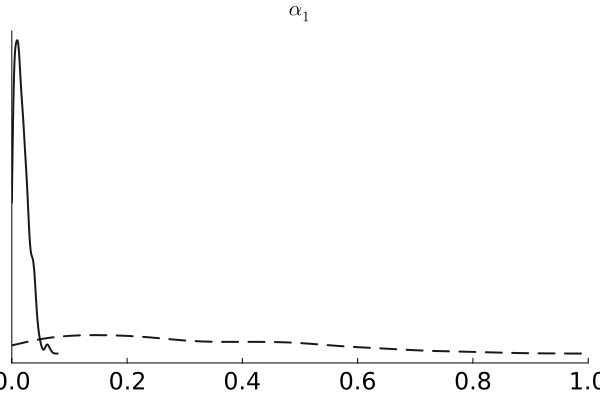}
		\end{subfigure}
		\begin{subfigure}[b]{\plotSizeQuadruple\textwidth}
				\centering
				\includegraphics[width=1\textwidth, clip]{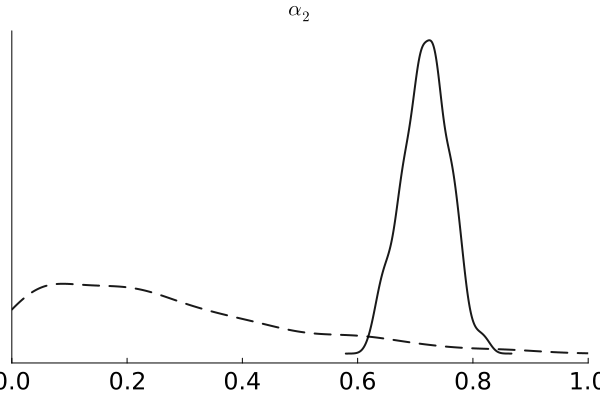}
		\end{subfigure}
		\begin{subfigure}[b]{\plotSizeQuadruple\textwidth}
				\centering
				\includegraphics[width=1\textwidth, clip]{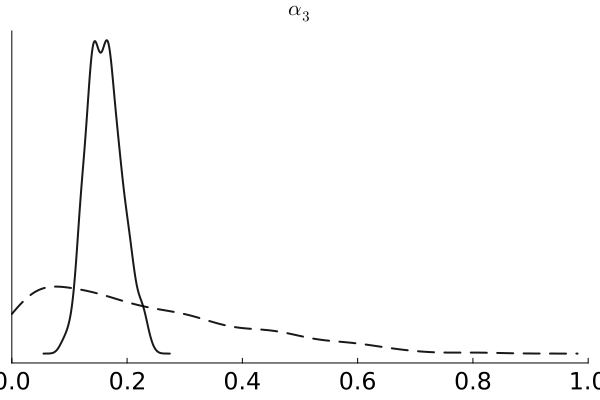}
		\end{subfigure}
		\begin{subfigure}[b]{\plotSizeQuadruple\textwidth}
				\centering
				\includegraphics[width=1\textwidth, clip]{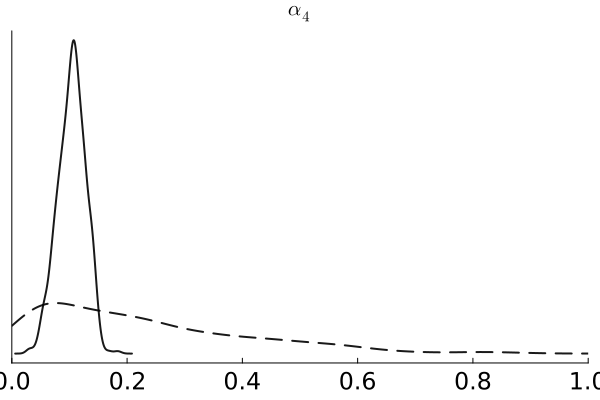}
		\end{subfigure}
		\figspaceAfterPlotArea

	\end{center}
	\caption{Marginal Prior- and Posterior-Distributions of $\alpha_l$, $l=1:p$\figspaceBetweenCaptionAndNotes
	\scriptsize {\em Notes:} 
	Each panel shows the marginal prior- (dashed line) and posterior-distribution (solid line) for $\alpha_l$, $l=1:p$, respectively.}
	\label{plot_app1_prpo_alphas}

\end{figure}

\begin{figure}[H]
	\captionsetup[subfigure]{justification=centering}
	\begin{center}
		\figspaceBeforePlotArea
		\begin{subfigure}[b]{\plotSizeQuadruple\textwidth}
				\centering
				\includegraphics[width=1\textwidth, clip]{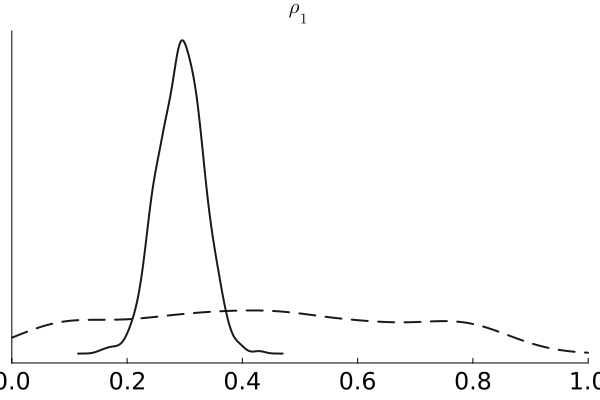}
		\end{subfigure}
		\begin{subfigure}[b]{\plotSizeQuadruple\textwidth}
				\centering
				\includegraphics[width=1\textwidth, clip]{\plotPath/plot_Post_rho1.png}
		\end{subfigure}
		\begin{subfigure}[b]{\plotSizeQuadruple\textwidth}
				\centering
				\includegraphics[width=1\textwidth, clip]{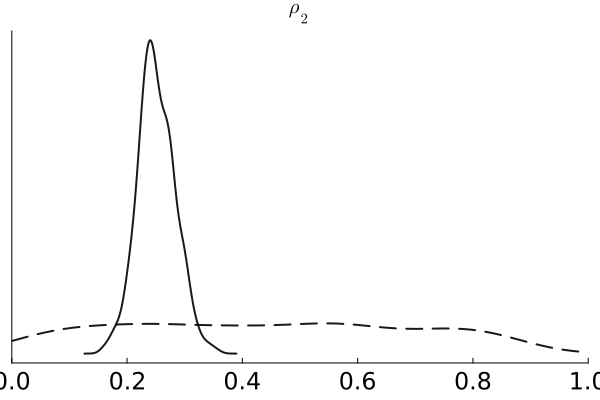}
		\end{subfigure}
		\begin{subfigure}[b]{\plotSizeQuadruple\textwidth}
				\centering
				\includegraphics[width=1\textwidth, clip]{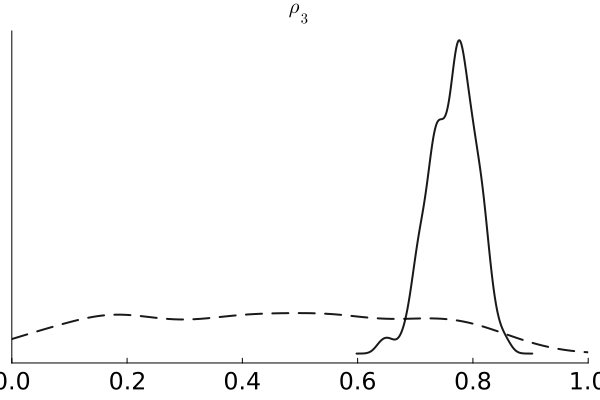}
		\end{subfigure}
		\figspaceAfterPlotArea

	\end{center}
	\caption{Marginal Prior- and Posterior-Distributions of $\rho_a$ and $\rho_i$, $i=1:3$\figspaceBetweenCaptionAndNotes
	\scriptsize {\em Notes:} 
	The first panel shows the marginal prior- (dashed line) and posterior-distribution (solid line) of $\rho_a$. The other panels do the same for $\rho_i$, $i=1:3$, respectively.}
	\label{plot_app1_prpo_rhos}

\end{figure}

\begin{figure}[H]
	\begin{center}
		\figspaceBeforePlotArea
		\begin{subfigure}[b]{\plotSizeDouble\textwidth}
				\centering
				\includegraphics[width=1\textwidth, clip]{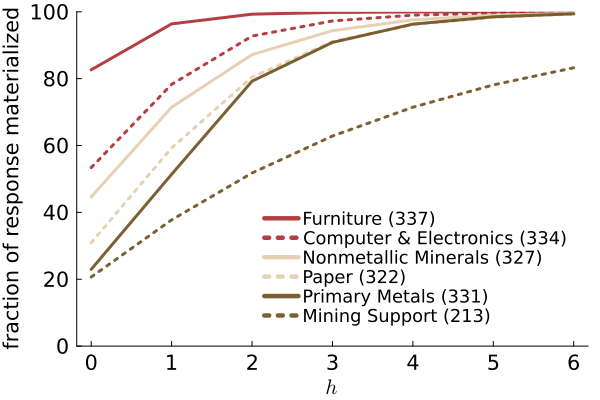}
		\end{subfigure}
		\begin{subfigure}[b]{\plotSizeDouble\textwidth}
				\centering
				\includegraphics[width=1\textwidth, clip]{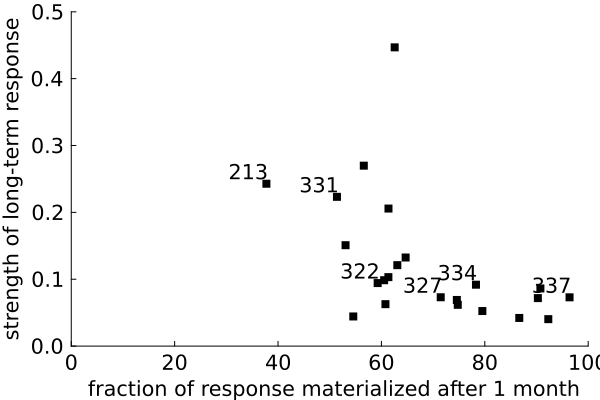}
		\end{subfigure}
		\figspaceAfterPlotArea
	\end{center}
\caption{Size and Timing of Aggregate Output Growth Responses to Sectoral TFP Shocks\figspaceBetweenCaptionAndNotes
\scriptsize {\em Notes:} The left panel shows the time profile of the effect of sectoral TFP shocks on aggregate IP for a few selected sectors. The right panel relates the strength of the effects to their timing. The shock sizes are equal to one standard deviation of the respective sectoral TFP shock.}
\label{plot_app1_aggIRFs_size_timing}

\end{figure}

\begin{figure}[H]
	\begin{center}
		\figspaceBeforePlotArea
		\begin{subfigure}[b]{\plotSizeDouble\textwidth}
			\subcaption*{\scriptsize{Oil and Gas Extraction}}
				\centering
				\includegraphics[width=1\textwidth, clip]{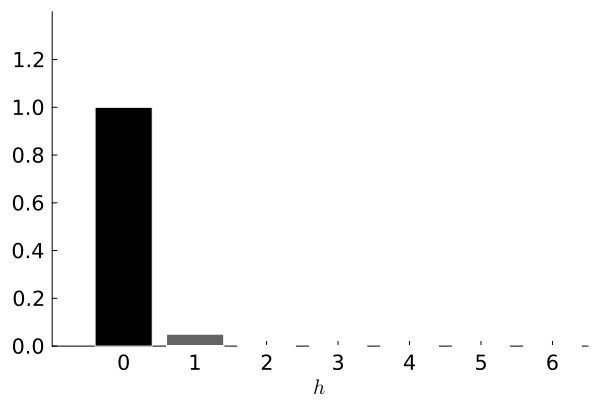}
		\end{subfigure}
		\begin{subfigure}[b]{\plotSizeDouble\textwidth}
			\subcaption*{\scriptsize{Machinery}}
				\centering
				\includegraphics[width=1\textwidth, clip]{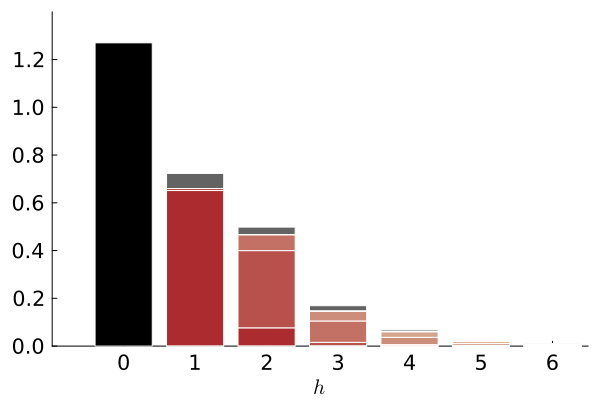}
		\end{subfigure}
		\figspaceAfterPlotArea
	\end{center}
	\caption{Impulse-Responses of Sectoral Output Growth to Common TFP Shocks\figspaceBetweenCaptionAndNotes
	\scriptsize {\em Notes:} The left panel shows the impulse-response of the production growth in \myquote{Oil and Gas Extraction} to a one standard deviation common TFP shock. The right panel does the same for the sector \myquote{Machinery}.}
	\label{plot_app1_sectorIRFsToAgg}

\end{figure}

